\pdfoutput=1
\documentclass[11pt,a4paper]{article}
\usepackage{geometry}
\usepackage[utf8]{inputenc}	
\usepackage{framed}
\usepackage[sort&compress,numbers,colon,merge]{natbib}

\usepackage[utf8]{inputenc}
\usepackage{caption}
\usepackage{subcaption}

\usepackage{amsmath}
\usepackage{amssymb}
\usepackage{amscd}
\usepackage{mathrsfs}
\usepackage{graphicx}
\usepackage{slashed}
\usepackage{mathbbol}
\usepackage{multirow}
\usepackage{array}

\numberwithin{equation}{section}

\newcommand{\FMDG}[1]{\includegraphics{#1}}

\newcommand{\braket}[1]{\ensuremath{\left\langle#1\right\rangle}}

\usepackage[colorlinks]{hyperref}
\hypersetup{
pdfauthor={Paul W. Angel, Yi Cai, Nicholas L. Rodd, Michael A. Schmidt, and Raymond R. Volkas},
pdftitle={Testable two-loop radiative neutrino mass model based on an LLQd^cQd^c effective operator}
}

\newcommand{\diag}{\mathop{\mathrm{diag}}}

\newcommand{\Eqref}[1]{Eq.~(\ref{#1})}

\newcommand{\Figref}[1]{Fig.~\ref{#1}}
\newcommand{\Tabref}[1]{Tab.~\ref{#1}}
\newcommand{\Secref}[1]{Sec.~\ref{#1}}
\newcommand{\Appref}[1]{Appendix \ref{#1}}

\newcommand{\eV}{\ensuremath{\,\mathrm{eV}}}

\newcommand{\GeV}{\ensuremath{\,\mathrm{GeV}}}
\newcommand{\TeV}{\ensuremath{\,\mathrm{TeV}}}
\newcommand{\ev}[1]{\ensuremath{\left\langle#1\right\rangle}}
\newcommand{\hc}{\ensuremath{\mathrm{h.c.}}}

\begin{document}

\allowdisplaybreaks[1]
\begin{titlepage}

\begin{center}
{\Large\textbf{Testable two-loop radiative neutrino mass model based on an $LLQd^cQd^c$ effective operator}}
\\[10mm]
{\large
Paul W. Angel\footnote{\texttt{pangel@student.unimelb.edu.au}},
Yi Cai\footnote{\texttt{yi.cai@unimelb.edu.au}},
Nicholas L. Rodd\footnote{\texttt{nrodd@student.unimelb.edu.au}},
Michael A.~Schmidt\footnote{\texttt{michael.schmidt@unimelb.edu.au}},
Raymond R. Volkas\footnote{\texttt{raymondv@unimelb.edu.au}}}
\\[5mm]
{\small\textit{ARC Centre of Excellence for Particle Physics at the Terascale,\\
School of Physics, The University of Melbourne, Victoria 3010, Australia\\
}}
\end{center}

\begin{abstract}
\noindent
A new two-loop radiative Majorana neutrino mass model is constructed from the gauge-invariant effective operator $L^i L^j Q^k d^c Q^l d^c \epsilon_{ik} \epsilon_{jl}$ that violates lepton number conservation by two units.  The ultraviolet completion features two scalar leptoquark flavors and a color-octet Majorana fermion.  We show that there exists a region of parameter space where the neutrino oscillation data can be fitted while simultaneously meeting flavor-violation and collider bounds.  The model is testable through lepton flavor-violating processes such as $\mu \to e \gamma$, $\mu \to eee$, and $\mu N \to e N$ conversion, as well as collider searches for the scalar leptoquarks and color-octet fermion. We computed and compiled a list of necessary Passarino-Veltman integrals up to boxes in the approximation of vanishing external momenta and made them available as a Mathematica package, denoted as \texttt{ANT}.
\end{abstract}

\end{titlepage}

\setcounter{footnote}{0}


\section{Introduction}
\label{sec:intro}
Uncovering the origin of neutrino masses and mixing angles is one of the most important problems in physics beyond the standard model (SM).  The minimal scenario is simply to add three right-handed neutrino fields to the SM and then write down all the new gauge invariant terms in the Lagrangian.  One class of new terms is a set of electroweak Yukawa couplings of the right-handed neutrinos to the lepton doublets and the Higgs doublet, which by themselves would lead to Dirac neutrino masses, putting neutrinos on par with the other quarks and leptons in terms of mass generation.  However, neutrinos have been observed to be extremely light, with cosmological constraints requiring the sum of their masses to be sub-eV.  While this can be accommodated by having suitably tiny neutrino Yukawa coupling constants, this ignores the likely truth that the anomalously small neutrino masses are an indication of a different origin.  The most obvious difference is revealed by recognizing the second class of new Lagrangian terms: the bare Majorana masses for the gauge-singlet right-handed neutrinos.  In the see-saw limit where these Majorana masses $M$ are far above the electroweak scale $v$, the neutral lepton sector breaks up into a sub-sector of three very light left-handed Majorana neutrinos of mass scale $v^2/M$ and a second sub-sector of three very massive neutral leptons at the scale $M$.

This type-I see-saw scenario is economical, elegant, compelling, and extremely difficult to test experimentally~\cite{Minkowski:1977sc,*Yanagida:1980,*Glashow:1979vf,*Gell-Mann:1980vs,*Mohapatra:1980ia}.\footnote{Note that there is some region of parameter space which might be testable at colliders. See e.g.~Ref.~\cite{Kersten:2007vk,Pilaftsis:1991ug,*Dev:2012sg}.}  The difficulties are that the favored scale for $M$ is far above the TeV scale being explored at the LHC, and even if the scale is brought down to a TeV the heavy neutral fermions have highly suppressed gauge interactions and thus are difficult to produce and detect.  The type-II~\cite{Magg:1980ut,*Schechter:1980gr,*Wetterich:1981bx,*Lazarides:1980nt,*Mohapatra:1980yp,Cheng:1980qt} and type-III~\cite{Foot:1988aq} see-saw variations are more testable because the required scalar triplet and fermion triplets, respectively, at least have electroweak gauge interactions.  But the favored see-saw scale is still far higher than a TeV.

Alternative neutrino mass generation schemes are worth exploring.  One motivation is that nature is not always minimal.  For example, when Pauli introduced the neutrino to solve the apparent energy conservation problem in beta decay, he also proposed that same particle to perform the role now known to be played by the neutron in nuclear structure.  While minimal models will always, justifiably, attract the greater attention, we should devote some effort to non-minimal models unless experimental results tell us not to bother.  On the origin of neutrino mass, experiment has yet to speak.  Now, many non-minimal neutrino mass models are also more testable than the see-saw models, so the new physics they predict should be searched for.  By constraining non-minimal models, or perhaps ruling some of them out, we increase the circumstantial evidence in favor of the simpler but less-testable schemes such as the type-I see-saw model.  Of course, we could also find to our surprise that one of the more involved theories is actually correct.

An important class of non-minimal theories are the radiative neutrino mass models. It can be arranged for nonzero neutrino masses to first arise at loop level rather than at tree level, and this may be a (partial) explanation for why neutrino masses are so small.  The first radiative neutrino mass models at one-loop level were proposed in Refs.~\cite{Zee:1980ai}, at two-loop order in Refs.~\cite{Cheng:1980qt,Zee:1985id,*Babu:1988ki} and with three loops in Ref.~\cite{Krauss:2002px}. 
Three-loop models with large couplings like the ones proposed in Ref.~\cite{Krauss:2002px,Aoki:2008av,*Aoki:2009vf,*Gustafsson:2012vj} as well as two-loop models with new colored states as proposed in Ref.~\cite{Dey:2008ht,*Babu:2010vp,*Babu:2011vb} are particularly interesting phenomenologically, because they promise to be easier to test experimentally. 
Several authors tried to combine neutrino mass generation with other unsolved problems of physics beyond the SM, e.g.\ the first papers simultaneously addressing dark matter were Refs.~\cite{Krauss:2002px,Ma:2006km,Cai:2011qr}, and for baryogenesis they were Refs.~\cite{Ma:2006fn,*Hambye:2006zn,Babu:2007sm}. The proposed models vastly differ in their complexity as well as predictive power. 
In recent years, several groups systematically studied radiative neutrino mass generation: e.g.\ one-loop radiative neutrino mass models~\cite{Ma:1998dn,*Bonnet:2012kz}, simple models with only two new particles~\cite{FileviezPerez:2009ud,*Law:2013dya}, classes of models with a discrete symmetry and a DM candidate~\cite{Farzan:2012ev,*Law:2013saa}, the use of effective operators of the type $LLHH(H^\dagger H)^n$~\cite{Bonnet:2009ej,*Bonnet:2012kh} (where $L$ is a lepton doublet and $H$ is the Higgs doublet) as well as a general classification in terms of $\Delta L=2$ operators~\cite{Babu:2001ex,deGouvea:2007xp,Angel:2012ug}.
Radiative neutrino mass generation is certainly a logical alternative to the see-saw procedure, and should be thoroughly explored. 

The purpose of this paper is not only to propose another testable radiative neutrino mass model at two-loop order, but also to exemplify that the new physics due to the exotic particles and their interactions cannot be arbitrarily weak.
We demonstrate that there is a region of parameter space where the neutrino oscillation data can be accommodated while simultaneously complying with constraints from the null observations of rare flavor-changing processes and the absence of exotic particles in collider searches.  In this specific model, the exotic particles in the theory are two copies of a certain scalar leptoquark as well as a color-octet Majorana fermion.  When these heavy exotics are integrated out, a $\Delta L = 2$ effective operator with flavor content $LLQd^cQd^c$ is generated.  The model can be tested by performing high precision searches for lepton flavor-violating processes such as $\mu \to e \gamma$, $\mu \to eee$, and $\mu N \to e N$ conversion, as well as through collider searches for the leptoquarks and the colored exotic fermion, while the requirement to generate the correct neutrino oscillation parameters (in particular, one neutrino must be heavier than the ``atmospheric'' lower bound of $0.05$ eV~\cite{Fukuda:1998mi}) imposes an upper bound on the new physics scale.

The next section defines the new model and places it in the context of the ``theory space'' of radiative neutrino mass models, 
using underlying effective operators as the structuring principle.  
Section~\ref{sec:constraints} then computes the neutrino masses and mixing angles. 
Constraints from flavor-violating processes induced by the couplings required by the generation of neutrino mass are examined in \Secref{sec:flavor1}, while the remaining ones and neutrinoless double beta-decay, which both give no constraints to the parameters relevant to the generation of neutrino mass,  are discussed in \Secref{sec:flavor2}.
Then collider constraints are discussed in \Secref{sec:collider}. Finally, we discuss naturalness in the context of this model and give the preferred region of parameter space in \Secref{sec:naturalness}.  A publicly available Mathematica code, denoted as \texttt{ANT}\footnote{It can be downloaded from \href{http://ant.hepforge.org}{http://ant.hepforge.org}.}, for calculating relevant loop integrals is described in \Secref{sec:mathematica}.
In \Secref{sec:conc}, we summarize all the constraints and draw the conclusions. 


\section{The Model}
\label{sec:model}

A very useful organizing principle for radiative Majorana neutrino mass models is the effective operator analysis pioneered by Babu and Leung (BL)~\cite{Babu:2001ex}, and followed-up by de Gouv\^{e}a and Jenkins (GJ)~\cite{deGouvea:2007xp}.  These operators are invariant under the SM gauge group and respect baryon-number conservation, but violate lepton-number conservation by two units.  They are constructed from the SM quark and lepton multiplets (with no right-handed neutrinos) and a single Higgs doublet.  

Such operators exist at odd mass-dimension.  At dimension-5, there is a unique operator (up to family replication): the Weinberg operator, denoted ${\cal O}_1 \equiv LLHH$ in an efficient notation.  The $LL$ structure is shorthand for  $\overline{(L_L)^c} L_L$.  This non-renormalizable operator may be ``opened up" -- derived from an underlying renormalizable or ultra-violet (UV) complete theory -- in three minimal ways at tree-level.  These three possibilities are precisely the type-I, -II and -III see-saw models.  The advantage of the effective operator perspective is that it encourages you to systematically construct all sensible UV completions, so no possibilities are missed.

This technique can be extended to radiative neutrino mass models.  All the $\Delta L = 2$ effective operators of dimension-7 and higher that are not of the form ${\cal O}_1 (\overline{H} H)^n$ necessarily contain some fields different from left-handed neutrinos and neutral Higgs bosons.  To turn such operators into self-energy diagrams for neutrinos, those other particles have to be closed off through loops.  Therefore the study of how such operators can generate neutrino masses is the study of radiative neutrino mass models.  The see-saw models exist at the tree-level end of an extensive family of Majorana neutrino-mass models founded upon SM $\Delta L = 2$ operators.  Some of the present authors used this approach in~\cite{Angel:2012ug} to systematically classify the neutrino self-energy diagram topologies and the exotic scalars and fermions they contain.  That work builds on the analyses of BL and GJ in providing a guide to the construction of all radiative neutrino mass models that obey certain conditions:  the gauge group is that of the SM and no larger, the left-handed neutrinos are Majorana, right-handed neutrinos are absent, there is a single Higgs doublet, and the exotic particles are scalars and fermions only.

In this paper we use the foundations just described to construct a new 2-loop neutrino-mass model based on the operator
\begin{equation}
{\cal O}_{11b} \equiv L^i L^j Q^k d^c Q^l d^c \epsilon_{ik} \epsilon_{jl}\ ,
\end{equation}
where we adopt the notation of~\cite{Babu:2001ex,deGouvea:2007xp,Angel:2012ug}.  The label 11b tells us that this operator is number 11 in the BL list, and the $i,j,k,l=1,2$ SU(2) index structure is of type b.  Written with explicit Lorentz structure it is
\begin{equation}
{\cal O}_{11b} \equiv \overline{(L_L)^{ci}} L_L^j\, \overline{(Q_L)^{ck}} (d_R)^c\, \overline{(Q_L)^{cl}} (d_R)^c\, \epsilon_{ik}\, \epsilon_{jl} = \overline{(L_L)^{ci}} L_L^j\, \overline{d}_R Q_L^k\, \overline{d}_R Q_L^l \, \epsilon_{ik}\, \epsilon_{jl}\ .
\end{equation}
Of course it is understood that there is a set of such operators because of the family structure of quarks and leptons.

We now use the procedure of~\cite{Angel:2012ug} to construct an underlying theory.  Table IV of~\cite{Angel:2012ug} tells us to look at Fig.~10 for completions involving scalars only, and Figs.~14 B-D for completions using both scalars and fermions.  We choose the latter option, partly because the existing well-studied radiative neutrino mass models use exotic scalars exclusively, and so it is interesting to examine a different possibility.  The text discussing Fig.~14 then informs us that diagrams B, C and D are applicable to exotic vector-like Dirac fermions, while only the diagrams in D allow the fermion to be Majorana.  We choose the Majorana option, which limits us to diagram D1 and D2, from which we select D2.  Making the only allowed identifications of the fermion lines with the fields in ${\cal O}_{11b}$, we arrive at almost a unique model.  The remaining choices are the weak isospin assignments of the scalar and Majorana fermion, and the color of the Majorana fermion.

We choose the exotic scalar and fermion multiplets to be, respectively,
\begin{equation}
\phi \sim (3^*,1, 1/3), \quad f_L \sim (8,1,0)\ ,
\end{equation}
where the first entry is color, the second weak isospin and the third hypercharge (normalized so that electric charge $Q = I_L + Y$).  
We shall see shortly that two copies of $\phi$ are required to produce two neutrino mass eigenvalues of appropriate magnitude, while one
copy of $f$ suffices. In \Secref{sec:constraints} we shall see that the model with three copies of $\phi$, which produces three neutrino masses, is disfavored by constraints from flavor physics.  

The Yukawa couplings of these exotics to SM fields and the various bare masses are given in
\begin{subequations}
\label{eq:exotic-Lag}
\begin{align}
-\Delta {\cal L} =  &\left( \lambda^{LQ}_{ij\alpha}\, \overline{L}^c_i \, Q_j \, \phi_\alpha 
+ \lambda^{df}_{i\alpha}\, \overline{d}_i \, f \, \phi_\alpha^*
 + \frac{1}{2}\, m_f \, \overline{f}^c f + H.c. \right) 
+ m_{\phi_\alpha}^2\, \phi^\dagger_\alpha\, \phi_\alpha
\label{eq:numassrequired}\\
&- \left( \lambda^{eu}_{ij\alpha}\, \overline{e}^c_i \, u_j \, \phi_\alpha 
+H.c. \right)\ ,
\label{eq:numassnotrequired}\end{align}
\end{subequations}
where $i,j = 1,2,3$ are quark-lepton family indices, $\alpha = a,b$ denotes the two copies of $\phi$, and chirality labels and gauge indices have been suppressed to reduce clutter.  Note that the $\phi$'s are scalar leptoquarks, and the choice of isospin singlet over triplet is made for simplicity.  The fermion $f$ can be either a color singlet or octet, with the latter chosen to prevent it having the quantum numbers of a right-handed neutrino.  The bare Majorana mass $m_f$ must be nonzero for this Lagrangian to be $\Delta L = 2$ and thus capable of inducing Majorana masses for the neutrinos. 

Besides the SM gauge symmetry group, we have to demand baryon-number conservation, in order to forbid the operators
$\lambda^{QQ}_{ij\alpha}\, \overline{Q}_i \, Q_j^c \, \phi_\alpha$ and $\lambda^{ud}_{ij\alpha}\, \overline{u}_i \, d_j^c \, \phi_\alpha$,
which induce proton decay, as discussed in e.g.\ Refs.~\cite{Bowes:1996xy,Baldes:2011mh}. Applying the discussion in~\cite{Baldes:2011mh}, we find the following estimates for proton decay.
If $\phi_\alpha$ couples to the first generation of quarks directly via the couplings $\lambda^{ud}$, there is a tree-level contribution to the proton decay channel $p\to\pi^0 e^+$. The decay rate can be estimated to be 
\begin{equation}
\Gamma\sim\mathcal{O}\left(\frac{|\lambda^{ud}_{11\alpha}\lambda_{11\alpha}^{LQ,eu}|^2 M_p^5}{m_{\phi_\alpha}^4}\right)\ ,
\end{equation}
which leads to a strong upper bound on the product of the couplings to the first generation of quarks $|\lambda^{ud}_{11\alpha}\lambda_{11\alpha}|\lesssim \left(m_{\phi_\alpha}/10^{16}\GeV\right)^2$.
If $\phi_\alpha$ does not couple to the first generation directly, there is a loop-level contribution to the proton decay channel $p\to K^+ \bar\nu$ leading to a partial decay width
\begin{equation}
\Gamma\sim\sum_m\frac{|\lambda^{ud}_{33\alpha}\lambda^{LQ,eu}_{m3\alpha}|^2 M_p^5 g^8}{m_{\phi_\alpha}^4}\left(\frac{|V_{ub}V_{td}V_{ts}|M_b M_{\ell_m}}{M_t^2}\ln\left(\frac{M_t}{M_b}\right)\ln\left(\frac{M_t}{M_{\ell_m}}\right)\right)^2\ ,
\end{equation}
where $g$ denotes the electroweak gauge coupling constant, $M_b$ ($M_t$) the bottom (top) mass and $M_{\ell_m}$ the mass of the charged lepton $\ell_m$. This translates into a strong upper bound on $\lambda^{ud}_{33\alpha}$. Similar bounds can be derived for the couplings $\lambda^{QQ}_{ij\alpha}$. Therefore, it makes sense to forbid these couplings by imposing baryon-number conservation.

In the following, we will perform all calculations in the full theory for simplicity using dimensional regularization in the $\overline{\text{MS}}$ scheme.\footnote{As there is no manifest decoupling, a more precise treatment would require an explicit decoupling of heavy particles at their given mass thresholds in order to be able to resum the logarithms. See \Appref{app:decoupling} for some comments on the treatment of neutrino masses in this approach.}


\section{Neutrino Masses}
\label{sec:constraints}

\begin{figure}
\centering
\FMDG{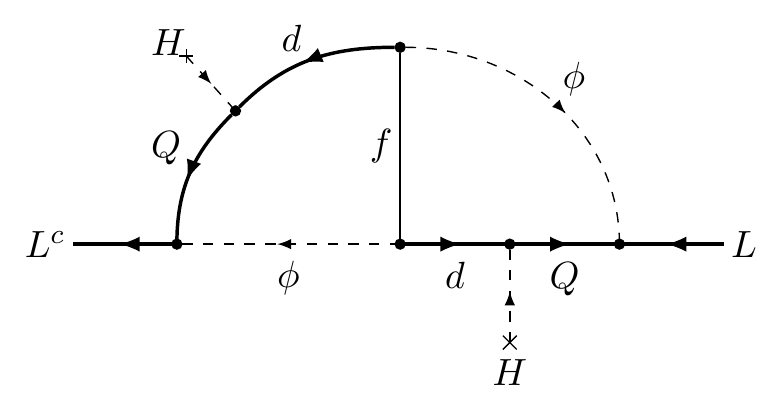}
\begin{minipage}{12cm}
\caption{The neutrino self-energy Feynman diagram. Note that the color-octet fermion $f$ is Majorana and therefore we do not add an arrow to the line.\label{fig:selfenergy}}
\end{minipage}
\end{figure}

The neutrino self-energy Feynman diagram depicted in Fig.~\ref{fig:selfenergy} leads to the Majorana neutrino mass matrix
\begin{equation}
(M_\nu)_{ij}  = 4 \frac{m_f}{(2\pi)^8} \sum_{k,l=1}^3\sum_{r,s=1}^3\sum_{\alpha,\beta=1}^{N_\phi} 
\left(\lambda_{ik\alpha}^{LQ} \lambda_{l\alpha}^{df} V_{kr}\right)
\left( m_{d_r}  I_{rs\alpha\beta}  m_{d_s}\right)
\left(\lambda_{jl\beta}^{LQ} \lambda_{k\beta}^{df}  V_{ls}\right)\ ,
\label{eq:Mnu-exact}
\end{equation}
where 4 is a color factor, $N_\phi$ is the number of leptoquarks $\phi_\alpha$, the integral $I_{ij\alpha\beta}$ is defined as
\begin{equation}
I_{ij\alpha\beta}\equiv \int d^4 p \int d^4 q \frac{1}{p^2 - m_{d_i}^2} \frac{1}{p^2-m_{\phi_\alpha}^2}
\frac{1}{q^2-m_{d_j}^2}\frac{1}{q^2-m_{\phi_\beta}^2}\frac{1}{(p-q)^2-m_f^2}\ ,
\label{eq:loopint-exact}
\end{equation}
and $m_{d_i}$ denote the $d$-, $s$- and $b$-quark masses for $i=1,2,3$, respectively.  Note that this is the exact loop integral, using the full propagators for the massive fermions, with the fermion mass factors in the numerator of Eq.~\ref{eq:Mnu-exact} arising from the chiral projection operators at the vertices.  The matrix $M_\nu$ corresponds to the effective term $\overline{(\nu_L)^c} M_\nu \nu_L + \overline{\nu}_L M^*_\nu (\nu_L)^c$.

For what follows it will be convenient to rewrite Eq.~\ref{eq:loopint-exact} in terms of the dimensionless parameters 
\begin{equation}
r_i \equiv \frac{m_{d_i}^2}{m_f^2}\quad\quad\mathrm{and}\quad \quad
t_\alpha \equiv \frac{m_{\phi_\alpha}^2}{m_f^2}\;.
\label{eq:deft}
\end{equation}
Specifically, factoring out $m_f^2$ and rescaling the momenta we have:
\begin{equation}
I_{ij\alpha\beta} = \frac{1}{m_f^2} \int d^4 p \int d^4 q \frac{1}{p^2 - r_i} \frac{1}{p^2-t_{\alpha}}
\frac{1}{q^2-r_j}\frac{1}{q^2-t_{\beta}}\frac{1}{(p-q)^2-1}\ .
\label{eq:loopint-exact2}
\end{equation}
In App.~\ref{sec:appint} we exactly evaluate this integral in general, and in the situation where the quark masses are much smaller than the leptoquark $\phi$ and color-octet fermion $f$ masses -- specifically where $r_{i/j} \to 0$.

We now make some useful approximations.  First, since the quark masses are much smaller than those of the leptoquark $\phi$ and color-octet fermion $f$, they may be neglected in the denominator.  This is equivalent to treating the internal quark lines in the mass-insertion approximation.  As mentioned, this case is evaluated in App.~\ref{sec:appint} and is denoted 
\begin{equation}
I_{\alpha\beta} \equiv \lim_{r_{i/j}\to 0} I_{ij\alpha\beta}= \frac{1}{m_f^2} \int d^4 p \int d^4 q \frac{1}{p^2q^2} \frac{1}{p^2-t_{\alpha}}
\frac{1}{q^2-t_{\beta}}\frac{1}{(p-q)^2-1}\ .
\label{eq:loopint-app2}
\end{equation}
Also, we only consider the parameter space region where the $b$-quark mass dominates the numerator, so that the $d$- and $s$-quark masses may be put to zero.  This requires the Yukawa couplings $\lambda^{LQ}\lambda^{df}$ to not be strongly hierarchical in the sense of being able to compensate for the hierarchy in the quark masses.  So, the formula for the neutrino mass matrix simplifies to
\begin{equation}
(M_\nu)_{ij}  \simeq 4 \frac{m_f m_b^2 V_{tb}^2}{(2\pi)^8} \sum_{\alpha,\beta=1}^{N_\phi}
\left(\lambda_{i3\alpha}^{LQ} \lambda_{3\alpha}^{df} \right)
\left( I_{\alpha\beta} \right)
\left(\lambda_{j3\beta}^{LQ} \lambda_{3\beta}^{df}  \right)\ .
\label{eq:Mnu-approx}
\end{equation}
Equation~\ref{eq:Mnu-approx} may be simply re-expressed in matrix notation as
\begin{equation}
M_{\nu} \simeq {\rm const.} \times \Lambda I \Lambda^T\ ,
\label{eq:Mnu-approx-alt}
\end{equation}
where
\begin{equation}
\Lambda_{i\alpha} \equiv \lambda_{i3\alpha}^{LQ} \lambda_{3\alpha}^{df}\quad\quad\mathrm{and}\quad\quad I\equiv \left(I_{\alpha\beta}\right)\;.
\label{eq:Lambda-defn}
\end{equation}
If there were only one leptoquark flavor, $I$ would be just a number and $\Lambda$ a $3 \times 1$ column vector. The neutrino mass matrix would then be of the form of an outer product of a column vector with its transpose, and hence of rank one.  With the $d$- and $s$-quark masses switched back on, the exact neutrino mass matrix of Eq.~\ref{eq:Mnu-exact} is (in general) rank-three, but the two smallest eigenvalues are generally too small to fit the neutrino oscillation data.\footnote{As stated earlier, we do not consider the unnatural possibility of a strong hierarchy in the $\Lambda$ values that just happens to offset $m_d \ll m_s \ll m_b$.}   The necessity of two sufficiently large neutrino mass eigenvalues therefore requires two leptoquark flavors to exist.  In that case the third eigenvalue is extremely small and for all practical purposes can be set to zero. Because we choose to adopt the natural parameter space regime where the bottom-quark mass dominates the neutrino mass-matrix formula, from now on we assume that there are only couplings to the third generation in the flavor basis, with the only mixing coming from the SM down-type Yukawa couplings. For most calculations the CKM induced couplings to the first two generations of down-type quarks can be neglected. Hence, we do not discuss the effects of the CKM mixing except in \Secref{sec:flavor2}. Given that the smallest neutrino mass almost vanishes, we know that the largest of the neutrino masses is just the square root of the ``atmospheric'' squared mass difference, namely about 0.05 eV.  The new physics in our model must be sufficiently strong to produce this eigenvalue.

It is convenient to solve Eq.~\ref{eq:Mnu-approx} for the $3 \times 2$ matrix $\Lambda$, which can be easily done through a Casas-Ibarra procedure~\cite{Casas:2001sr}.  One obtains
\begin{equation}
\Lambda_{i\alpha} = 
\sum_{j,k=1}^3\sum_{\beta=1}^2
\frac{(2\pi)^4}{2V_{tb}m_b\sqrt{m_f} } \left(V^*_{\nu} \right)_{ij} \left(\hat{M}_\nu^\frac{1}{2}\right)_{jk} O_{k\beta}   \left( \hat I^{-\frac{1}{2}} S\right)_{\beta\alpha}\ ,
\label{eq:CasasIbarra}
\end{equation}
where $V_{\nu}$ and $S$ diagonalize the neutrino mass matrix,
\begin{equation}
\hat{M}_\nu = V^T_\nu M_\nu V_\nu\ ,
\label{eq:VPMNS}
\end{equation}
and the matrix of integrals $I$, 
\begin{equation}
\hat I = S^T I S\ ,
\end{equation}
respectively, with $\hat I$ being a real and positive diagonal matrix.
The leptonic mixing or PMNS matrix is defined by $V_{PMNS} = V_e^{\dagger} V_\nu$.  In the case of a normal [inverted] mass ordering, the first [third] neutrino is massless, i.e. the (1,1) [(3,3)] element of $\hat{M}_\nu$ vanishes 
and the 2-3 [1-2] sub-block of the matrix $O$ is given by a general complex orthogonal matrix, while the first [third] row of $O$ is arbitrary. 
In order to understand the flavor structure of the neutrino mass matrix as well as the bounds on the leptoquark masses in more detail, we have to study the flavor structure of the matrix of integrals $I$, which demands a hierarchy in $\Lambda$ in order to explain a small hierarchy in the neutrino masses.

\mathversion{bold}
\subsection{Flavor Structure of the Matrix of Integrals $I$}
\mathversion{normal}
\label{sec:Iflavor}

The main feature of $I$ we wish to demonstrate here is that the matrix develops a hierarchy in its eigenvalues.  We will discuss the case with two leptoquarks in detail and generalize it in the end to an arbitrary number of leptoquarks. This hierarchy emerges for a wide range of values of $t_1$ and $t_2$ (where $1$ and $2$ distinguish the two leptoquarks as defined in \Eqref{eq:deft}), however it does not have a uniform origin.  Firstly if one of the leptoquarks and the octet fermion are much heavier than the other leptoquark, without loss of generality we can take $t_1 \to 0$ and leave $t_2$ constant.  One can check using the analytic expression in App.~\ref{sec:appint} that both $I_{11}$ and $I_{12}=I_{21}$ diverge, the former doing so faster, whilst $I_{22}$ remains constant.  Accordingly the matrix $I$ becomes singular and thus a hierarchy develops in its eigenvalues.
The origin of the hierarchy in this limit is relatively straightforward.  Sending $t_1 \to 0$ is equivalent to sending the corresponding leptoquark mass to the mass of the down-type quark that appears in the same loop.  In the limit that the two masses are equal a Landau singularity appears and the amplitude associated with the diagram becomes IR divergent.  For definiteness, we give an explicit description assuming that $t_1\ll t_2$ and therefore $I_{11}\gg I_{12}\gg I_{22}$:
\begin{equation}
I\approx I_{11}\begin{pmatrix}
1 & \epsilon\\
. &  0
\end{pmatrix}\qquad\mathrm{with}\qquad \epsilon\equiv\frac{I_{12}}{I_{11}}
\end{equation}
neglecting the small $I_{22}$ and we obtain for the inverse square root of the matrix $I$,
\begin{equation}
I^{-\frac12}\equiv \hat I^{-\frac12} S=
I_{11}^{-\frac12}
\left(
\begin{array}{cc}
 -i  & \frac{i}{\epsilon} \\
 1 & \epsilon \\
\end{array}
\right)\ .
\end{equation}
Through Eq.~\ref{eq:CasasIbarra}, this expression feeds into the determination of the required Yukawa couplings $\Lambda$.  Note that the flavor structure above implies that the mixing from the matrix $I$ is small, so the eventual neutrino mixing is mainly determined by the Yukawa couplings in $\Lambda$. 

At the other end of parameter space, consider the case when $t_1 \simeq t_2$.  It is easy to check using the analytic expressions in App.~\ref{sec:appint} that $I_{\alpha\beta}$ shows a weak dependence on the leptoquark masses.  In particular,
\begin{equation}
\left|\frac{t_1-t_2}{t_2}\left[\frac{\partial I_{12}}{\partial t_1} \right]_{t_1=t_2}\right|\ll \left|I_{22}\right|\ .
\end{equation}
As the integral is symmetric in the leptoquark masses, we can expand $I$ as follows
\begin{equation}
I\approx I_{22}
\begin{pmatrix}
1+2 \epsilon & 1+\epsilon\\
. & 1\\
\end{pmatrix}\qquad\mathrm{with}\qquad \epsilon\equiv\frac{t_1-t_2}{t_2}\frac{\left.\frac{\partial I_{12}}{\partial t_1}\right|_{t_1=t_2}}{I_{22}}\ .
\end{equation}
Expanding the ratio of the eigenvalues results in a quadratic hierarchy in $\epsilon$, given by
$
{\epsilon^2}/{4}+\mathcal{O}(\epsilon^3)
$.
This shows that there is a hierarchy in the matrix $I$ for $t_1\simeq t_2$ as well. Using this expansion, we obtain a simple expression for the square root of the matrix $I$,
\begin{equation}
I^{-\frac12}\equiv\hat I^{-\frac{1}{2}} S=I_{22}^{-\frac12}\begin{pmatrix}
-\frac{i}{|\epsilon|}-\frac{i \left|\epsilon\right|}{8} & \frac{i}{\left|\epsilon\right|}-i-\frac{3\,i \left|\epsilon\right|}{8}\\
\frac12&\frac12\left(1+\left|\epsilon\right|\right)\\
\end{pmatrix}+\mathcal{O}(\epsilon^2)\ .
\end{equation}
Note that the leading contribution, which sets the scale of the couplings $\Lambda$, is entirely determined by the first row of $I^{-\frac12}$.

\begin{figure}\centering
\includegraphics[width=8cm]{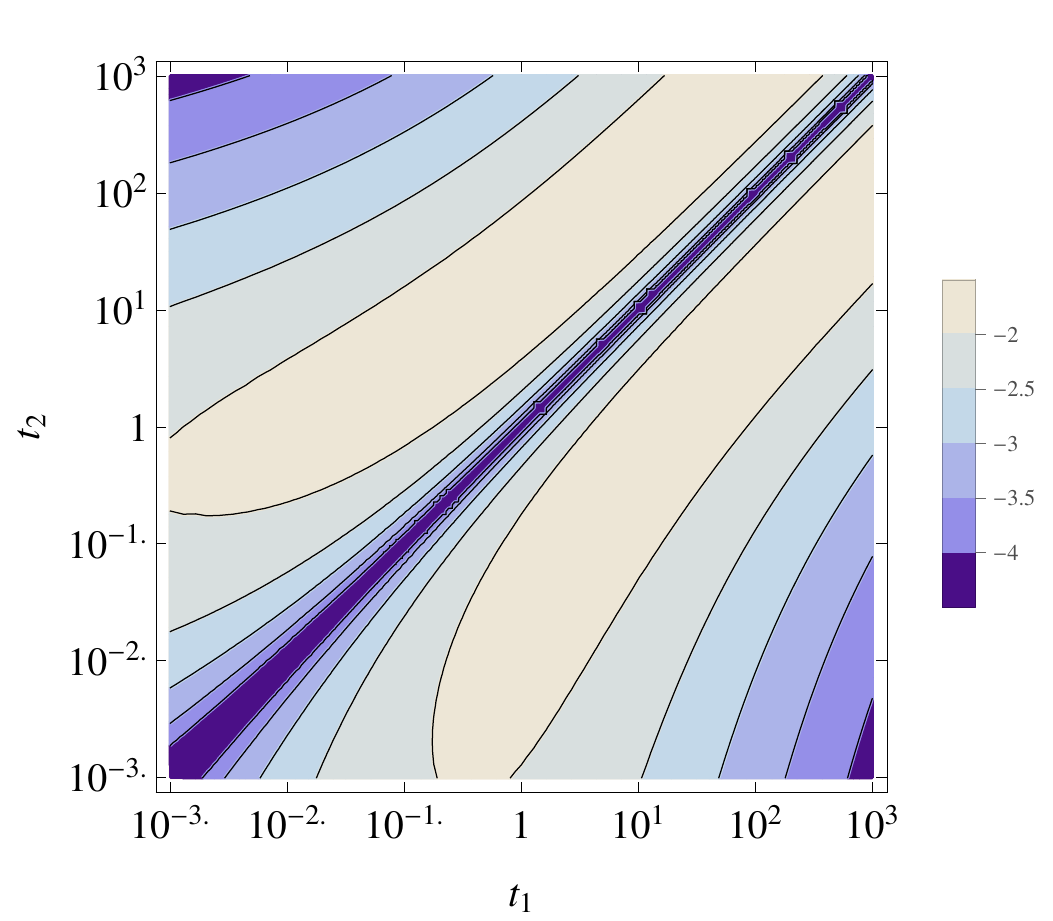}
\begin{minipage}{10cm}
\caption{Contour plot of the hierarchy $\log_{10}\lambda_1/\lambda_2$ in the eigenvalues $\lambda_{1,2}$ of the matrix $I$ as a function of $t_1$ and $t_2$ on a logarithmic scale. \label{fig:hierarchyPlot}}
\end{minipage}
\end{figure}

Having understood these two limiting cases, we now draw our attention to the general case, which we analyze numerically. As the absolute mass scale of the integral can be factored out, we see that the hierarchy of the eigenvalues can only depend on the two ratios of masses $t_\alpha$. Hence, we show the hierarchy of the eigenvalues in the contour plot in \Figref{fig:hierarchyPlot}. 
The mixing angle of the integral matrix is maximal ($\pm \pi/2$) for the degenerate case, vanishes for the hierarchical case and can be interpolated between those two limiting cases.

The generalization to more generations of leptoquarks is straightforward by grouping the leptoquark masses with a similar mass scale. There is a large hierarchy between the blocks with similar masses due to the mentioned Landau singularity. Additionally, there is a hierarchy between the integrals involving similar leptoquark masses as discussed for the degenerate leptoquark mass case.
So if there are three copies of leptoquarks, large hierarchies among the Yukawa couplings  $\Lambda=\lambda^{LQ}\lambda^{df}$ are required to 
compensate for large hierarchies among the three eigenvalues of the matrix of integral $I$, which then induces flavor-changing processes that 
exceed the current limits.

\subsection{Understanding the Neutrino Flavor Structure}
\label{sec:nuflavor}

We can use the approximations from the previous subsection and obtain analytic expressions for the leading order contribution to the neutrino mass matrix. 

\subsubsection{Normal Mass Ordering}
As mentioned above, 
the $2-3$ block of the matrix $O$ is given by a complex orthogonal matrix parameterized by the complex angle $\theta$:
\begin{equation}
O=\begin{pmatrix}
\cos\theta & -\sin\theta\\
\sin\theta & \cos\theta\\
\end{pmatrix}\ .
\end{equation}
Consider the expression $\left(\hat M_\nu^{\frac12}\right)_{jk}O_{k\beta} \left(I^{-\frac12}\right)_{\beta\alpha}$ which appears in Eq.~\ref{eq:CasasIbarra}. In the case of a strong normal hierarchy with $m_1=0$, the first row vanishes and the resulting leading order approximation to the columns of $\Lambda$ are given by a linear combination of the second and third column of $V_\nu^*$.

We discuss the different limits in turn starting with the strongly hierarchical leptoquark masses, namely $t_1\ll t_2$.
Here, the leading order contribution is given by
\begin{equation}
\left(\hat M_\nu^{\frac12}\right)_{jk}O_{k\beta} \left(I^{-\frac12}\right)_{\beta\alpha}\simeq
I_{11}^{-\frac12}\begin{pmatrix}
 0 & 0 \\
 -i \sqrt{m_2}\, e^{-i \theta }  & i\frac{\sqrt{m_2}}{\epsilon}\left(\cos\theta +i \epsilon^2  \sin\theta\right) \\
\sqrt{m_3}\, e^{-i \theta }  & \frac{\sqrt{m_3}}{\epsilon} \left(\epsilon^2  \cos\theta +i \sin \theta \right) \\\end{pmatrix}\ .
\end{equation}

Similarly, in the case of degenerate leptoquark masses, $t_1\simeq t_2$, the leading order contribution results in 
\begin{equation}
\left(\hat M_\nu^{\frac12}\right)_{jk}O_{k\beta} \left(I^{-\frac12}\right)_{\beta\alpha}\simeq
I_{22}^{-\frac12}\begin{pmatrix}
 0 & 0 \\
 -\frac{i \sqrt{m_2} \cos \theta }{\left|\epsilon\right|} & \frac{i \sqrt{m_2} \cos \theta }{\left|\epsilon\right|} \\
 -\frac{i \sqrt{m_3} \sin \theta }{\left|\epsilon\right|} & \frac{i \sqrt{m_3} \sin \theta }{\left|\epsilon\right|} \\
\end{pmatrix}\ .
\end{equation}
Note that a non-vanishing mixing angle $\theta$ can lead to substantially larger couplings because $m_2^2=\Delta m_{21}^2\ll \Delta m_{31}^2=m_3^2$. 

\subsubsection{Inverted Mass Ordering}
In the case of inverted mass ordering, the $1-2$ block of the matrix $O$ is given by a complex orthogonal matrix.
Following the discussion for normal mass ordering, we consider the expression $\left(\hat M_\nu^{\frac12}\right)_{jk}O_{k\beta} \left(I^{-\frac12}\right)_{\beta\alpha}$. In the case of a strong inverted hierarchy with $m_3=0$, the third row vanishes and the resulting leading order approximation to the columns of $\Lambda$ are given by a linear combination of the first and second columns of $V_\nu^*$.

In the limit of a large hierarchy among the leptoquark masses, $t_1\ll t_2$, the leading order contribution is given by
\begin{equation}
\left(\hat M_\nu^{\frac12}\right)_{jk}O_{k\beta} \left(I^{-\frac12}\right)_{\beta\alpha}\simeq
I_{11}^{-\frac12}\begin{pmatrix}
 -i\,\sqrt{m_1}\, e^{-i \theta }  & \frac{i \sqrt{m_1}}{\epsilon }\left( \cos\theta +i \epsilon^2  \sin \theta \right) \\
\sqrt{m_2}\, e^{-i \theta }  & \frac{\sqrt{m_2}}{\epsilon}\left( \epsilon^2  \cos \theta +i \sin \theta\right) \\
 0 & 0 \\
\end{pmatrix}\ .
\end{equation}
In the case of quasi-degenerate leptoquark masses, the expression becomes
\begin{equation}
\left(\hat M_\nu^{\frac12}\right)_{jk}O_{k\beta} \left(I^{-\frac12}\right)_{\beta\alpha}\simeq
I_{22}^{-\frac12}\begin{pmatrix}
-\frac{i \sqrt{m_1} \cos (\theta )}{\left|\epsilon\right|} & \frac{i \sqrt{m_1} \cos (\theta )}{\left|\epsilon\right|} \\
 -\frac{i \sqrt{m_2} \sin (\theta )}{\left|\epsilon\right|} & \frac{i \sqrt{m_2} \sin (\theta )}{\left|\epsilon\right|} \\
 0 & 0 \\
\end{pmatrix}\ .
\end{equation}



\section{Constraints from Flavor Physics Relevant to Neutrino Mass Generation}
\label{sec:flavor1}

The Lagrangian in \Eqref{eq:exotic-Lag} violates the family lepton numbers and the total lepton number explicitly.  Thus we expect rare processes, such as $\mu \to e \gamma$, to place limits on the model parameters.  Many of these processes proceed through one loop Feynman diagrams and the associated amplitudes can be reduced into Passarino-Veltman integrals \cite{Passarino:1979jh}. For the convenience of the reader and for future reference, we
have compiled our results in \Appref{sec:appPV} and made a Mathematica package denoted as \texttt{ANT} publicly available, which is briefly introduced in \Secref{sec:mathematica}. 

In this section, we restrict ourselves to the contributions which are required by the generation of neutrino mass, i.e.\ those in \Eqref{eq:numassrequired}. 
For the coupling $\lambda^{eu}$ in \Eqref{eq:numassnotrequired}, which does not enter the neutrino mass formula, we only give an estimation of the constraints in \Secref{sec:flavor2}.
However, we compiled some of the most relevant full expressions in \Appref{app:flavor}. We extended the FeynRules~\cite{Christensen:2008py} SM implementation to include the new particles and used FeynArts~\cite{Hahn:2000kx} as well as FormCalc~\cite{Hahn:1998yk} to obtain analytical results for the different processes.

In this work, we fix the mixing angles and mass-squared differences to the experimental best fit values (v1.1) of the NuFIT collaboration~\cite{GonzalezGarcia:2012sz}\footnote{See~\cite{Tortola:2012te,*Fogli:2012ua} for other global fits to the neutrino oscillation data.}
\begin{align}
\sin^2\theta_{12}&=0.306\,, &
\Delta m_{21}^2&=7.45\times 10^{-5} \eV^2\,,\nonumber\\
\sin^2\theta_{13}&=0.0231\,, &
\Delta m_{31}^2& =2.421\times 10^{-3}\eV^2 (N)\,,\\\nonumber
\sin^2\theta_{23}&=0.437\,, &
\Delta m_{32}^2& =-2.410\times 10^{-3}\eV^2 (I)\ .
\end{align}
Furthermore we set the lightest neutrino mass to zero and assume vanishing CP phases in the PMNS matrix, i.e. $\delta=\varphi_1=\varphi_2=0$ as well as a vanishing mixing in the matrix $O$ in the Casas-Ibarra-type parameterization of the Yukawa couplings $\Lambda$. We expect stronger bounds for non-vanishing CP phases or non-vanishing mixing in the matrix $O$. We leave the discussion of non-vanishing CP phases in the PMNS matrix as well as a non-vanishing mixing in $O$ for future work. 
Currently there are no bounds on the Yukawa couplings $\lambda^{df}_{3\alpha}$ and, for simplicity, we will conservatively set them to a rather large value of $1$ in this work. A detailed study would require each $\lambda^{df}_{3\alpha}$ to be varied separately. However simple arguments are already enough to understand the qualitative behavior. The neutrino mass quadratically depends on the combination $\Lambda_{i\alpha}=\lambda^{LQ}_{i3\alpha} \lambda^{df}_{3\alpha}$ and thus a change in the coupling $\lambda^{df}_{3\alpha}$ by a factor $\xi$ has to be compensated by a factor $\xi^{-1}$ in the coupling $\lambda^{LQ}_{i3\alpha}$ in order to leave the neutrino mass matrix unchanged. The most important flavor-changing processes turn out to be the LFV decays $\mu\to e\gamma$, $\mu\to eee$ as well as $\mu\leftrightarrow e$ conversion in nuclei, which all depend on $\lambda^{LQ}_{i3\alpha}$. The dominant contributions to the amplitudes of $\mu\to eee$ and $\mu\leftrightarrow e $ conversion in nuclei originate from penguin diagrams which are proportional to $\lambda^{LQ}_{23\alpha}\lambda^{LQ*}_{13\alpha}$. Similarly the amplitude of $\mu\to e\gamma$ is proportional to $\lambda^{LQ}_{23\alpha}\lambda^{LQ*}_{13\alpha}$. Hence a decrease of $\lambda^{df}_{3\alpha}$ by a factor of $\xi$ leads to an enhancement of the relevant branching ratios by a factor of about $\xi^{-4}$ and consequently a stronger constraint. 

\mathversion{bold}
\subsection{Radiative Lepton Flavor Violating Decays $l_i^- \to l_j^-\gamma$}
\begin{figure}\centering
\FMDG{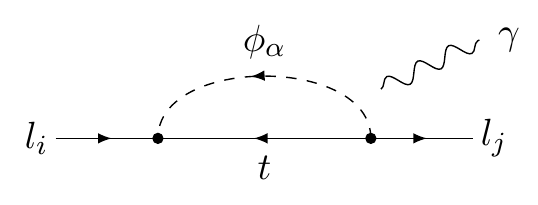}

\begin{minipage}{10cm}
\caption{Lepton-flavor violating rare decays. The photon can be attached to any of the four lines.\label{fig:LFVrareDecay}}
\end{minipage}
\end{figure}

\mathversion{normal}
The first constraints considered are those arising from lepton flavor violating (LFV) processes of the form $l_i^- \to l_j^-\gamma$, such as $\mu \to e \gamma$, which are shown in \Figref{fig:LFVrareDecay}.
From~\cite{Lavoura:2003xp} and the Lagrangian in \Eqref{eq:exotic-Lag}, the amplitude for such processes in this model can be written as
\begin{equation}
\mathcal{M}(l_i\to l_j \gamma) = e\epsilon_\mu^*\bar{u}(p_j) i\sigma^{\mu\nu}q_\nu(\sigma_{Lij} P_L +\sigma_{Rij} P_R) u(p_i)\ ,
\label{ampllg}
\end{equation}
where $e$ is the electric charge, $P_{L,R}\equiv \frac12\left(1 \mp \gamma_5\right)$ are the projection operators and the coefficients\footnote{The full expressions for $\sigma_{L,R}$ can be found in \Appref{app:flavor}.} $\sigma_{L,R}$ are given by
\begin{align}\label{eq:sigmaLR}
\sigma_{Lij} &= \frac{m_{l_j}}{16\pi^2} \sum_{m=1}^2
 \frac{\lambda_{i3m}^{LQ}\lambda_{j3m}^{LQ\dagger}}{m_{\phi_m}^2} F(t_{3m})\ ,&
\sigma_{Rij} &= \frac{m_{l_i}}{16\pi^2} \sum_{m=1}^2
 \frac{\lambda_{i3m}^{LQ}\lambda_{j3m}^{LQ\dagger}}{m_{\phi_m}^2} F(t_{3m})\ ,
\end{align}
where the mass ratio $t_{3m} =m_{t}^2 / m_{\phi_m}^2$ and the loop function $F$ is defined as
\begin{equation}
F(t_{3m})=\frac{1+4t_{3m}-5t_{3m}^2+2t_{3m}(2+t_{3m})\ln t_{3m}}{4 ( t_{3m}-1)^4}\ .
\end{equation} 
Then, the resulting partial decay width of $l_i\to l_j\gamma$ is
\begin{equation}\label{eq:DWliljgamma}
\Gamma(l_i\to l_j \gamma) =\frac{(m_{l_i}^2-m_{l_j}^2)^3 e^2\left(|\sigma_L|^2+|\sigma_R|^2\right)}{16\pi m_{l_i}^3}\ .
\end{equation}
A comparison to the dominant tree-level decay $l_i^-\to l_j^- \nu_i \bar\nu_j$, neglecting the final state lepton mass, results in a good analytic estimate of the branching ratio for $\mu\to e \gamma$,
\begin{equation}\label{eq:Brliljgamma}
\mathrm{Br}(\mu\to e \gamma) \simeq \frac{3 s_W^2}{8\pi^3 \alpha} \left(\sum_{m=1}^2
 \lambda_{23m}^{LQ}\lambda_{13m}^{LQ\dagger} \frac{m_W^2}{m_{\phi_m}^2}\,F(t_{3m})\right)^2
\end{equation}
with $s_W^2=\sin^2\theta_W$, which captures the main dependence on the leptoquark masses $m_{\phi_m}^2$ and couplings $\lambda^{LQ}_{i3m}$. In the numerical evaluation, we use the full expression of the partial decay width and the experimentally measured total decay width.

\begin{figure}[bthp]\centering
\begin{subfigure}{0.49\linewidth}
\includegraphics[width=\linewidth]{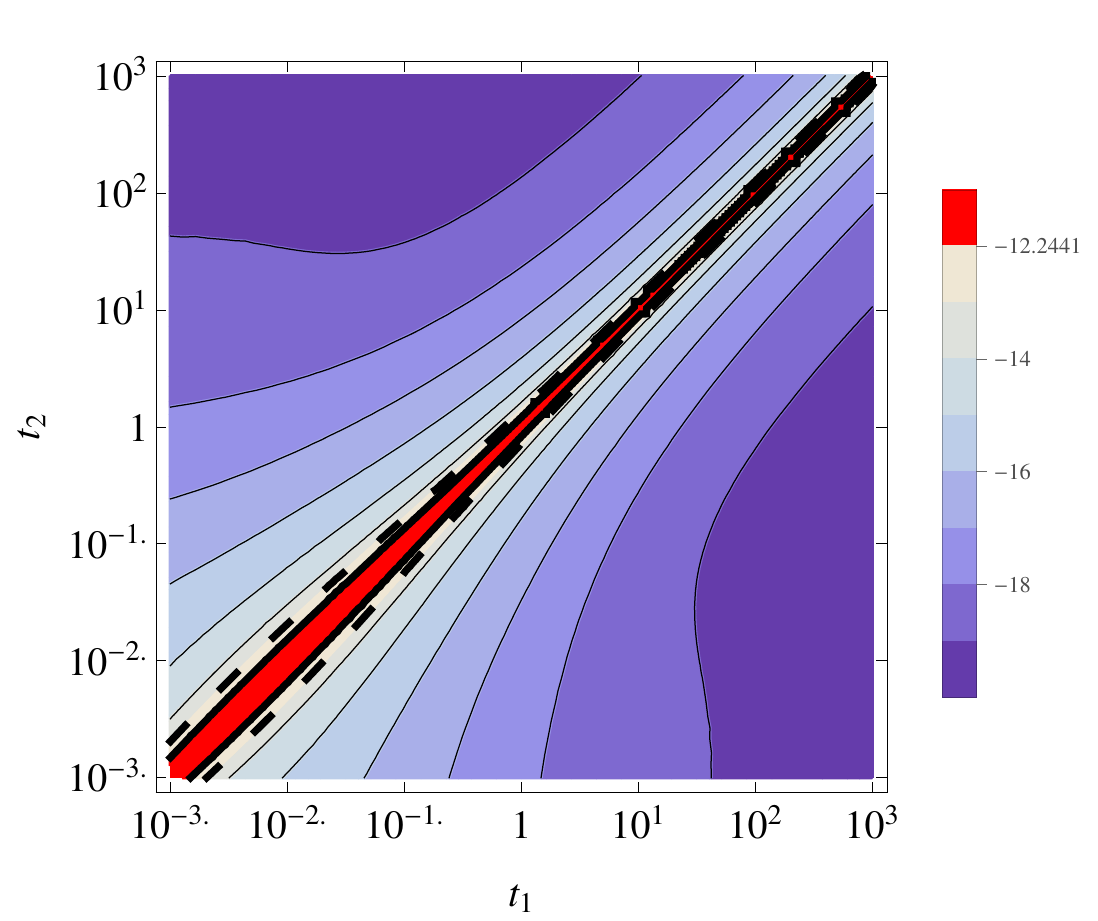}
\caption{$\log_{10} \mathrm{Br}(\mu\to e\gamma)$ for $m_f=1 \TeV$}
\label{fig:LFVdecaya}
\end{subfigure}
\hfill
\begin{subfigure}{0.49\linewidth}
\includegraphics[width=\linewidth]{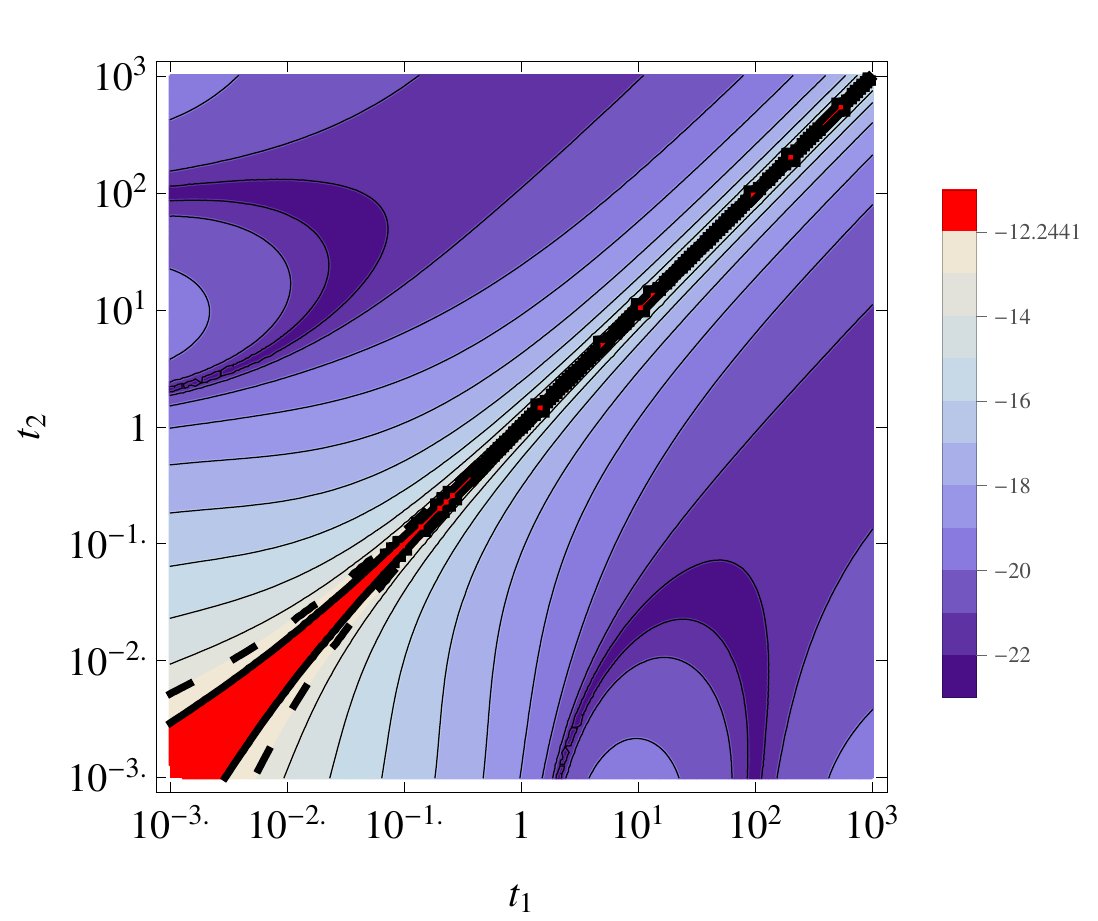}
\caption{$\log_{10} \mathrm{Br}(\mu\to e\gamma)$ for $m_f=10 \TeV$}
\label{fig:LFVdecayb}
\end{subfigure}

\begin{subfigure}{0.49\linewidth}
\includegraphics[width=\linewidth]{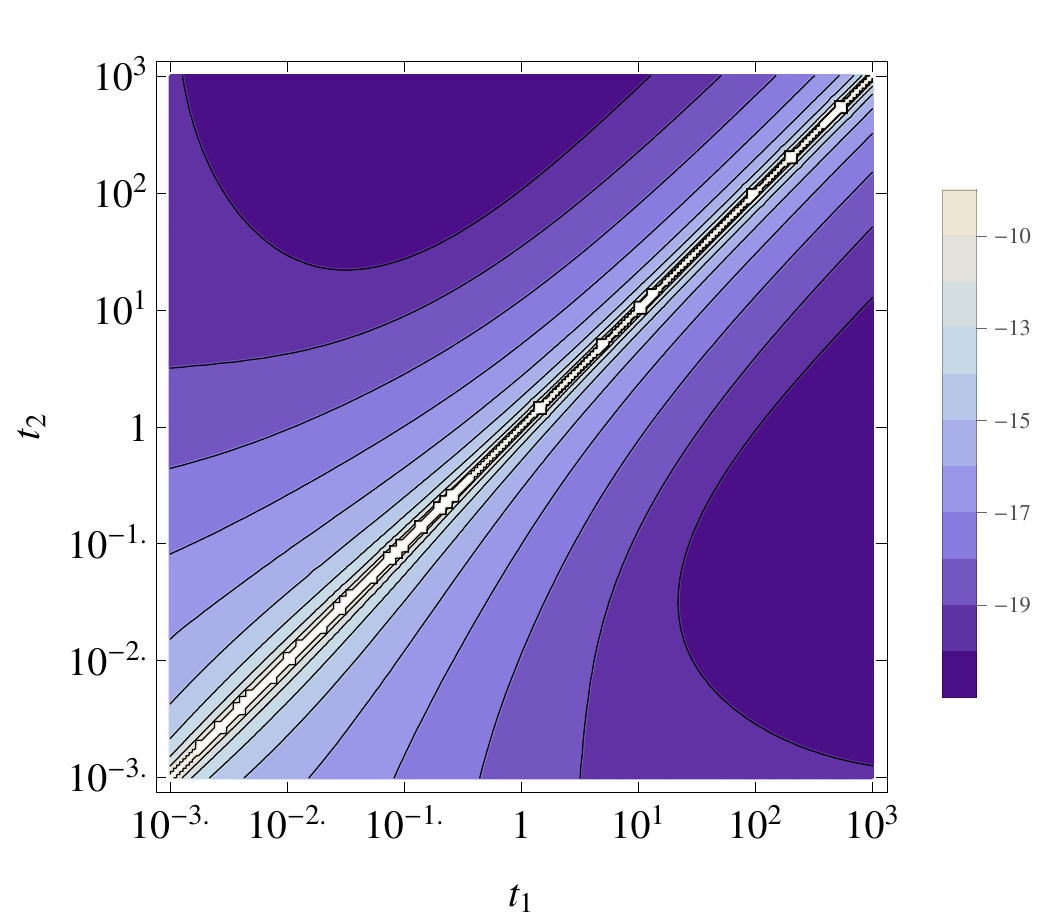}
\caption{$\log_{10} \mathrm{Br}(\tau\to e\gamma)$ for $m_f=1 \TeV$}
\label{fig:LFVdecayc}
\end{subfigure}
\hfill
\begin{subfigure}{0.49\linewidth}
\includegraphics[width=\linewidth]{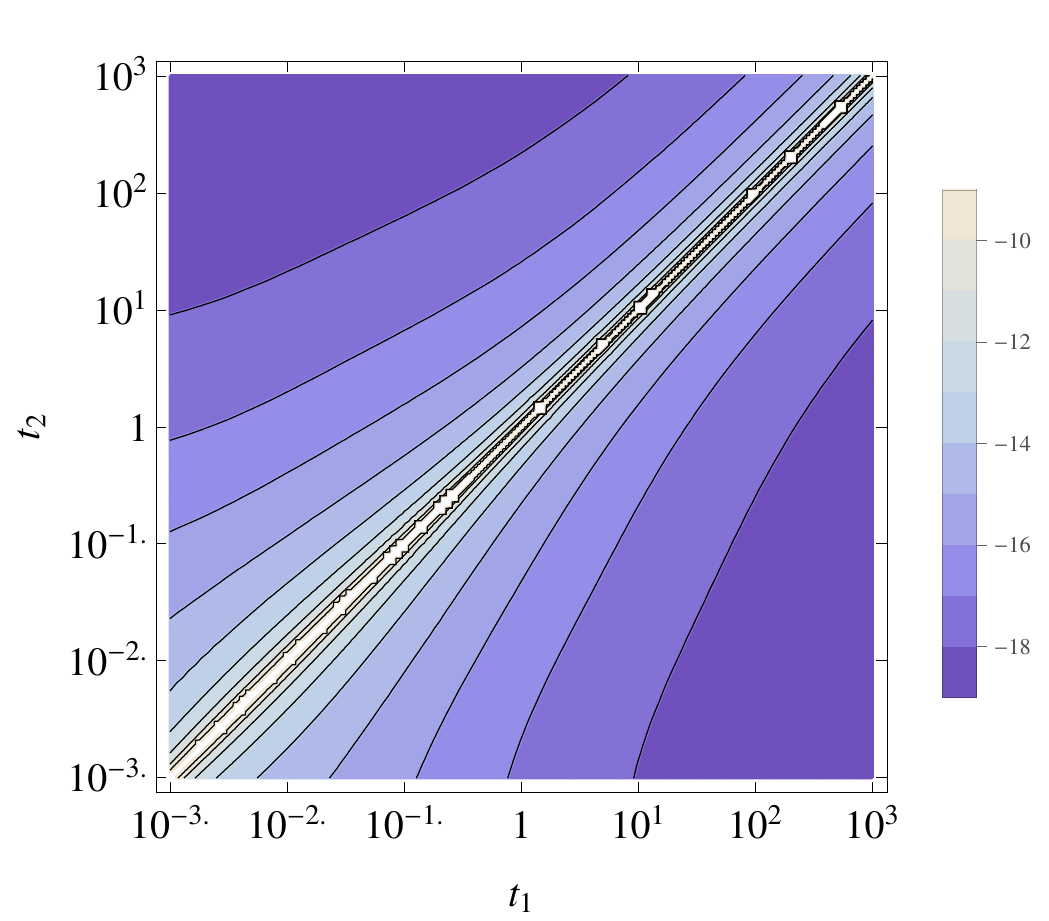}
\caption{$\log_{10} \mathrm{Br}(\tau\to \mu\gamma)$ for $m_f=1 \TeV$}
\label{fig:LFVdecayd}
\end{subfigure}
\caption{Contour plots of the branching ratio in $\mu\to e\gamma$ at $m_f=1\ \TeV$ (top left), $m_f=10\ \TeV$ (top right),
$\tau\to e\gamma$ at $m_f=1\ \TeV$ (bottom left) and $\tau\to \mu\gamma$ at $m_f=1\ \TeV$ (bottom right). 
The current exclusion region at Br$(\mu\to e\gamma)<5.7\times 10^{-13}$~\cite{Adam:2013mnn} as well as Br$(\tau\to e\gamma)<3.3\times 10^{-8}$ and  Br$(\tau\to\mu\gamma)<4.4\times 10^{-8}$~\cite{PDG:2012} at 90\% C.L. are plotted in red. The sensitivity of the proposed upgrade of MEG is $6\times 10^{-14}$~\cite{Baldini:2013ke} and it is shown as a dashed black line. The Yukawa couplings $\lambda^{df}_{3\alpha}$ are set to 1.}
\label{fig:LFVdecays}
\end{figure}
Here we present the contour plots of the branching ratio of $\mu\to e\gamma$ at $m_f=1\TeV$ and $m_f=10\TeV$ in \Figref{fig:LFVdecaya} and 
\Figref{fig:LFVdecayb}. As expected, the region where $t_1\simeq t_2$ is excluded, because the large hierarchy in the integral matrix $I$ 
needs to be compensated by a large hierarchy in the Yukawa couplings $\Lambda$, which then leads to a large decay width for $\mu\to e\gamma$.
For the same $m_f$, the hierarchy gets larger in the third quadrant when leptoquark masses get smaller as shown in \Figref{fig:hierarchyPlot}, which leads to a wider exclusion region. 
For the same $t_1$ and $t_2$, the branching ratio in the first quadrant scales as $m_f^{-1}$ when $t_1\simeq t_2\gg 1$. 
Thus the exclusion region in the first quadrant is narrower when $m_f$ gets larger.
In the third quadrant where $t_1\simeq t_2 \ll 1$, however, the branching ratio scales as $m_f^3 \ln m_f^2$ for fixed $t_1$ and $t_2$. 
So the exclusion region gets wider when $m_f$ gets larger. 
The results for $\tau\to e\gamma$ and $\tau\to \mu\gamma$ at $m_f=1\TeV$ are shown in \Figref{fig:LFVdecayc} and \Figref{fig:LFVdecayd} respectively,
which are similar but less stringent than the constraints from $\mu\to e\gamma$.

\subsection{Anomalous Magnetic Moment}
In terms of the functions defined in \Eqref{eq:sigmaLR}, the anomalous magnetic moment is simply  
\begin{equation}
\Delta a_i=2\,  e\, m_{l_i} \,(\sigma_{Lii} + \sigma_{Rii}) \simeq  \frac{e }{8\pi^2} \sum_{m=1}^2 \left|\lambda_{i3m}^{LQ}\right|^2\frac{m_{l_i}^2}{m_{\phi_m}^2} F(t_{3m})\ .
\end{equation}
The predicted values for all three flavors are substantially below current experimental limits, implying that these values do not provide meaningful limits on the model. 
For example, the maximum contribution to the muon anomalous magnetic moment from our model for $m_f=1\TeV$ is $\mathcal{O} (10^{-13})$, 
which is almost four orders of magnitude smaller than the experimental uncertainty of $\Delta a_\mu^{exp} = (1.1659209\pm0.0000006) \times 10^{-3}$~\cite{PDG:2012}.

\mathversion{bold}
\subsection{LFV Rare Decay $\mu^- \to e^-e^+e^-$}
\mathversion{normal}
The model has a number of different contributions to $\mu^- \to e^-e^+e^-$, including photon penguins, $Z$-penguins, Higgs penguins and box diagrams.  As the Higgs penguin diagrams are suppressed by the electron mass, we have neglected their contribution here.  We consider the contribution of the remaining diagrams in turn.\footnote{We do not discuss leptonic LFV $\tau$ decays, e.g. $\tau\to eee$, since they are less constraining like radiative LFV $\tau$ decays are less constraining than radiative LFV decays of $\mu$.}

To begin with the amplitude associated with the $\gamma$-penguin can be written as~\cite{Hisano:1995cp,*Hisano:1998fj,*Arganda:2005ji}
\begin{eqnarray}
\nonumber
\mathcal{M}_{\gamma} &=& \bar{u}(p_1)\left(q^2 \gamma_\mu (A_1^L P_L +A_1^R P_R)
+ i m_\mu \sigma_{\mu\nu}q^\nu\left( A_2^L P_L +A_2^R P_R \right)
\right) u(p) \\
&\times& \frac{e^2}{q^2} \bar{u}(p_2)\gamma^\mu v(p_3) -\left( p_1\leftrightarrow p_2 \right)\ ,
\end{eqnarray}
where the form factors $A_{1,2}^{L,R}$ are given by\footnote{The full expressions for $A_{1,2}^{L,R}$ can be found in \Appref{app:flavor}.}
\begin{subequations}
\begin{align}
A_1^{L} =& \sum_{a=1}^3\sum_{m=1}^2 \frac{\lambda^{LQ}_{2am}\lambda^{LQ\dagger}_{1am} }{ 384 \pi^2m_{\phi_m}^2} 
\frac{ \left(t_{am}-1\right) \left(10+\left(-17+t_{am}\right)t_{am}\right)-2\left(4-6t_{am}-t_{am}^3\right)\ln t_{am}}{(t_{am}-1)^4}\\ 
A_1^{R}=&0\\
A_2^{L,R}=&\frac{\sigma_{L21,R21}}{m_\mu}\label{eq:A2LR}\ ,
\end{align}
\end{subequations}
where $t_{am} =m_{u_a}^2 / m_{\phi_m}^2$ and $\sigma_{L,R}$ are as defined in Eq.~\ref{eq:sigmaLR}. Note that the external momentum as well as the electron mass $m_e$ have been set to zero and all color factors have been taken into account.

Next the contribution from the $Z$-penguin diagrams can be written as
\begin{eqnarray}
\nonumber
\mathcal{M}_Z&=&\frac{1}{m_Z^2}\bar{u}(p_1) \gamma_\mu \left( F_L P_L+F_R P_R \right) u(p) \\
&\times& \bar{u}(p_2) \gamma^\mu \left(Z_L P_L + Z_R P_R\right) v(p_3) - (p_1\leftrightarrow p_2)\ ,
\end{eqnarray}
where the form factors $F_{L,R}$\footnote{The full expressions for $F^{L,R}$ can be found in \Appref{app:flavor}.}  and $Z_{L,R}$ are given by
\begin{subequations}
\begin{eqnarray}
F_{L}&=& \sum_{a=1}^3\sum_{m=1}^2
-\frac{ 3 e \lambda^{LQ}_{2am}\lambda^{LQ\dagger}_{1am}}{32 \pi^3 \sin\theta_W\cos\theta_W}\frac{t_{am}(1-t_{am}+\ln t_{am})}{(t_{am}-1)^2}\\
F_R&=& 0\\
Z_L&=&-\frac{e}{\sin\theta_W\cos\theta_W}  \left(-\frac{1}{2}+\sin\theta_W^2\right)\\
Z_R&=&- \frac{e}{\sin\theta_W\cos\theta_W}\sin\theta_W^2 \ .
\end{eqnarray}
\end{subequations}
Finally for the amplitude of the contribution from box diagrams we have
\begin{eqnarray}
\mathcal{M}_{Box} &=& 
e^2 B_1^L \left[ \bar{u}_i(p_1) \left( \gamma^{\mu} P_L \right) u_j(p) \right] \left[ \bar{u}_i(p_2) \left( \gamma_{\mu} P_L \right) v_i(p_3) \right] \nonumber \\
&+& e^2 B_1^R \left[ \bar{u}_i(p_1) \left( \gamma^{\mu} P_R \right) u_j(p) \right] \left[ \bar{u}_i(p_2) \left( \gamma_{\mu} P_R \right) v_i(p_3) \right] \nonumber \\
&+& e^2 B_2^L \left\{ \left[ \bar{u}_i(p_1) \left( \gamma^{\mu} P_L \right) u_j(p) \right] \left[ \bar{u}_i(p_2) \left( \gamma_{\mu} P_R \right) v_i(p_3) \right] - (p_1 \leftrightarrow p_2) \right\} \nonumber \\
&+& e^2 B_2^R \left\{ \left[ \bar{u}_i(p_1) \left( \gamma^{\mu} P_R \right) u_j(p) \right] \left[ \bar{u}_i(p_2) \left( \gamma_{\mu} P_L \right) v_i(p_3) \right] - (p_1 \leftrightarrow p_2) \right\} \nonumber \\
&+& e^2 B_3^L \left\{ \left[ \bar{u}_i(p_1) P_L u_j(p) \right] \left[ \bar{u}_i(p_2) P_L v_i(p_3) \right] - (p_1 \leftrightarrow p_2) \right\} \nonumber \\
&+& e^2 B_3^R \left\{ \left[ \bar{u}_i(p_1) P_R u_j(p) \right] \left[ \bar{u}_i(p_2) P_R v_i(p_3) \right] - (p_1 \leftrightarrow p_2) \right\} \nonumber \\
&+& e^2 B_4^L \left\{ \left[ \bar{u}_i(p_1) \left( \sigma_{\mu \nu} P_L \right) u_j(p) \right] \left[ \bar{u}_i(p_2) \left( \sigma^{\mu \nu} P_L \right) v_i(p_3) \right] - (p_1 \leftrightarrow p_2) \right\} \nonumber \\
&+& e^2 B_4^R \left\{ \left[ \bar{u}_i(p_1) \left( \sigma_{\mu \nu} P_R u_j(p) \right) \right] \left[ \bar{u}_i(p_2) \left( \sigma^{\mu \nu} P_R \right) v_i(p_3) \right] - (p_1 \leftrightarrow p_2) \right\} \ ,
\end{eqnarray}
where the form factors $B_i$ are given by\footnote{The full expressions for $B_i$ are given in \Appref{app:flavor}.}
\begin{subequations}
\begin{eqnarray}
B_1^{L}&=& \sum_{i,j,m,n}-\frac{3}{16\pi^2 e^2} \lambda^{LQ}_{2im}\lambda^{LQ\dagger}_{1in} \lambda^{LQ\dagger}_{1jm} \lambda^{LQ}_{1jn} 
D_{00}\left[ m_{\phi_m}^2, m_{\phi_n}^2, m_{u_i}^2, m_{u_j}^2 \right]\ ,\\
B_1^R=B_{2,3,4}^{L,R}&=& 0,
\end{eqnarray}
\end{subequations}
where $3$ is the color factor and all the external momenta and masses have been neglected.  The function $D_{00}$ can be found in App.~\ref{sec:appPV}.
Using the form factors given above, we can write the decay width for  $\mu^- \to e^-e^+e^-$ as follows~\cite{Arganda:2005ji}:
\begin{align}
\Gamma(\mu^-\to e^- e^+ e^-)=&\frac{e^4}{512\pi^3} m_\mu^5\Bigg[ 
\left|A_1^L\right|^2+\left|A_1^R\right|^2-2 \left(A_1^L A_2^{R*}+A_2^L A_1^{R*}+ \hc\right)\\\nonumber
&+ \left( \left|A_2^L\right|^2 +\left|A_2^R\right|^2\right)\left(\frac{16}{3}\ln\frac{m_\mu}{m_e}-\frac{22}{3}\right)\\\nonumber
&+\frac16\left(\left|B_1^L\right|^2+\left|B_1^R\right|^2\right)+\frac13\left( \left| B_2^L\right|^2+\left| B_2^R\right|^2\right)\\\nonumber
&+\frac{1}{24}\left(\left| B_3^L\right|^2+\left| B_3^R\right|^2\right)+6\left(\left| B_4^L\right|^2+\left| B_4^R\right|^2\right)\\\nonumber
&-\frac12\left( B_3^L B_4^{L*} + B_3^R B_4^{R*} +\hc\right)\\\nonumber
&+\frac13\left(A_1^L B_1^{L*}+A_1^R B_1^{R*} +A_1^L  B_2^{L*}+A_1^R  B_2^{R*} +\hc\right)\\\nonumber
&-\frac23\left(A_2^R B_1^{L*}+A_2^L B_1^{R*} +A_2^L  B_2^{R*}+A_2^R  B_2^{L*} +\hc\right)\\\nonumber
&+\frac13\Big\{ 2\left(\left|F_{LL}\right|^2+\left|F_{RR}\right|^2\right)+\left|F_{LR}\right|^2+\left|F_{RL}\right|^2\\\nonumber
&+\left(B_1^L F_{LL}^*+ B_1^R F_{RR}^* + B_2^L F_{LR}^* + B_2^R F_{RL}^* +\hc\right)\\\nonumber
&+2\left(A_1^L F_{LL}^* +A_1^R F_{RR}^* +\hc\right) 
+ \left(A_1^L F_{LR}^* +A_1^R F_{RL}^* +\hc\right)\\\nonumber
&-4\left(A_2^R F_{LL}^* +A_2^L F_{RR}^* +\hc\right)
-2\left(A_2^L F_{RL}^*+A_2^R F_{LR}^*+\hc\right)
\Big\}\Bigg]\ ,
\end{align}
with 
\begin{align}
F_{LL}&=\frac{F_L Z_L}{g^2 s_W^2 m_Z^2}\ , &
F_{RR}&=\left.F_{LL}\right|_{L\leftrightarrow R}\ ,&
F_{LR}&=\frac{F_L Z_R}{g^2 s_W^2 m_Z^2} \ ,&
F_{RL}&=\left.F_{LR}\right|_{L\leftrightarrow R}\ ,
\end{align}
and an approximate expression for the branching ratio is obtained by dividing by the decay width of the dominant muon decay $\mu^-\to e^-\bar\nu_e\nu_\mu$, as for the rare radiative LFV decay $\mu\to e\gamma$.
\begin{figure}[bt]\centering
\begin{subfigure}{0.49\linewidth}
\includegraphics[width=\linewidth]{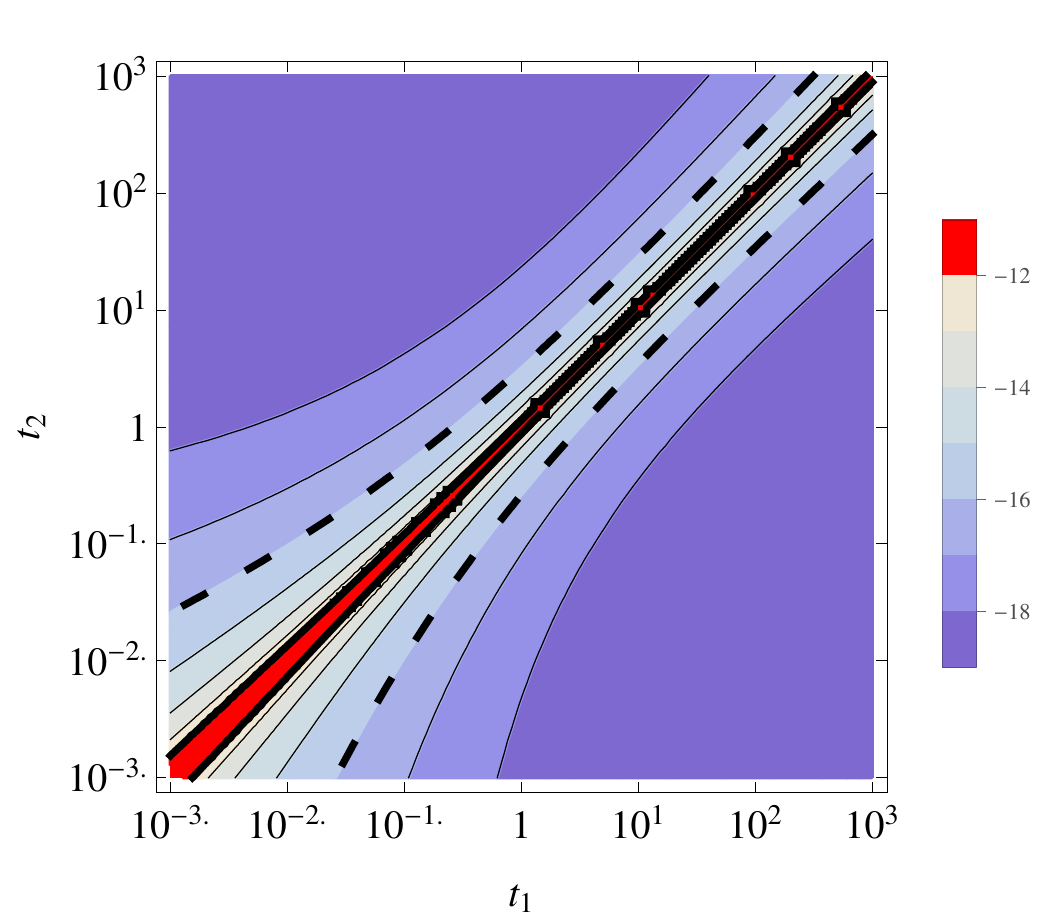}
\caption{$\log_{10} \mathrm{Br}(\mu\to eee)$ for $m_f=1\TeV$}
\end{subfigure}
\hfill
\begin{subfigure}{0.49\linewidth}
\includegraphics[width=\linewidth]{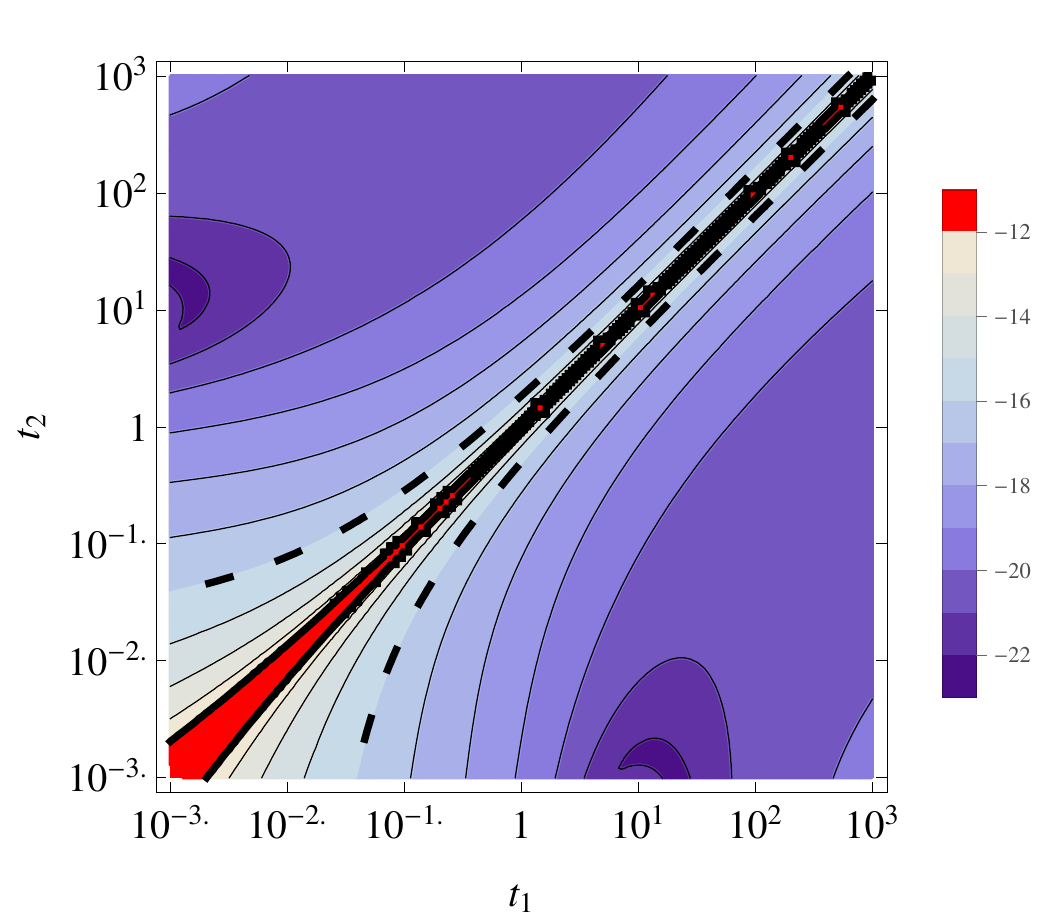}
\caption{$\log_{10} \mathrm{Br}(\mu\to eee)$ for $m_f=10\TeV$}
\end{subfigure}
\caption{Contour plots of the branching ratio in $\mu^-\to e^-e^+e^-$ at  $m_f=1\ \TeV$ (left) and $m_f=10\ \TeV$ (right).
The current exclusion region at Br$(\mu\to eee)<10^{-12}$ at 90\% C.L.~\cite{PDG:2012} is plotted in red. The sensitivity of measuring Br$(\mu\to eee)$ down to $10^{-16}$ of a proposed future experiment~\cite{Blondel:2013ia} is shown as a dashed black line. 
The Yukawa couplings $\lambda^{df}_{3\alpha}$ are set to 1.\label{fig:MuEEEdecay}}
\end{figure}

We show the numerical results in contour plots in \Figref{fig:MuEEEdecay}, where the experimental exclusion region is plotted in red.
When we increase the mass of the colored octet from $m_f=1\TeV$ to $m_f=10\TeV$, 
 the exclusion region gets smaller in both the first and the third quadrant, because the branching ratio is dominated by 
 the $Z$-penguin and $\gamma$-penguin respectively, which scales as $m_f^{-3/2} \ln m_f^2$ in the first quadrant and $m_f^{-3} \ln m_f^2$ in the third quadrant. 
 Compared with the constraints from $\mu\to e\gamma$, the ones from $\mu \to eee$ are more stringent in the first quadrant because of the slower decoupling of the contribution from 
 the $Z$-penguin, while in the third quadrant they are less stringent. 

\mathversion{bold}
\subsection{$\mu\leftrightarrow e$ Conversion in Nuclei}
\mathversion{normal}
\label{sec:mu2eConversion}
The effective Lagrangian contributing to $\mu\leftrightarrow e$ conversion in this model is~\cite{Kitano:2002mt}
\begin{eqnarray}
    {\cal L}_{\rm int} &=&
- \frac12
    \left(
    m_\mu A_2^L\, \bar{\mu}\, \sigma^{\mu \nu} P_L e F_{\mu \nu}
    + m_\mu A_2^R\, \bar{\mu}\, \sigma^{\mu \nu} P_R e F_{\mu \nu}
    + {\rm h.c.}
    \right) \nonumber \\
    &&
    - 
 \sum_{q = u,d,s}
    \left[ {\rule[-3mm]{0mm}{10mm}\ } \right.
    \left(
    g_{LS(q)} \bar{e} P_R \mu + g_{RS(q)} \bar{e} P_L \mu
    \right) \bar{q} q  \nonumber \\
    &&   \hspace*{2.4cm}
    +
    \left(
    g_{LP(q)} \bar{e} P_R \mu + g_{RP(q)} \bar{e} P_L \mu
    \right) \bar{q} \gamma_5 q \nonumber \\
    &&   \hspace*{2.4cm}
    +
    \left(
    g_{LV(q)} \bar{e} \gamma^{\mu} P_L \mu
    + g_{RV(q)} \bar{e} \gamma^{\mu} P_R \mu
    \right) \bar{q} \gamma_{\mu} q \nonumber \\
    &&   \hspace*{2.4cm}
    +
    \left(
    g_{LA(q)} \bar{e} \gamma^{\mu} P_L \mu
    + g_{RA(q)} \bar{e} \gamma^{\mu} P_R \mu
    \right) \bar{q} \gamma_{\mu} \gamma_5 q \nonumber \\
    &&   \hspace*{2.4cm}
    + \ \frac{1}{2}
    \left(
    g_{LT(q)} \bar{e} \sigma^{\mu \nu} P_R \mu
    + g_{RT(q)} \bar{e} \sigma^{\mu \nu} P_L \mu
    \right) \bar{q} \sigma_{\mu \nu} q
	+ {\rm h.c.}
    \left. {\rule[-3mm]{0mm}{10mm}\ } \right]\ .
    \label{eq:mue}
\end{eqnarray}
At one loop, there are several contributions to $\mu\leftrightarrow e$ conversion in nuclei. 
The long-range interaction is determined by the electromagnetic dipole contribution, which is described by the coefficients $A_2^{L,R}$ defined in \Eqref{eq:A2LR}. The remaining interactions are short-range interactions. Taking all SM particles lighter than the W boson massless, there is only a contribution to the Wilson coefficient of $(\bar e\gamma^\mu P_L\mu) (\bar q\gamma_\mu q)$ from box, $\gamma$- as well as $Z$-penguins:
\begin{subequations}
\begin{align}
g_{LV(d)}^{box}=&
\frac{|V_{td}|^2}{64\pi^2} \Bigg\{
2\sum_{m,n,i}
\lambda ^{LQ}_{23m}\lambda ^{LQ*}_{13n}\lambda^{LQ}_{i3n} \lambda ^{LQ*}_{i3m} D_{00}\left(0,m_{\phi
   _m}^2,m_{\phi _n}^2,m_t^2\right)\\\nonumber
&-\sum_m\lambda _{23m}^{LQ} \lambda _{13m}^{LQ*}
\Big[
m_t^2 y_t^2 D_0\left(m_W^2,m_{\phi _m}^2,m_t^2,m_t^2\right)
+g^2 \Big(C_0\left(m_W^2,0,m_{\phi _m}^2\right)\\\nonumber
&+m_t^2 D_0\left(m_W^2,0,m_{\phi_m}^2,m_t^2\right)
-2 \left(D_{00}\left(m_W^2,0,m_{\phi _m}^2,m_t^2\right)+D_{00}\left(m_W^2,m_{\phi_m}^2,m_t^2,m_t^2\right)\right)
\Big)
\Big]\Bigg\}\ ,
\\
g_{LV(d)}^{\gamma}=&-\frac{\alpha }{144 \pi }\sum_m\frac{\lambda ^{LQ}_{23m} \lambda ^{LQ*}_{13m}}{m_{\phi _m}^2 }\frac{t_{3m}^3-18 t_{3m}^2+27 t_{3m}+2 \left(t_{3m}^3+6 t_{3m}-4\right) \ln
   \left(t_{3m}\right)-10}{\left(t_{3m}-1\right)^4}\,
\\
g_{LV(d)}^{Z}=&\frac{g^2\left(4 s_W^2-3\right)}{128 \pi ^2 m_W^2}\sum_m \lambda ^{LQ}_{23m} \lambda^{LQ*}_{13m}
\frac{ t_{3m} \left(t_{3m}-\ln \left(t_{3m}\right)-1\right)}{\left(t_{3m}-1\right){}^2}\ ,
\end{align}
\end{subequations}
and
\begin{align}
g_{LV(u)}^{box}=&0\ ,
&
g_{LV(u)}^{\gamma}=&-2 g_{LV(d)}\ ,
&
g_{LV(u)}^{Z}=&-\frac{8 s_W^2-3}{4 s_W^2-3} g_{LV(d)}^Z\ .
\end{align}
The gluon penguin contribution vanishes due to its color structure. As the coherent conversion process dominates, i.e. the final state of the nucleon is the same as the initial state~\cite{Kitano:2002mt}, the vector coupling to the sea quarks vanishes and it is enough to consider
\begin{align}
g_{LV(u)}=&g_{LV(u)}^{box}+g_{LV(u)}^{\gamma}+g_{LV(u)}^{Z}\ ,
&
g_{LV(d)}=&g_{LV(d)}^{box}+g_{LV(d)}^{\gamma}+g_{LV(d)}^{Z}\ .
\end{align}
The coefficients of the vector interaction with protons and neutrons are defined by
\begin{align}
\tilde g^{(p)}_{LS,RS}&= \sum_q G_S^{(q,p)} g_{LS,RS(q)}\ , &
\tilde g^{(n)}_{LS,RS}&= \sum_q G_S^{(q,n)} g_{LS,RS(q)}\ , \\
\tilde g_{LV}^{(p)}&=2\, g_{LV(u)} + g_{LV(d)}\ , 
&
\tilde g_{LV}^{(n)}&= g_{LV(u)} + 2\, g_{LV(d)}\ ,
\end{align}
with the coefficients $G^{(u,p)}_S=G^{(d,n)}_S=5.1$, $G^{(u,n)}_S=G^{(d,p)}_S=4.3$ and $G^{(s,p)}_S=G^{(s,n)}_S=2.5$~\cite{Kosmas:2001mv,Kitano:2002mt}.
\begin{figure}[hbtp]
\centering
\begin{subfigure}{0.49\linewidth}
\includegraphics[width=\linewidth]{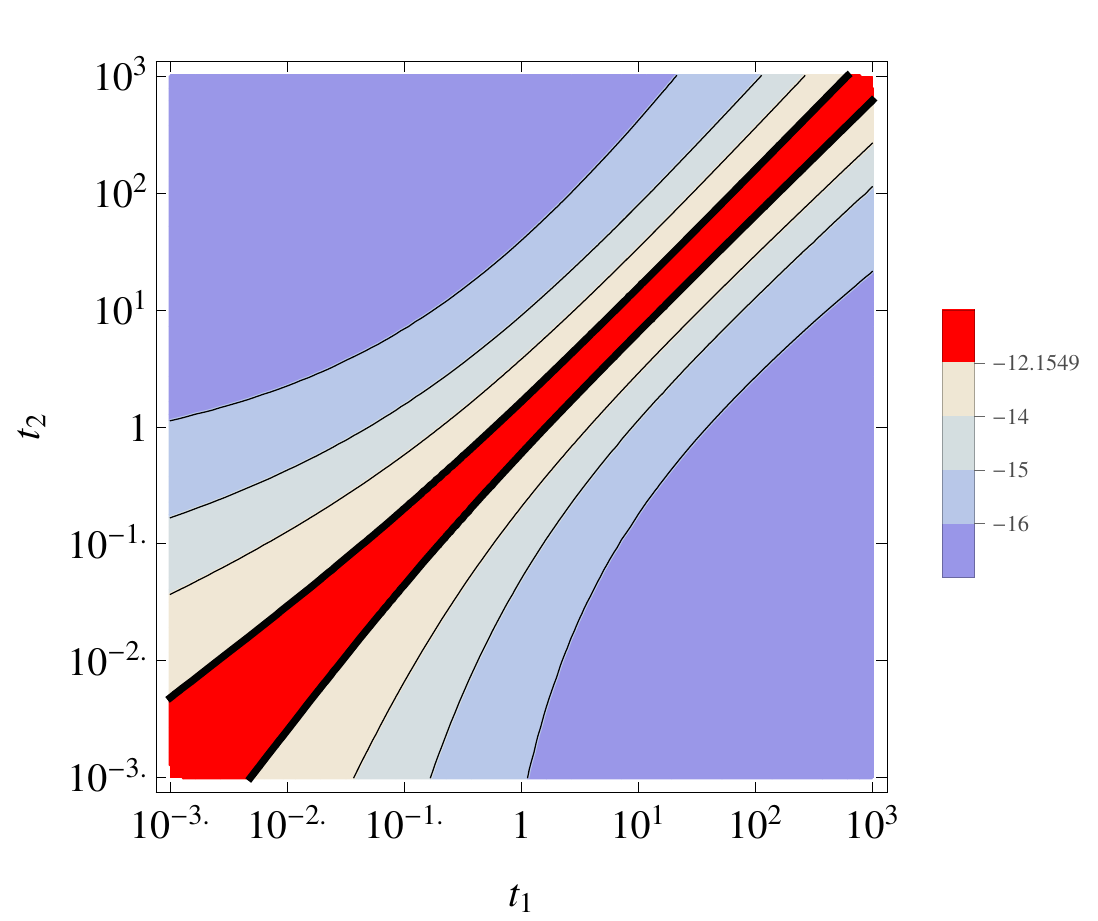}
\caption{$\log_{10} \mathrm{Br}(\mu\mathrm{Au}\to e \mathrm{Au})$ for $m_f=1\TeV$}
\label{fig:Mu2EAu1TeV}
\end{subfigure}
\hfill
\begin{subfigure}{0.49\linewidth}
\includegraphics[width=\linewidth]{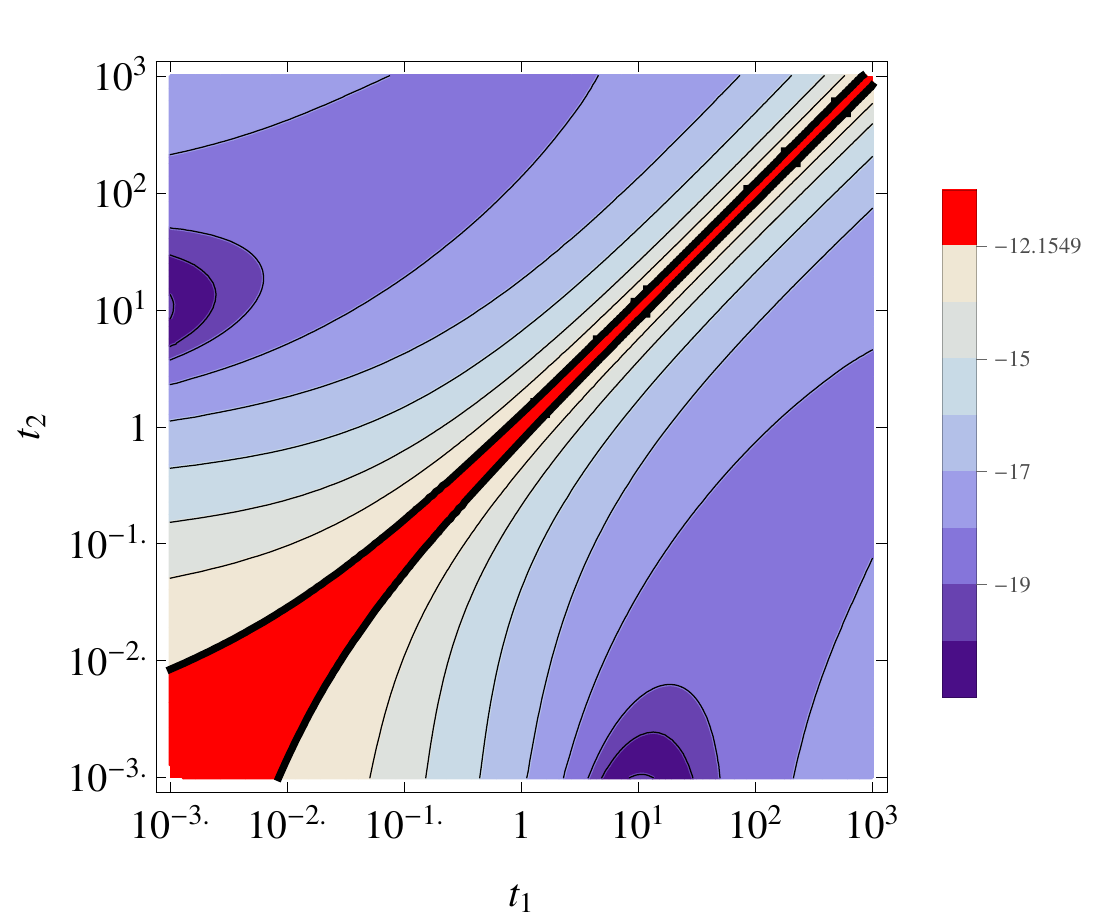}
\caption{$\log_{10} \mathrm{Br}(\mu\mathrm{Au}\to e\mathrm{Au})$ for $m_f=10\TeV$}
\label{fig:Mu2EAu10TeV}
\end{subfigure}

\begin{subfigure}{0.49\linewidth}
\includegraphics[width=\linewidth]{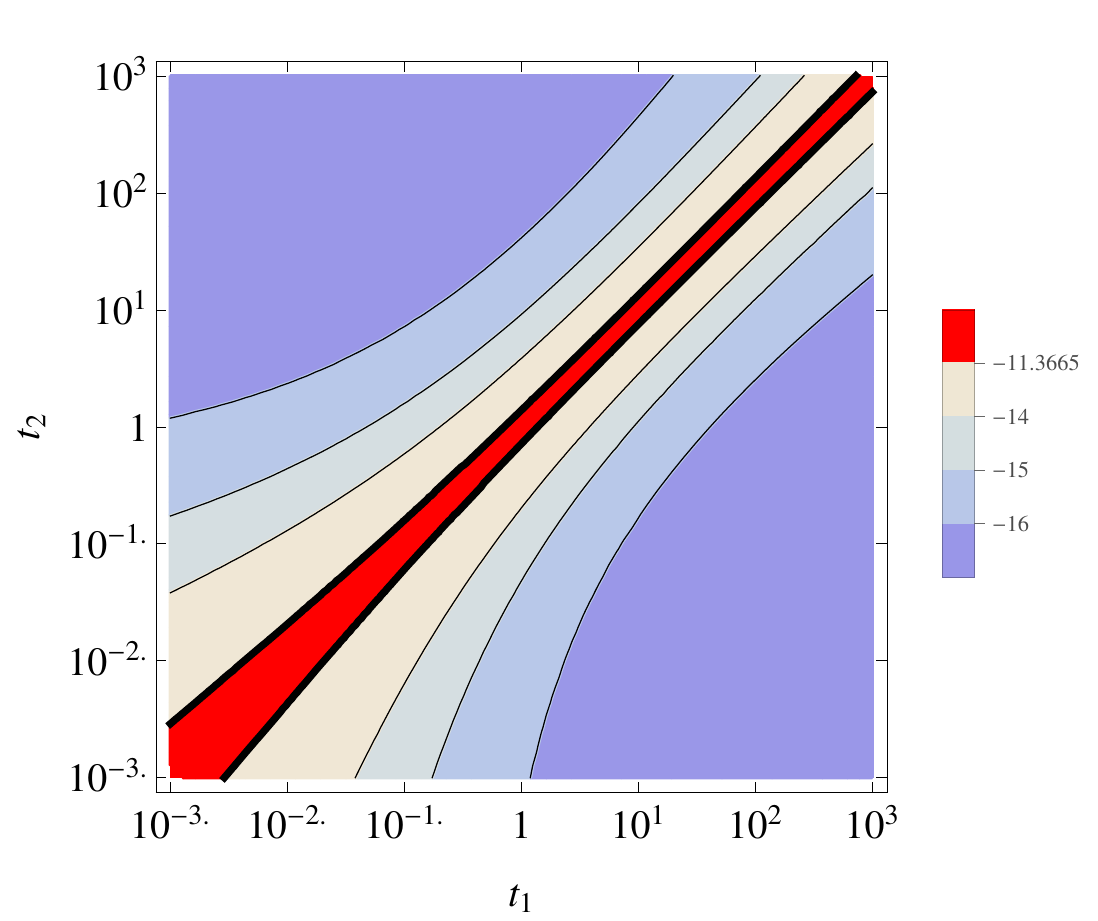}
\caption{$\log_{10} \mathrm{Br}(\mu\mathrm{Ti}\to e\mathrm{Ti})$ for $m_f=1\TeV$}
\label{fig:Mu2ETi1TeV}
\end{subfigure}
\hfill
\begin{subfigure}{0.49\linewidth}
\includegraphics[width=\linewidth]{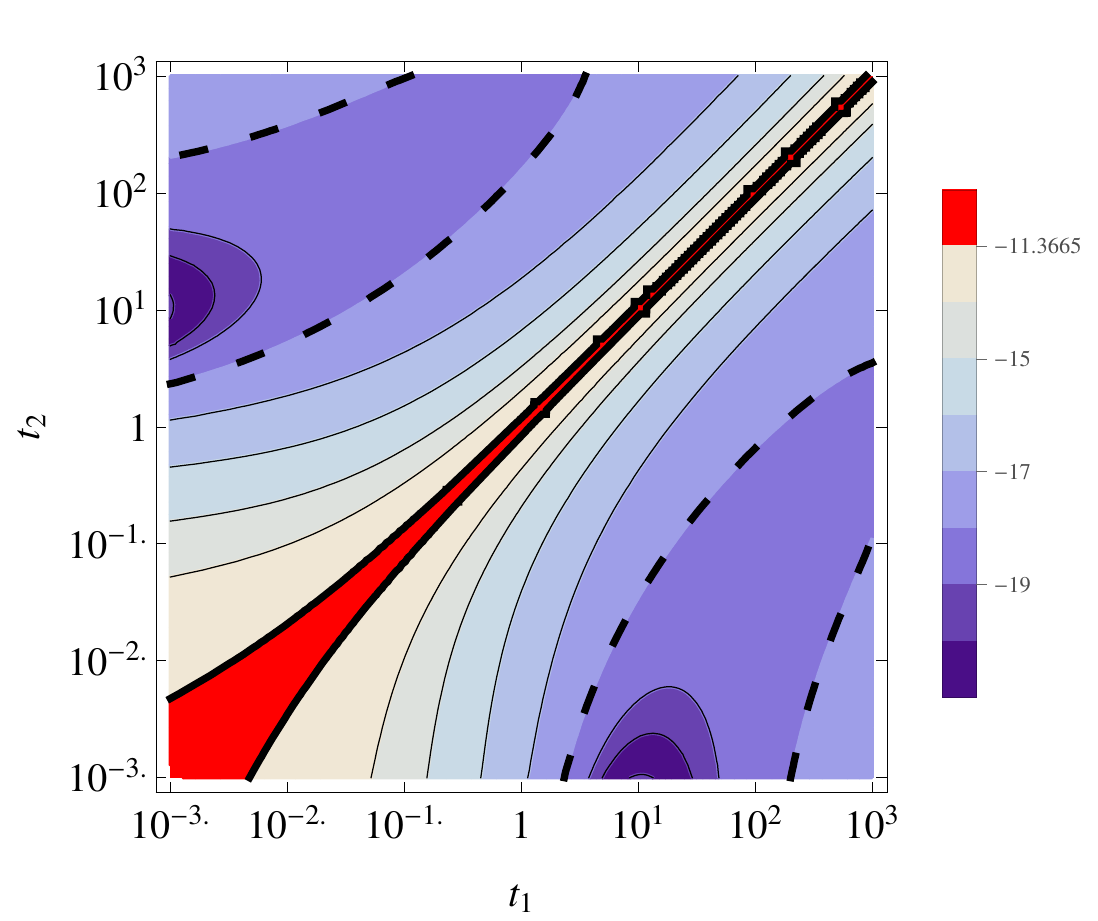}
\caption{$\log_{10} \mathrm{Br}(\mu\mathrm{Ti}\to e\mathrm{Ti})$ for $m_f=10\TeV$}
\label{fig:Mu2ETi10TeV}
\end{subfigure}

\begin{subfigure}{0.49\linewidth}
\includegraphics[width=\linewidth]{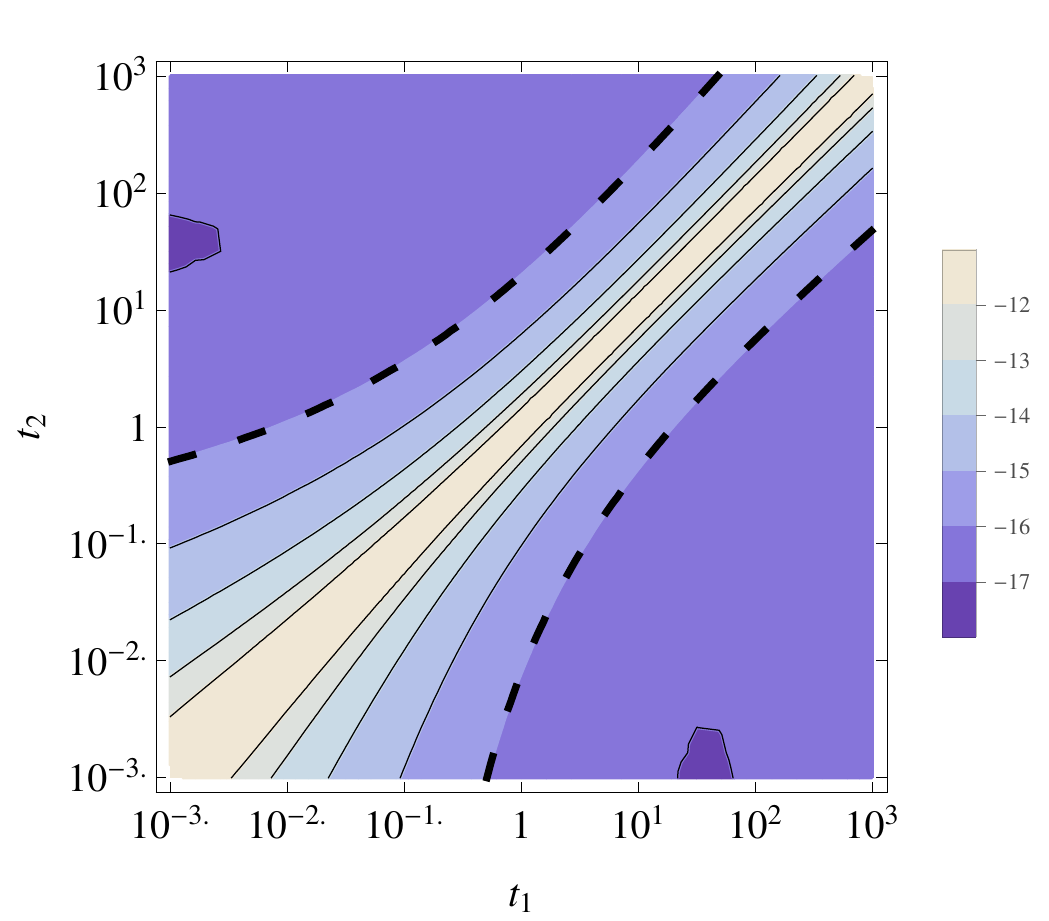}
\caption{$\log_{10} \mathrm{Br}(\mu\mathrm{Al}\to e\mathrm{Al})$ for $m_f=1\TeV$}
\label{fig:Mu2ETi100TeV}
\end{subfigure}
\hfill
\begin{subfigure}{0.49\linewidth}
\includegraphics[width=\linewidth]{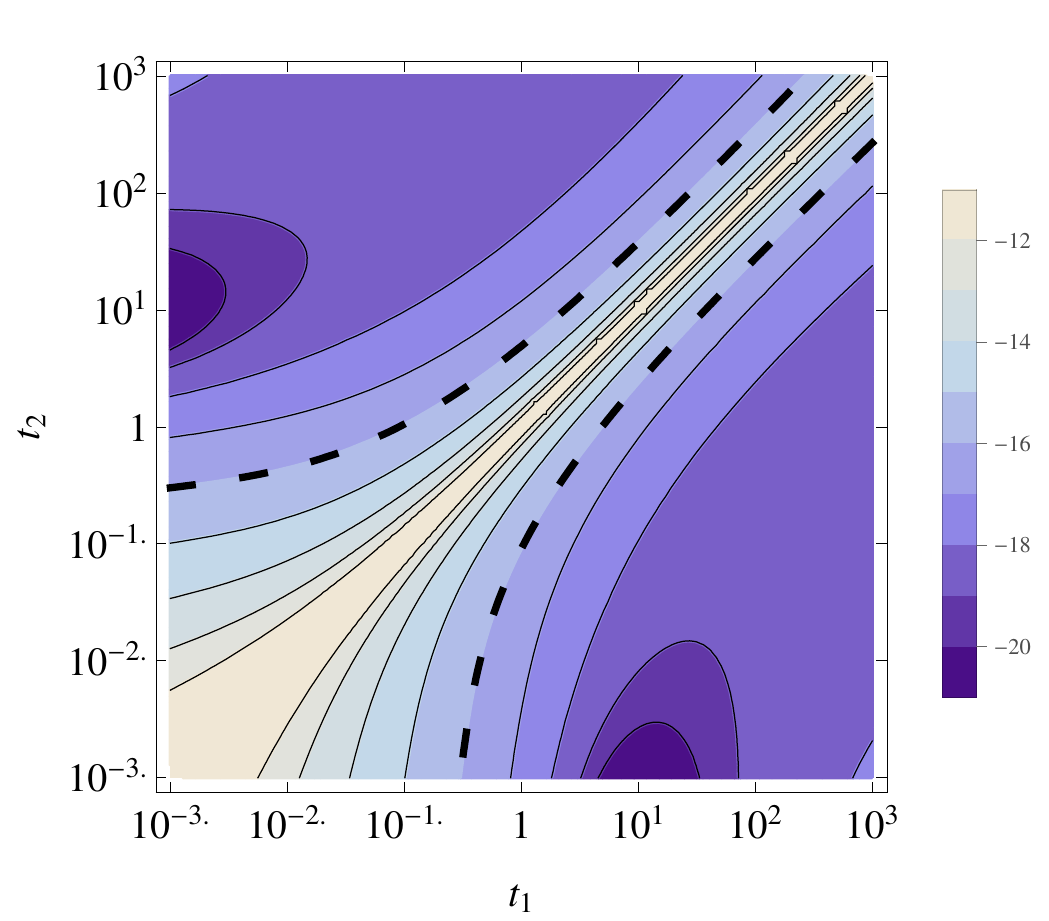}
\caption{$\log_{10} \mathrm{Br}(\mu\mathrm{Al}\to e\mathrm{Al})$ for $m_f=10\TeV$}
\label{fig:Mu2ETi1000TeV}
\end{subfigure}
\caption{Contour plots of the branching ratio of $\mu\leftrightarrow e$ conversion, current experimental bounds (in red) and prospects (thick dashed lines) in $^{197}_{79}\text{Au}$ (top), $^{48}_{22}\text{Ti}$ (center), and $^{27}_{13}\text{Al}$ (bottom) at $m_f=1, 10\TeV$. The Yukawa couplings $\lambda^{df}_{3\alpha}$ are set to 1. The current most stringent bound comes from Br$(\mu \mathrm{Au} \to e \mathrm{Au})<7\times 10^{-13}$ at 90 \% C.L.~\cite{PDG:2012} and the best experimental prospects are from Br$(\mu\mathrm{Al}\to e\mathrm{Al})\lesssim10^{-16}$~\cite{Hungerford:2009zz,*Cui:2009zz,Carey:2008zz,*Kurup:2011zza,*Kutschke:2011ux} as well as Br$(\mu\mathrm{Ti}\to e\mathrm{Ti})\lesssim10^{-18}$~\cite{Hungerford:2009zz,*Cui:2009zz} improving the current limit  Br$(\mu \mathrm{Ti}\to e\mathrm{Ti})<4.3\times 10^{-12}$ at 90\% C.L.~\cite{PDG:2012}.\label{fig:Mu2E}}
\end{figure}
In terms of these expressions, we can express the conversion rate as
\begin{align}\label{eq:mu2eOmega}
	\omega_{\rm conv} =&
4 \left| \frac18 A_2^{R*} D+ \tilde g_{LS}^{(p)} S^{(p)}+ \tilde g_{LS}^{(n)}S^{(n)}+ \tilde{g}_{LV}^{(p)} V^{(p)} + \tilde{g}_{LV}^{(n)} V^{(n)}\right|^2 \nonumber\\
+&4 \left|  \frac18 A_2^{L*} D+ \tilde g_{RS}^{(p)} S^{(p)}+ \tilde g_{RS}^{(n)}S^{(n)}+ \tilde{g}_{RV}^{(p)} V^{(p)} + \tilde{g}_{RV}^{(n)} V^{(n)}\right|^2  \ ,
\end{align}
where the overlap integrals $D$,  $S^{(p)}$, $ S^{(n)}$, $V^{(p)}$ and $V^{(n)}$ take values  as shown in \Tabref{tab:overlap}. The branching ratio is defined by Br$(\mu N \to e N) \equiv \omega_{conv}/\omega_{capt}$.
\begin{table}[tbp]
\centering\setlength{\extrarowheight}{3pt}
\begin{tabular}{c|ccccc|c}
\hline
\hline
 & $S^{(p)}$& $S^{(n)}$ & $V^{(p)}$ & $V^{(n)}$ & $D$ & $\omega_{capt}(10^6 s^{-1})$\\
 \hline
$^{197}_{79}\text{Au}$& 0.167 &0.0523&0.0610& 0.0859 & 0.108  & 13.07 \\
$^{48}_{22}\text{Ti}$  & 0.0870  &0.0371&0.0462 & 0.0399 & 0.0495  & 2.59\\
$^{27}_{13}\text{Al}$   & 0.0169&0.0153&0.0163  & 0.0357 & 0.0159  & 0.7054\\
\hline
\hline
\end{tabular}

\vspace{1ex}

\begin{minipage}{12cm}
\caption{The overlap integrals in the unit of $m_{\mu}^{5/2}$ and the total capture rates for different nuclei. The overlap integrals of $^{197}_{79}\text{Au}$ as well as $^{27}_{13}\text{Al}$ are taken from Table 2 and 
$^{48}_{22}\text{Ti}$ are taken from Table 4 of~\cite{Kitano:2002mt}, while the total capture rates are both from Table 8. \label{tab:overlap}}
\end{minipage}
\end{table}
In \Figref{fig:Mu2E} we present the contour plots of the branching ratio for $\mu\leftrightarrow e$ conversion in $^{197}_{79}\text{Au}$, $^{48}_{22}\text{Ti}$ and $^{27}_{13}\text{Al}$,
where the first gives the most stringent current constraints and the last two may give the most stringent future ones.\footnote{The bound from $\mu\mathrm{Pb}\to e\mathrm{Pb}$ conversion by the SINDRUM II experiment~\cite{Honecker:1996zf} is less competitive than $\mu\leftrightarrow e$ conversion in $^{197}_{79}\text{Au}$.} 
The current bounds from $\mu\leftrightarrow e$ conversion in $^{197}_{79}\text{Au}$ at $m_f=1 \TeV$ and $m_f=10\TeV$ are plotted in red \Figref{fig:Mu2EAu1TeV} and \Figref{fig:Mu2EAu10TeV},
which are obviously the most stringent ones so far compared with those from $\mu\to e\gamma$ and $\mu \to eee$.
When we increase the mass of the colored octet, the exclusion region gets smaller because the dominant contribution, the $Z$-penguin, scales as $m_f^{-3/2}\ln m_f^2$.
For $\mu\leftrightarrow e$ conversion in $^{48}_{22}\text{Ti}$, although the current data gives weaker bounds,
the COMET experiment~\cite{Hungerford:2009zz,*Cui:2009zz} will be able to explore a much larger region of the parameter space or give
the strongest bounds in the future, i.e. the dashed thick line in \Figref{fig:Mu2ETi10TeV} as well as the whole region shown in \Figref{fig:Mu2ETi1TeV}. 

\subsection{Other Constraints}
In the previous subsection, we discussed the most stringent constraints. In this subsection, we discuss constraints which turn out to be not competitive compared to the constraints from LFV processes. We start with dimension-6 operators generated at tree-level using the list of constraints compiled in Ref.~\cite{Carpentier:2010ue}, then discuss bounds from meson mixing, the $b\to s$ transition using the constraints in~\cite{Altmannshofer:2011gn,Altmannshofer:2012az} and finally generic dimension-6 operators $LLQQ$ generated at the one-loop order using the list of constraints in Ref.~\cite{Carpentier:2010ue}.

\mathversion{bold}
\subsubsection{Processes from $\Delta F=0$ Operators Generated at Tree Level}
\mathversion{normal}
Integrating out the leptoquarks leads to a few new operators at tree-level involving only the third generation of quarks. In particular, we generate the operators
\begin{align}
C_{ijtt}^{\ell u,LL} &2\sqrt{2} G_F  (\bar{\ell_i} \gamma^\mu P_L \ell_j)(\bar t \gamma_\mu  P_L t)\ , & 
C_{ijtb}^{CC,LL} &2\sqrt{2} G_F (\bar \ell_i \gamma^\mu P_L \nu_j) (\bar t \gamma_\mu P_L b)\ , \\\nonumber
C_{ijkl}^{\nu d,LL} &2\sqrt{2} G_F  (\bar \nu_i \gamma^\mu P_L \nu_j)(\bar d_k \gamma_\mu  P_L d_l)\ ,&
C_{ijkl}^{\nu d,RL} &2\sqrt{2} G_F (\bar \nu_i^C \gamma^\mu P_R \nu_j^C) (\bar d_k \gamma_\mu  P_L d_l)\ , 
\end{align}
with their Wilson coefficients given by
\begin{align}
C_{ijtt}^{\ell u,LL} & = -\sum_m \frac{\lambda ^{LQ}_{i3m} \lambda^{LQ*}_{j3m}}{4\sqrt{2} G_F\,m_{\phi _m}^2}\ ,&   
C_{ijtb}^{CC,LL} & = V_{tb}\sum_m\frac{ \lambda ^{LQ}_{j3m} \lambda^{LQ*}_{i3m}}{4\sqrt{2} G_F\, m_{\phi _m}^2}\ ,\\\nonumber
C_{ijkl}^{\nu d,LL} & =- V_{tk}^* V_{tl} \sum_m\frac{\lambda^{LQ*}_{j3m}\lambda^{LQ}_{i3m}}{4\sqrt{2} G_F\, m_{\phi_m}^2}\ ,&
C_{ijkl}^{\nu d,RL} & =V_{tk}^* V_{tl} \sum_m \frac{ \lambda^{LQ*}_{j3m}\lambda^{LQ}_{i3m}}{4\sqrt{2} G_F\,m_{\phi_m}^2}     
\ .
\end{align}
There are already weak constraints on the Wilson coefficient $C_{ijtt}^{\ell u,LL}$, mainly from $Z$-decay~\cite{Carpentier:2010ue}. The most stringent constraint is $C_{\mu\mu tt}^{\ell u,LL}<0.061$ from $Z\to \mu\bar\mu$. It translates into 
\begin{equation}
m_{\phi_m} \gtrsim \left(\frac{\sqrt{\lambda ^{LQ}_{23m} \lambda^{LQ*}_{23m}}}{0.1}\right) \times 50\GeV
\end{equation}
and currently does not pose any competitive constraint. The bound on $C_{ijtt}^{\ell u,LL}$ with $i,j=e,\mu$ could be improved by measuring the top pair-production cross section at a lepton collider precisely, like a future linear collider. The Wilson coefficient $C_{ijtb}^{CC,LL}$, leads to an additional contribution to top decay. We can estimate its relative contribution to the dominant top decay, $t\to W^+ b \to \ell_i^+ \nu b$, as
\begin{equation}
\sum_{l,m}\frac{C_{ijtb}^{CC,LL}}{V_{tb}}=\sum_{j,m}\frac{\lambda_{i3m}^{LQ*}\lambda_{j3m}^{LQ}}{4\sqrt{2} G_F\,m_{\phi_m}^2}\ .
\end{equation}
However, the resulting constraint is not as stringent as the constraints from LFV processes.
It would require a big improvement of the precision or a dedicated analysis of the angular distribution due to the different angular dependence compared to $t\to Wb$.
The Wilson coefficients $C_{ijbb}^{\nu b,LL (RL)}$ contribute to the invisible decay of $\Upsilon(nS)$ competing with the SM decay mediated by an $s$-channel $Z$-boson, but there are no limits and the experimental sensitivity is not good enough to measure even the SM decay. The bounds from semi-leptonic meson decays collected in Tab.~12 in~\cite{Carpentier:2010ue} lead to a lower bound on $m_{\phi_m}/\sqrt{\lambda^{LQ*}_{i3m}\lambda^{LQ}_{j3m}}$ of the order of $770\GeV$ for $C_{ijds}^{\nu b,LL (RL)}$, $52\GeV$ for $C_{ijbd}^{\nu b,LL (RL)}$, $800\GeV$ for $C_{ijbs}^{\nu b,LL (RL)}$. A comparison with realistic values, $m_{\phi_m}\gtrsim100\GeV$ and $|\lambda^{LQ}_{i3m}|\lesssim 0.1$, demonstrates that there is currently no competitive constraint from this class of operators. 

\mathversion{bold}
\subsubsection{Meson Mixing}
\mathversion{normal}

The most general effective Hamiltonian has the form, following the notation of~\cite{Bona:2007vi},
\begin{equation}
\mathcal{H}_{\rm eff}^{ij} =\sum_{m=1}^5 C_m^{ij} Q_m^{ij} + \sum_{m=1}^3 \tilde C_m^{ij} \tilde Q_m^{ij}\ ,
\end{equation}
where the superscripts $i,j$ denote the quark flavors, $C_m^{ij}$ ($\tilde C_m^{ij}$) the Wilson coefficients and $Q_m^{ij}$ ($\tilde Q_m^{ij}$) the following operators
\begin{align}
Q_1^{ij} &= (\bar q_i^\alpha \gamma_\mu P_L q_j^\alpha)(\bar q_i^\beta \gamma^\mu P_L q_j^\beta)\ ,&
Q_2^{ij} &= (\bar q_i^\alpha P_L q_j^\alpha)(\bar q_i^\beta P_L q_j^\beta)\ ,&
Q_3^{ij} &= (\bar q_i^\alpha P_L q_j^\beta)(\bar q_i^\beta P_L q_j^\alpha)\ ,\\\nonumber
Q_4^{ij} &= (\bar q_i^\alpha P_L q_j^\alpha)(\bar q_i^\beta P_R q_j^\beta)\ ,&
Q_5^{ij} &= (\bar q_i^\alpha P_L q_j^\beta)(\bar q_i^\beta P_R q_j^\alpha)\ , 
\end{align}
the Greek letters being color indices. The operators $\tilde Q_m^{ij}$ are obtained from the operators $Q_m^{ij}$ by the exchange $L\leftrightarrow R$.

The leptoquarks contribute to meson mixing via one-loop box diagrams with neutrinos and leptoquarks in the loop. These box diagrams only induce the Wilson coefficient
\begin{equation}
C_1^{ij}=-\frac{V_{ti}^2 V_{tj}^{*2} }{48 \pi ^2}
\sum_{i,j,m,n} 
\frac{\lambda _{i3m}^{LQ} \lambda _{j3n}^{LQ} \lambda_{i3n}^{LQ*} \lambda _{j3m}^{LQ*}  }{m_{\phi _m}^2-m_{\phi _n}^2}\ln \left(\frac{m_{\phi _m}^2}{m_{\phi_n}^2}\right)
\end{equation}
in the approximation of vanishing SM fermion masses. In the SM meson mixing is induced by one-loop box diagrams with tops and $W$-bosons in the loop (see e.g.~\cite{Fleischer:2008uj}):
\begin{equation}
C_{1,SM}^{ij}=\frac{G_F^2 m_W^2}{12\pi^2} V_{ti}^2 V_{tj}^{*2}S_0\left(\frac{m_t^2}{m_W^2}\right)
\end{equation}
with the well-known Inami-Lim function~\cite{Inami:1980fz}
\begin{equation}
S_0(x)=x \left[\frac14+\frac94\frac{1}{1-x}-\frac32\frac{1}{(1-x)^2}\right]-\frac32\left[\frac{x}{1-x}\right]^3\ .
\end{equation}
There is no contribution from the leptoquarks to $D_0-\bar D_0$ mixing.

As the different operators are mixed by renormalization group running, we have to evolve the Wilson coefficients from the matching scale to the scale of the relevant meson.
Hence, the meson mixing amplitudes for $\bar B_q-B_q$ with $q=d,s$ as well as $\bar K_0-K_0$ mixing are given by
\begin{subequations}
\begin{align}
\braket{K_0|\mathcal{H}_{\rm eff}^{NP}(\mu_K)|\bar K_0} & =\sum_i (b_i^{r,1} + \eta\, c_i^{r,1})\, \eta^{a_i}  C_1^{ds} \braket{ K_0| Q_r^{ds}|\bar K_0}\ ,\\
\braket{B_d|\mathcal{H}_{\rm eff}^{NP}(\mu_b)|\bar B_d} & = \sum_i (b_i^{r,1} + \eta\, c_i^{r,1})\, \eta^{a_i}  C_1^{db} \braket{ B_d| Q_r^{db}|\bar B_d}\ ,\\
\braket{B_s|\mathcal{H}_{\rm eff}^{NP}(\mu_b)|\bar B_s} & =  \sum_i (b_i^{r,1} + \eta\, c_i^{r,1})\, \eta^{a_i}  C_1^{sb} \braket{ B_s| Q_r^{db}|\bar B_s}\ ,
\end{align}
\end{subequations}
where $\eta=\alpha_s(\Lambda)/\alpha_s(m_t)$, the magic numbers $a_i, b_i^{r,1}, c_i^{r,1}$ as well as the matrix elements for $K$ ($B$) meson mixing can be found in Ref.~\cite{Ciuchini:1998ix} (\cite{Becirevic:2001jj}). As we are only interested in an order of magnitude estimate, we will take $\Lambda=m_t$, i.e.\ $\eta=1$, which means we are neglecting the running from the leptoquark mass to the top-quark mass. In this approximation, the meson matrix elements are
\begin{subequations}
\begin{align}
\braket{K_0|\mathcal{H}_{\rm eff}^{NP}(\mu_K)|\bar K_0} & =\eta_K  C_1^{ds} \,\frac13 M_K f_K^2 B_1^K(\mu_K)\ ,\\
\braket{B_d|\mathcal{H}_{\rm eff}^{NP}(\mu_b)|\bar B_d} & = \eta_B  C_1^{db} \,\frac13 M_{B_d} f_{B_d}^2 B_1^{B}(\mu_b)\ ,\\
\braket{B_s|\mathcal{H}_{\rm eff}^{NP}(\mu(b)|\bar B_s} & = \eta_B  C_1^{sb} \,\frac13 M_{B_s} f_{B_s}^2 B_1^{B}(\mu_b)\ ,
\end{align}
\end{subequations}
where $\mu_K=2\GeV$ and $\mu_b=4.6\GeV$. The QCD correction factors are $\eta_K=0.804$ and $\eta_B=0.848$, while the bag parameters are $B_1^K(\mu_K)=0.60$ and $B_1^B(\mu_b)=0.87$. The SM contribution at the low-energy scale looks analogous, since it trivially satisfies $\Lambda=m_t$, which means  the following ratios are independent of the meson masses, bag parameter and QCD running:
\begin{equation}
\frac{\braket{M|\mathcal{H}_{\rm eff}^{NP}(\mu)|\bar M} }{\braket{M|\mathcal{H}_{\rm eff}^{SM}(\mu)|\bar M} }=\frac{C_1^{ij}}{C_{1,SM}^{ij}}=
-\sum_{i,j,m,n} \frac{
\lambda _{i3m}^{LQ} \lambda _{j3n}^{LQ} \lambda_{i3n}^{LQ*} \lambda _{j3m}^{LQ*}  \ln \left(\frac{m_{\phi _m}^2}{m_{\phi_n}^2}\right)}{4 (m_{\phi _m}^2-m_{\phi _n}^2) G_F^2 m_W^2 S_0\left(\frac{m_t^2}{m_W^2}\right)}
\approx -\sum_{i,j,m} \frac{
|\lambda _{i3m}^{LQ} \lambda _{j3m}^{LQ}|^2}{8.6 G_F^2 m_W^2 m_{\phi _m}^2}\ ,
\end{equation}
where we assumed that the leptoquarks have similar masses in the last equation and inserted $S_0\left(m_t^2/m_W^2\right)\approx 2.3$.

The UTfit collaboration~\cite{Bona:2007vi,UTfit:2013} simultaneously determines the CKM parameters and constraints on $\Delta F=2$ processes in terms of ratios of the meson mixing amplitudes $\braket{ M|\mathcal{H}_{\rm eff}|\bar M}$ normalized to the SM prediction:
\begin{subequations}
\begin{align}
C_{B_q}e^{2i\phi_{B_q}} &= \frac{\braket{B_q|\mathcal{H}_{\rm eff}^{full}|\bar B_q}}{\braket{B_q|\mathcal{H}_{\rm eff}^{SM}|\bar B_q}}=1+\frac{\braket{B_q|\mathcal{H}_{\rm eff}^{NP}|\bar B_q}}{\braket{B_q|\mathcal{H}_{\rm eff}^{SM}|\bar B_q}}\ ,\\
C_{\Delta m_K} &= \frac{\mathrm{Re} \left[\braket{B_q|\mathcal{H}_{\rm eff}^{full}|\bar B_q}\right]}{\mathrm{Re}\left[\braket{B_q|\mathcal{H}_{\rm eff}^{SM}|\bar B_q}\right]}=1+\frac{\mathrm{Re} \left[\braket{B_q|\mathcal{H}_{\rm eff}^{NP}|\bar B_q}\right]}{\mathrm{Re}\left[\braket{B_q|\mathcal{H}_{\rm eff}^{SM}|\bar B_q}\right]}\ ,\\
C_{\epsilon_K} &= \frac{\mathrm{Im} \left[\braket{B_q|\mathcal{H}_{\rm eff}^{full}|\bar B_q}\right]}{\mathrm{Im}\left[\braket{B_q|\mathcal{H}_{\rm eff}^{SM}|\bar B_q}\right]}= 1+\frac{\mathrm{Im} \left[\braket{B_q|\mathcal{H}_{\rm eff}^{NP}|\bar B_q}\right]}{\mathrm{Im}\left[\braket{B_q|\mathcal{H}_{\rm eff}^{SM}|\bar B_q}\right]}\ .
\end{align}
\end{subequations}
The current best fit values for these parameters are given by~\cite{Bona:2007vi,UTfit:2013,Silvestrini:2013private}
\begin{align}
C_{B_d}&=1.01\pm0.15\ ,&
\phi_{B_d}&=(2.2\pm3.7)^\circ\ ,\nonumber\\\label{eq:utfit}
C_{B_s}&=1.03\pm0.10\ ,&
\phi_{B_s}&=(-0.84\pm2.47)^\circ\ ,\\\nonumber
C_{\Delta m_K}&=0.978\pm0.331\ , &
C_{\epsilon_K}&=1.08\pm0.18\ .
\end{align}
Hence, $B$ meson mixing constrains the new physics contribution to be less than $\sim10-15\%$ of the SM contribution and the absorptive part of the $K$ meson mixing\footnote{As the SM contribution to $\Delta m_K$ is affected by long-distance contributions and can not be calculated reliably~\cite{Silvestrini:2013private}, we do not attempt to use it as constraint. The constraint shown in \Eqref{eq:utfit} is obtained under the assumption that the long-distance to $\Delta m_K$ can be as large as to saturate the experimental value.~\cite{Silvestrini:2013private} } constrains new physics to be less than $\sim20\%$ compared to the SM. Therefore, we can na\"{i}vely estimate
\begin{equation}
10^{-3} \,\sum_{i,j,m} \frac{|\lambda _{i3m}^{LQ} \lambda _{j3m}^{LQ}|^2}{0.1^4} \left(\frac{100\GeV}{m_{\phi _m}}\right)^2\ll0.1 \ ,
\end{equation}
which demonstrates that meson mixing does not lead to new competitive constraints.

\mathversion{bold}
\subsubsection{Constraints from the $b\to s$ Transition}
\mathversion{normal}
Recently a model-independent analysis~\cite{Altmannshofer:2011gn,*Altmannshofer:2012az} constrained the operators relevant for the $b\to s$ transition. We will apply the constraints from this analysis to our model. The new physics contributions to the $b\to s$ transition are described by the following effective Hamiltonian~\cite{Bobeth:1999mk,*Bobeth:2001jm}
\begin{equation}
\mathcal{H}_\text{eff}=-\frac{4G_F}{\sqrt{2}}V_{tb}V_{ts}^*\frac{e^2}{16\pi^2}\sum_i\left(C_i O_i+C_i^\prime O_i^\prime\right)+H.c.\ ,
\end{equation}
where the most sensitive operators to new physics are given by
\begin{align}
O_7^{(\prime)}&=\frac{m_b}{e}\left(\bar s\sigma_{\mu\nu} P_{R(L)} b\right) F^{\mu\nu}, & 
O_8^{(\prime)}&=\frac{g_s m_b}{e^2}\left(\bar s\sigma_{\mu\nu} T^\alpha P_{R(L)} b\right) G^{\mu\nu\,\alpha}, \nonumber\\
O_9^{(\prime)}&=\left(\bar s\gamma_\mu P_{L(R)} b\right) \left(\bar \ell \gamma^\mu \ell \right), & 
O_{10}^{(\prime)}&=\left(\bar s\gamma_\mu P_{L(R)} b\right) \left(\bar \ell \gamma^\mu \gamma_5\ell \right), \\\nonumber
O_S^{(\prime)}&=\frac{m_b}{m_{B_s}}\left(\bar s P_{R(L)} b\right) \left(\bar \ell \ell \right), & 
O_P^{(\prime)}&=\frac{m_b}{m_{B_s}}\left(\bar s P_{R(L)} b\right) \left(\bar \ell\gamma_5 \ell \right)\ .
\end{align}
The analysis in~\cite{Altmannshofer:2011gn,Altmannshofer:2012az} assumed $C_{9,10}^{(\prime)}$ to be independent of lepton flavor. In our model we expect all couplings $\lambda^{LQ}_{i3m}$ to be of the same order of magnitude and therefore the decays are almost independent of lepton flavor.  Although LFV decays are allowed in our case, this analysis should give bounds of the correct order of magnitude.
Reference~\cite{Altmannshofer:2011gn,Altmannshofer:2012az} also assumed the (pseudo-) scalar operators $C_{S,P}^{(\prime)}$ to be proportional to the lepton Yukawa coupling as well as conservation of lepton flavor.  They defined all Wilson coefficients at a matching scale $\mu_h=160\GeV$, with the operators $O_{7}^{(\prime)}$ and $O_8^{(\prime)}$ mixing under renormalization.  At the low-scale $\mu_b=2.8\GeV$, only the operator $O_7^{(\prime)}$ was important and therefore only the linear combination~\footnote{We denote the new physics contribution by $C_{i}^{(\prime)\mathrm{NP}}$.} $C_7^{(\prime)\mathrm{NP}}(\mu_b)=0.623\, C_7^{(\prime)\mathrm{NP}}(\mu_h)+0.101 \, C_8^{(\prime)\mathrm{NP}}(\mu_h)$ was constrained by the different $B$-decays, and all results were shown in the limit of vanishing $C_8^{(\prime)\mathrm{NP}}(\mu_h)$~\cite{Altmannshofer:2011gn,Altmannshofer:2012az}. In our case, there is a contribution to $C_8^{(\prime)}$, so we have to interpret all constraints on $C_7^{(\prime)\mathrm{NP}}(\mu_h)$ as constraints on $C_{7,eff}\equiv C_7^{(\prime)\mathrm{NP}}(\mu_h)+0.162\,  C_8^{(\prime)\mathrm{NP}}(\mu_h)$ as discussed in Ref.~\cite{Altmannshofer:2011gn}.

In our model, the Wilson coefficients in the approximation of massless lepton and light quark masses are given by 
\begin{align}
C_7&=\sum_{i,m}\frac{|\lambda _{i3m}^{LQ}|^2}{144 \sqrt{2} G_F m_{\phi _m}^2}\ ,&
C_7^\prime&=\frac{m_s}{m_b}\sum_{i,m}\frac{|\lambda _{i3m}^{LQ}|^2}{144 \sqrt{2} G_F m_{\phi _m}^2}\ ,\\
C_8&=-\sum_{i,m}\frac{|\lambda _{i3m}^{LQ}|^2}{48 \sqrt{2} G_F m_{\phi _m}^2}\ ,&
C_8^\prime&=-\frac{m_s}{m_b}\sum_{i,m}\frac{|\lambda _{i3m}^{LQ}|^2}{48 \sqrt{2} G_F m_{\phi _m}^2}\ ,
\end{align}
for the photonic and gluonic dipole operators, and therefore the combinations constrained by the analysis are
\begin{align}
C_{7,eff}&=0.514 \sum_{i,m}\frac{|\lambda _{i3m}^{LQ}|^2}{144 \sqrt{2} G_F m_{\phi _m}^2} \sim 2.2\times 10^{-4} \sum_{i,m} \left(\frac{|\lambda_{i3m}|}{0.1}\right)^2 \left(\frac{100\GeV}{m_{\phi_m}}\right)^2\ ,\\
C_{7,eff}^{\prime}&=0.514 \frac{m_s}{m_b}\sum_{i,m}\frac{|\lambda _{i3m}^{LQ}|^2}{144 \sqrt{2} G_F m_{\phi _m}^2}\sim 4.9\times 10^{-6} \sum_{i,m} \left(\frac{|\lambda_{i3m}|}{0.1}\right)^2 \left(\frac{100\GeV}{m_{\phi_m}}\right)^2
\ .
\end{align}
Neglecting the phase of $C_{7,eff}^{(\prime)}$ , the strongest constraint from Tab.~3 in~\cite{Altmannshofer:2012az} translates into $|C_{7,eff}|<0.017$ and $|C_{7,eff}^\prime|<0.20$ and therefore the constraints on $C_{7,eff}^{(\prime)}$ do not give competitive constraints on the leptoquarks $\phi_m$ compared to those from LFV processes.

The non-vanishing Wilson coefficients of the dimension-6 operators are evaluated as
\begin{align}\label{eq:C9C10}
C_9=-C_{10}=&\sum_m \frac{|\lambda^{LQ}_{i3 m}|^2}{32\sqrt{2} \pi \alpha_{em} G_F} \Bigg\{
m_t^2 y_t^2 D_0\left(m_W^2,m_{\phi _m}^2,m_t^2,m_t^2\right)
+g^2 \Big[C_0\left(m_W^2,0,m_t^2\right)
\\\nonumber
&+m_{\phi _m}^2 D_0\left(m_W^2,0,m_{\phi
   _m}^2,m_t^2\right)
-2 D_{00}\left(m_W^2,0,m_{\phi _m}^2,m_t^2\right)-2 D_{00}\left(m_W^2,m_{\phi
   _m}^2,m_t^2,m_t^2\right)\Big]
\Bigg\}\\\nonumber
&-\sum_{i,m,n}
\frac{\lambda _{13m}^{LQ} \lambda _{13n}^{LQ*} \lambda_{i3n}^{LQ} \lambda _{i3m}^{LQ*} D_{00}\left(0,m_{\phi _m}^2,m_{\phi
   _n}^2,m_t^2\right)}{16 \sqrt{2} \pi  \alpha _{em} G_F}\ ,
\end{align}
and we can estimate the magnitude for one leptoquark, $\phi_m$, as
\begin{equation}
C_9=-C_{10}\sim 0.011\left[1+0.012 \sum_i \left(\frac{|\lambda_{i3m}|}{0.1}\right)^2\right] \left(\frac{\lambda^{LQ*}_{k3m}\lambda^{LQ}_{k3m}}{0.01}\right) \left(\frac{100\GeV}{m_{\phi_m}}\right)^2\ .
\end{equation}
The right-handed counterparts, $C_{9,10}^\prime$, are proportional to $m_s$ and vanish in the limit of vanishing  quark masses for the first two generations. The Wilson coefficients $C_{S,P}^{(\prime)}$ vanish in the limit of massless neutrinos as well as the first two generations of leptons.
Again, a comparison with the constraints in Tab.~3 of~\cite{Altmannshofer:2012az} shows that the constraints on $C_{9,10}$ do not lead to competitive constraints on the leptoquarks $\phi_m$.

Finally, let us comment on the newly measured decay $B_s\to\mu^+\mu^-$ at the LHCb experiment~\cite{Aaij:2012nna,*Aaij:2013aka} as well as CMS experiment~\cite{Chatrchyan:2013bka}. This measurement improves the constraint on $C_9$ and $C_{10}$. Naively rescaling the bound by the square root of the uncertainty in the measured branching ratio over the upper bound, we expect at most an improvement by one order of magnitude compared to the previous constraint. However, the constraints from LFV processes are still much stronger even assuming a very optimistic improvement of one order of magnitude over the old constraint.

\mathversion{bold}
\subsubsection{$\Delta F=1$ Flavor Changing Neutral Current Processes at One-Loop Level}
\mathversion{normal}
In the previous subsection we have discussed constraints from the $b\to s$ transition and in this subsection we broaden our view in this direction. Specifically we consider $\Delta F=1$ operators, i.e.\ operators with a change of quark flavor. This brings in constraints from processes like (semi-) leptonic meson decays, precision measurements of $Z$-decay as well as collider searches for contact interactions. 
Neglecting all lepton/quark masses except for that of the top quark, the FCNC processes are determined by the following effective operator:
\begin{equation}
D_{ijkl} \;2\sqrt{2} G_F (\bar d_i \gamma^\mu P_L d_j) (\bar \ell_k \gamma_\mu P_L \ell_l)\ .
\end{equation}
Note that the corresponding operator with an up-type quark current vanishes in this limit. Evaluating the expression, we arrive at an analogous expression to \Eqref{eq:C9C10},
\begin{align}
D_{ijkl}&= \frac{V_{ti}^* V_{tj}}{32\pi^2} \Bigg\{
2 \sum_{i,m,n}  \lambda ^{LQ}_{l3m} \lambda ^{LQ*}_{k3n}\lambda^{LQ}_{i3n} \lambda ^{LQ*}_{i3m} D_{00}\left(0,m_{\phi_m}^2,m_{\phi _n}^2,m_t^2\right)\\\nonumber
&-\sum_m \lambda^{LQ*}_{k3m}\lambda^{LQ}_{l3m}\Big[
m_t^2 y_t^2 D_0\left(m_w^2,m_{\phi _m}^2,m_t^2,m_t^2\right)
+g^2 \Big(C_0\left(m_w^2,0,m_t^2\right)\\\nonumber
&+m_{\phi _m}^2 D_0\left(m_w^2,0,m_{\phi
   _m}^2,m_t^2\right)
-2   D_{00}\left(m_w^2,0,m_{\phi _m}^2,m_t^2\right)-2 D_{00}\left(m_w^2,m_{\phi
   _m}^2,m_t^2,m_t^2\right)\Big)
\Big]
\Bigg\}\ .
\end{align}
We can estimate the magnitude of the Wilson coefficient as
\begin{equation}
|D_{ijkl}|\sim 1.3\times 10^{-5}\;\left[1+ 0.012 \sum_i \left(\frac{|\lambda_{i3m}|}{0.1}\right)^2\right] V_{ti}^*V_{tj} \left(\frac{\lambda^{LQ*}_{k3m}\lambda^{LQ}_{l3m}}{0.01}\right) \left(\frac{100\GeV}{m_{\phi_m}}\right)^2\ .
\end{equation}
A comparison with the list of constraints on two-lepton, two-quark operators in Tab.~2 of Ref.~\cite{Carpentier:2010ue} shows that the most constrained operators are 
$D_{1212}<3\times 10^{-7}=8\times 10^{-4} \,|V_{td}^* V_{ts}|$, $D_{2312}<8\times 10^{-5}=1.9\times 10^{-3} \,|V_{ts}^* V_{tb}|$, $D_{2311}<1.8\times 10^{-4}=4.3\times 10^{-3} \,|V_{ts}^*V_{tb}|$, and $D_{2322}<7.0\times10^{-5}=1.7\times 10^{-3}\,|V_{ts}^* V_{tb}|$, where we already weighted each constraint by the CKM mixing. Most constraints originate from (semi-)leptonic $B$-decays as well as leptonic $K_L$ decays. We already discussed the constraints from the $b\to s$ transition in a previous section.
None of these constraints are competitive to the constraints from LFV processes.


\section{Constraints from Other Flavor- and Lepton-Number Violating Processes}
\label{sec:flavor2}
In the previous section, we have discussed the constraints from flavor physics for the Yukawa couplings required by the generation of neutrino masses. In the next two subsections, we address the remaining couplings, which will be generated radiatively in any case. We then discuss neutrinoless double beta-decay.

\subsection{Constraints Induced by Right-handed Coupling of the Third Generation}
Now we consider the constraints from processes induced by non-zero $\lambda^{eu}_{i3m}$.  Suppose this coupling constant is set equal to zero at some scale $\mu_0$.  At other scales, it will be nonzero, and can be estimated by looking at the non-diagonal contribution to the renormalization group equation from a vertex correction,
\begin{equation}
16\pi^2\frac{d\lambda^{eu}_{ikm}}{dt}\sim \left(Y_e^T\right)_{ij} \lambda^{LQ}_{jlm} (Y_u)_{lk} \ ,
\end{equation}
where $Y_u$ and $Y_e$ are the up-quark and charged-lepton Yukawa couplings of the standard model in the $LR$ convention, and $t=\ln\mu$. This leads to the estimate
\begin{equation}
\lambda^{eu}_{i3m}(\mu)\sim \frac{y_i y_t \lambda^{LQ}_{i3m}(\mu_0)}{16\pi^2}\ln\frac{\mu}{\mu_0}\ ,
\label{eq:lambdaeuapprox}
\end{equation}
i.e.~the couplings to the different charged leptons are suppressed by their respective Yukawa couplings $(y_e,\, y_\mu,\,y_\tau)=(2.9\times 10^{-6},\,6.1\times 10^{-4},\,0.01)$. 
The couplings of the first two right-handed generations of up-type quarks are even further suppressed. We will discuss them in the following subsection together with the left-handed couplings to the first two generations.

These couplings will lead to additional contributions to LFV processes. In \Appref{app:flavor}, we summarize the formulae of the additional contributions for radiative LFV decays, $\mu^-\to e^- e^+ e^+$ as well as $\mu\leftrightarrow e$ conversion in nuclei in order to illustrate their effects.
Generally, we expect similar constraints for the right-handed couplings (neglecting the left-handed couplings) as for the left-handed couplings. However, there might be cancellations and the bounds weakened if both couplings are present.
In order to illustrate the constraints on the couplings $\lambda^{eu}$ in presence of the couplings $\lambda^{LQ}$, we will consider six benchmark points and calculate the constraints for these points. The results are shown in \Tabref{tab:lamEUconstraints} and the bounds on $\lambda^{eu}_{33m}$ are of the order of $10^{-5} - 10^{-3}$ assuming the mentioned constraints. Obviously, the bounds get weaker the larger the overall mass scale.  The upper bounds are all higher than the estimates of Eq.~(\ref{eq:lambdaeuapprox}), so it is phenomenologically consistent to set these Yukawa couplings to zero at $\mu_0 \sim \GeV$.
\begin{table}[tb]\centering\setlength{\extrarowheight}{3pt}
\begin{tabular}{l|ccc|ccc}\hline\hline
& $m_f$ & $m_{\phi_1}$ &$m_{\phi_2}$ & \multicolumn{3}{c}{$\max|\lambda^{eu}_{33m}|$} \\
& [GeV] & [GeV] & [GeV]& [Br($\mu^-\to e^-\gamma$)] & [Br($\mu^-\to e^-e^+ e^-$)] & [Br($\mu N \to e N$)] \\\hline
1 & $10^3$ & $100$ & $200$ & $1.3\times 10^{-5}$ & $2.0\times 10^{-4}$ & $6.7\times 10^{-4}$\\
2 & $10^3$ & $200$ & $900$ & $7.7\times10^{-5}$&$1.3\times 10^{-3}$&$4.3\times 10^{-3}$\\
3 &$10^3$ & $900$ & $10^4$ & $9.8\times 10^{-5}$ & $1.7\times 10^{-3}$& $5.4\times 10^{-3}$\\
4 &$10^4$ & $10^3$ & $2\times 10^3$ & $7.6\times10^{-6}$ & $1.2\times 10^{-4}$&$3.9\times 10^{-4}$\\
5 &$10^4$ & $2\times 10^3$ &$9\times 10^3$ & $2.2\times 10^{-5}$ &  $3.8\times 10^{-4}$ & $1.2\times 10^{-3}$\\
6 &$10^4$ & $9\times 10^3$ &$10^5$ &$7.2\times 10^{-4}$ & $1.2\times 10^{-2}$& $4.0\times 10^{-2}$\\
\hline\hline
\end{tabular}

\caption{Experimental constraints on $\lambda^{eu}_{33m}$ from the three most constraining LFV processes under the assumption that both matrix elements are equal and $\lambda^{df}_{3m}=1$.}
\label{tab:lamEUconstraints}
\end{table}

\subsection{Constraints from Flavor-Violating Processes Induced by Coupling to the First Two Generations of Quarks}
Similarly to the right-handed coupling $\lambda^{eu}$, the couplings to the first two generations are generated radiatively as well. The wave-function renormalization of the left-handed quark doublet induces the following flavor-non-diagonal contributions to the coupling $\lambda^{LQ}$:
\begin{equation}
16\pi^2\frac{d\lambda^{LQ}_{ikm}}{dt}\sim \lambda^{LQ}_{ilm} \left( Y_u Y_u^\dagger + Y_d Y_d^\dagger \right)_{lk}=\lambda^{LQ}_{ilm} \left( \diag(y_u^2,\,y_c^2,\,y_t^2) + V \diag(y_d^2,\,y_s^2,y_b^2) V^\dagger \right)_{lk}\ ,
\end{equation}
where $Y_u$ and $Y_d$ are the up- and down-type SM quark Yukawa couplings.  Hence the RG contribution to the left-handed coupling to the first two generations of quarks can be estimated as
\begin{equation}
\lambda_{ikm}^{LQ}(\mu)\sim V_{tk}\frac{V_{tb}y_b^2\lambda^{LQ}_{i3m}(\mu_0)}{16\pi^2}\ln\frac{\mu}{\mu_0}\ ,
\end{equation}
or explicitly in terms of the Wolfenstein parameterization as
\begin{subequations}
\begin{align}
\lambda_{i1m}^{LQ}(\mu)&\sim A \lambda^3 (\rho+ i \eta)\, \frac{y_b^2\lambda^{LQ}_{i3m}(\mu_0)}{16\pi^2} \sim 10^{-8}\lambda^{LQ}_{i3m}(\mu_0)\ln\frac{\mu}{\mu_0}\ ,\\
\lambda_{i2m}^{LQ}(\mu)&\sim A \lambda^2\, \frac{ y_b^2\lambda^{LQ}_{i3m}(\mu_0)}{16\pi^2} \sim 10^{-7}\lambda^{LQ}_{i3m}(\mu_0)\ln\frac{\mu}{\mu_0}\ .
\end{align}
\end{subequations}
The couplings of the first two generations to the leptoquark $\phi$ and the fermionic color-octet $f$, described by $\lambda^{df}$, is not induced at one loop if the right-handed mixing in the down sector vanishes
\begin{equation}
16\pi^2\frac{d\lambda^{df}_{km}}{dt}\sim \lambda^{df}_{lm} \left( Y_d^\dagger Y_d \right)_{lk}=\lambda^{df}_{lm} \diag(y_d^2,\,y_s^2,y_b^2)_{lk}\ ,
\end{equation}
where the second equality holds for vanishing right-handed mixing. A small mixing is however induced at the two-loop order.

We will restrict ourselves to the discussion of the tree-level processes, since it is not the main focus of our paper, but would like to point out that there are also constraints from quark FCNCs, like meson mixing and $b\to s$ transitions, as we discussed in the previous section. In comparison to the previous section, these processes are not suppressed by the smallness of the CKM mixing angles. At tree-level, the leptoquark induces several operators of the type $LLQQ$
\begin{subequations}
\begin{align}
C_{ijkl}^{\ell u, LR} \, 2\sqrt{2} G_F (\bar \ell_i P_L \ell_j)(\bar u_k P_R u_l)\ ,&&
C_{ijkl}^{\ell u, RL} \, 2\sqrt{2} G_F (\bar \ell_i P_R \ell_j)(\bar u_k P_L u_l)\ ,\\
C_{ijkl}^{\ell u, VLL} \, 2\sqrt{2} G_F (\bar \ell_i \gamma^{\mu} P_L \ell_j)(\bar u_k \gamma_{\mu} P_L u_l)\ ,&&
C_{ijkl}^{\ell u, VRR} \, 2\sqrt{2} G_F (\bar \ell_i \gamma^{\mu} P_R \ell_j)(\bar u_k \gamma_{\mu} P_R u_l)\ ,\\
C_{ijkl}^{\ell u, TLR} \, 2\sqrt{2} G_F (\bar \ell_i \sigma^{\mu\nu} P_L \ell_j)(\bar u_k \sigma_{\mu\nu} P_R u_l)\ ,&&
C_{ijkl}^{\ell u, TRL} \, 2\sqrt{2} G_F (\bar \ell_i \sigma^{\mu\nu} P_R \ell_j)(\bar u_k \sigma_{\mu\nu} P_L u_l)\ ,\\
C_{ijkl}^{\nu d, VLL} \, 2\sqrt{2} G_F (\bar \nu_i \gamma^{\mu} P_L \nu_j)(\bar d_k \gamma_{\mu} P_L d_l)\ ,&&
C_{ijkl}^{\nu d, VRL} \, 2\sqrt{2} G_F (\bar \nu_i^C \gamma^{\mu} P_R \nu_j^C)(\bar d_k \gamma_{\mu} P_L d_l)\ ,\\
C_{ijkl}^{CC , RR} \, 2\sqrt{2} G_F (\bar \nu_i P_R \ell_j)(\bar d_k P_R u_l)\ ,&&
C_{ijkl}^{CC , VLL} \, 2\sqrt{2} G_F (\bar \nu_i \gamma^\mu  P_L \ell_j)(\bar d_k \gamma_\mu P_L u_l)\ ,\\
C_{ijkl}^{CC , TRR} \, 2\sqrt{2} G_F (\bar \nu_i \sigma^{\mu\nu} P_R \ell_j)(\bar d_k \sigma_{\mu\nu}P_R u_l)\ .
\end{align}
\end{subequations}
The Wilson coefficients are given by
\begin{subequations}
\begin{align}
C_{ijkl}^{\ell u, LR} &=-\frac{\lambda _{jkm}^{LQ} \lambda _{ilm}^{eu*}}{4\sqrt{2} G_F m_{\phi _m}^2}\ ,&
C_{ijkl}^{\ell u, RL}&=-\frac{\lambda _{jkm}^{eu} \lambda _{ilm}^{LQ*}}{4 \sqrt{2} G_F m_{\phi _m}^2}\ ,\\
C_{ijkl}^{\ell u, VLL}&=-\frac{\lambda _{jkm}^{LQ} \lambda _{ilm}^{LQ*}}{4 \sqrt{2} G_F m_{\phi _m}^2}\ ,&
C_{ijkl}^{\ell u, VRR}&=-\frac{\lambda _{jkm}^{eu} \lambda _{ilm}^{eu*}}{4 \sqrt{2} G_F m_{\phi _m}^2}\ ,\\
C_{ijkl}^{\ell u, TLR} &=-\frac{\lambda _{jkm}^{LQ} \lambda _{ilm}^{eu*}}{16 \sqrt{2} G_F m_{\phi _m}^2}\ ,&
C_{ijkl}^{\ell u, TRL} &=-\frac{\lambda _{jkm}^{eu} \lambda _{ilm}^{LQ*}}{16 \sqrt{2} G_F m_{\phi _m}^2}\ ,\\
C_{ijkl}^{\nu d, VLL}&=\frac{\lambda _{\hat j\hat km}^{LQ} \lambda _{\hat i\hat lm}^{LQ*}}{4 \sqrt{2} G_F m_{\phi _m}^2}V_{\hat i i}^*V_{\hat j j} U_{\hat k k} U_{\hat l l}^*\ ,&
C_{ijkl}^{\nu d, VRL}&=-\frac{\lambda _{\hat i\hat km}^{LQ} \lambda _{\hat j\hat lm}^{LQ*}}{4 \sqrt{2} G_F m_{\phi _m}^2}V_{\hat i i}V_{\hat j j}^* U_{\hat k k} U_{\hat l l}^*\ ,\\
C_{ijkl}^{CC , RR} &=\frac{\lambda _{jlm}^{eu} \lambda _{\hat i\hat km}^{LQ*}}{4 \sqrt{2} G_F m_{\phi _m}^2}V_{\hat i i}^*U_{\hat k k}^*\ ,&
C_{ijkl}^{CC , VLL}&=-\frac{\lambda _{jlm}^{LQ} \lambda _{\hat i\hat km}^{LQ*}}{4 \sqrt{2} G_F m_{\phi _m}^2}V_{\hat i i}^*U_{\hat k k}^*\ ,\\
C_{ijkl}^{CC, TRR} &=-\frac{\lambda _{jlm}^{eu} \lambda _{\hat i\hat km}^{LQ*}}{16 \sqrt{2} G_F m_{\phi _m}^2}V_{\hat i i}^*U_{\hat k k}^*\ .
\end{align}
\end{subequations}
Using the analysis in~\cite{Carpentier:2010ue}, we can derive bounds for the different operators.\footnote{Note that the analysis in~\cite{Carpentier:2010ue} assumed no accidental cancellations between the different processes.} The tensor operators have not been studied in~\cite{Carpentier:2010ue}, neither possible scalar operators, where the color indices are not contracted within the same fermion chain. However, similar bounds should apply as for the scalar operators. Therefore, we will use the bound on the corresponding scalar operator to obtain an order of magnitude estimate for the bound of the tensor operator. As the CKM (PMNS) mixing angles are small (large), we approximate them by the identity (democratic, $U_{ij}\sim 1/\sqrt{3}$) matrix. Under these assumptions, we obtain constraints on the products of the couplings $\lambda^{LQ,eu} \lambda^{LQ,eu*}$. We will only highlight the most important constraints and refer the interested reader to the tables in~\cite{Carpentier:2010ue}, particularly Tabs.~3,6,7 as well as Tabs.~12-15. 
The most stringent constraints originate from leptonic meson decays, especially charged meson decays, as well as $\mu\leftrightarrow e$ conversion in nuclei. The constraints from leptonic charged meson decays especially constrain the operator $C_{ijkl}^{CC , RL} \sim 2.3\,\lambda^{eu}_{jlm}\lambda_{ikm}^{LQ*}m_W^2/m_{\phi_m}^2$ neglecting mixing. A selection of constraints from Tab. 14 in~\cite{Carpentier:2010ue}:
\begin{itemize}
\item The measurement of the ratio of pions decaying to electrons vs muons, $R_\pi$, constrains $\lambda^{eu}_{i1m}\lambda_{11m}^{LQ*}\lesssim 6.8\times 10^{-6} m_{\phi_m}^2/m_W^2$ as well as $\lambda^{eu}_{i1m}\lambda_{21m}^{LQ*}\lesssim 1.4\times 10^{-3} m_{\phi_m}^2/m_W^2$.
\item The measurement of $K^+\to \bar e\nu_i$ leads to $\lambda^{eu}_{i2m}\lambda_{11m}^{LQ*}\lesssim 6.4\times 10^{-6} m_{\phi_m}^2/m_W^2$.
\item The measurement of $R_K$ leads to $\lambda^{eu}_{i2m}\lambda_{21m}^{LQ*}\lesssim 1.3\times 10^{-3} m_{\phi_m}^2/m_W^2$.
\item The measurement of $B^+\to \bar e \nu$ leads  $\lambda^{eu}_{i3m}\lambda_{11m}^{LQ*}\lesssim 7.7\times 10^{-5} m_{\phi_m}^2/m_W^2$.
\item The measurement of $B^+\to \bar \mu \nu$ leads to $\lambda^{eu}_{i3m}\lambda_{21m}^{LQ*}\lesssim 4.3\times 10^{-5} m_{\phi_m}^2/m_W^2$.
\item The measurement of $B^+\to \bar \tau \nu$ leads to $\lambda^{eu}_{i3m}\lambda_{31m}^{LQ*}\lesssim 3.5\times 10^{-4} m_{\phi_m}^2/m_W^2$.
\item The measurement of $D^+\to \bar \mu \nu$ leads to $\lambda^{eu}_{i1m}\lambda_{22m}^{LQ*}\lesssim 1.6\times 10^{-3} m_{\phi_m}^2/m_W^2$.
\end{itemize} 
Furthermore, we would like to highlight the constraint from $\mu\leftrightarrow e$ conversion
in nuclei via a leptoquark exchange $\mu u \to \phi^{\star} \to e u$. 
In this case the only relevant nonzero couplings in the effective Lagrangian \Eqref{eq:mue} are
\begin{subequations}
\begin{align}
g_{LS(u)}&=-\sum_m\frac{\lambda _{21m}^{LQ} \lambda _{11m}^{eu*}}{2 m_{\phi _m}^2}\ ,&
g_{RS(u)}&=-\sum_m\frac{\lambda _{21m}^{eu} \lambda _{11m}^{LQ*}}{2 m_{\phi _m}^2}\ ,\\
g_{LV(u)}&=\sum_m \frac{\lambda _{21m}^{LQ} \lambda _{11m}^{LQ*}}{2 m_{\phi _m}^2}\ ,&
g_{RV(u)}&=\sum_m\frac{\lambda _{21m}^{eu} \lambda _{11m}^{eu*}}{2 m_{\phi _m}^2}\ . 
\end{align}
\end{subequations}
Plugging them in \Eqref{eq:mu2eOmega}, we obtain for the conversion rate
\begin{align}
\omega_\mathrm{conv}=&\left|\sum_m m_{\phi_m}^{-2}\left[ \left(G_S^{u,p} S^{(p)} +G_S^{u,n} S^{(n)}\right) \lambda _{21m}^{LQ} \lambda _{11m}^{eu*} + \left(2V^{(p)}+V^{(n)}\right) \lambda_{21m}^{LQ}\lambda_{11m}^{LQ*}\right]\right|^2 \nonumber\\
+&\left|\sum_m m_{\phi_m}^{-2}\left[ \left(G_S^{u,p} S^{(p)} +G_S^{u,n} S^{(n)}\right) \lambda _{21m}^{eu} \lambda _{11m}^{LQ*} + \left(2V^{(p)}+V^{(n)}\right) \lambda_{21m}^{eu}\lambda_{11m}^{eu*}\right]\right|^2\ .
\end{align}
Assuming that there is no accidental cancellation between the different terms, we obtain bounds on the different coupling combinations of the order of 
\begin{subequations}
\begin{align}
\lambda _{21m}^{LQ} \lambda _{11m}^{eu*}&\lesssim \left(4\times 10^{-9}-7\times 10^{-8}\right)\frac{m_\phi^2}{m_W^2}\ , &
\lambda _{21m}^{eu} \lambda _{11m}^{LQ*}&\lesssim  \left(4\times 10^{-9}-7\times 10^{-8}\right)\frac{m_\phi^2}{m_W^2}\ ,\\
\lambda _{21m}^{LQ} \lambda _{11m}^{LQ*}&\lesssim \left(10^{-8} - 10^{-7}\right)\frac{m_\phi^2}{m_W^2}\ , &
\lambda _{21m}^{eu} \lambda _{11m}^{eu*} &\lesssim \left(10^{-8} - 10^{-7}\right) \frac{m_\phi^2}{m_W^2}\ . 
\end{align}
\end{subequations}

We conclude that there are already strong constraints on the couplings to the first two generations. In particular the coupling to the first generation of quarks is constrained by $\pi^\pm$ decays via the measurement of $R_\pi$ and the constraints from $\mu\leftrightarrow e$ conversion in nuclei.

\subsection{Neutrinoless Double Beta-Decay}
\begin{figure}\centering
\begin{subfigure}{5cm}\centering
\FMDG{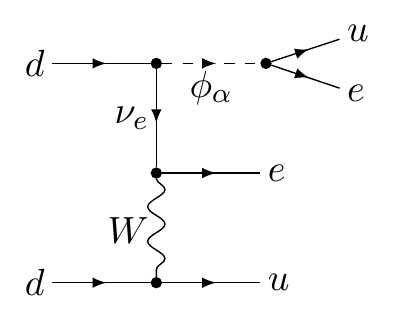}
\caption{Long-range contribution}
\end{subfigure}
\hspace{1cm}
\begin{subfigure}{5cm}\centering
\FMDG{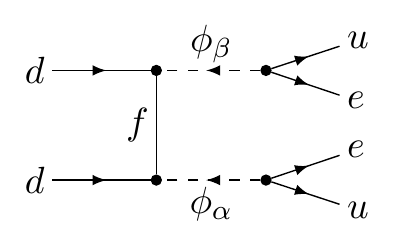}
\caption{Short-range contribution}
\end{subfigure}
\caption{New contributions to neutrinoless double beta-decay.\label{fig:0nu2beta}}
\end{figure}

As well as constraints
from flavor violation, there are also, in principle, constraints from
total lepton-number violating processes such as neutrinoless double-beta
decay.
A complete classification of all possible tree-level contributions to the relevant effective dimension-9 operators of neutrinoless double beta-decay has been given in Ref.~\cite{Bonnet:2012kh}. There are two possible contributions in this model besides the standard contribution from light neutrinos; they are shown in \Figref{fig:0nu2beta}, the short-range contribution 5-i in the notation of~\cite{Bonnet:2012kh} as well as the long-range contribution 2-i-b.
The couplings of the leptoquark contributing to neutrinoless double beta-decay, i.e.\ $\lambda^{LQ}_{i1\alpha}$, $\lambda^{eu}_{i1\alpha}$ as well as $\lambda^{df}_{1\alpha}$, do not enter the expression for neutrino masses and thus are not constrained from below and can be arbitrary without affecting the neutrino mass contribution. Hence, we are left with the standard contribution from light neutrinos. The magnitude of the effective mass controlling the light neutrino contribution to neutrinoless double beta decay is given by
\begin{equation}
\ev{m_{ee}}=\sum U_{ei}^2m_i\ .
\end{equation}
As the mass of the lightest neutrino almost vanishes, we read off $\ev{m_{ee}}\simeq (0.2 - 4)\times 10^{-3}\eV$  [$(1-4)\times 10^{-2}\eV$] for $m_1\simeq 0\eV$ from Fig.~2 in~\cite{Rodejohann:2012xd} in case of a normal [inverted] mass ordering varying the other neutrino mass parameters in their $3\sigma$ ranges. Concluding, there are currently no constraints from neutrinoless double beta decay on the parameter space of this model.


\section{Collider Constraints}
\label{sec:collider}
As there are several new particles in the model, which might have masses close to the electroweak scale, they can be searched for at colliders. In particular, hadron colliders, like the LHC, seem promising, because all new particles carry color charge and therefore couple to gluons. We will not perform a detailed collider study, since it is beyond the scope of this paper and thus left for future work, but we will discuss the different search channels for the scalar leptoquark and the fermionic colored octet in the following subsections.


\subsection{Scalar Leptoquark}
\begin{figure}\centering
\begin{subfigure}{0.24\linewidth}
\FMDG{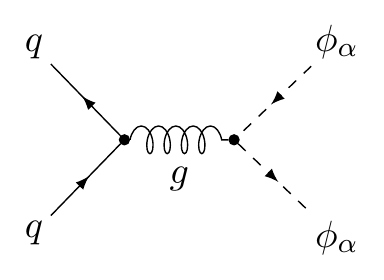}
\caption{}
\end{subfigure}
\begin{subfigure}{0.24\linewidth}
\FMDG{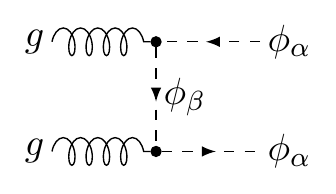}
\caption{}
\end{subfigure}
\begin{subfigure}{0.24\linewidth}
\FMDG{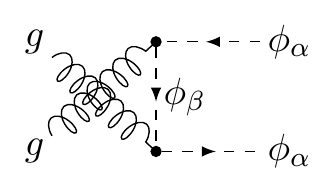}
\caption{}
\end{subfigure}
\\
\begin{subfigure}{0.24\linewidth}
\FMDG{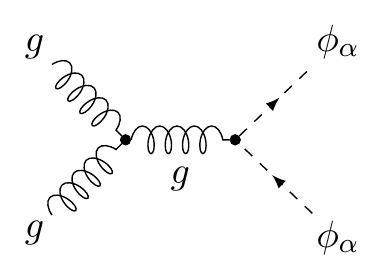}
\caption{}
\end{subfigure}
\begin{subfigure}{0.24\linewidth}
\FMDG{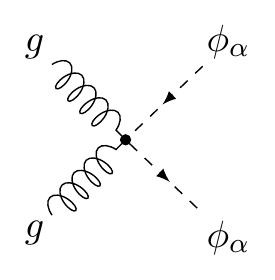}
\caption{}
\end{subfigure}
\caption{Main pair production channels for leptoquarks\label{fig:LQprod}}
\end{figure}

The scalar leptoquark $\phi_\alpha$ is mainly pair-produced in gluon fusion as well as $q\bar q$ annihilation, while the production via the exchange of leptons through Yukawa interactions is suppressed. The main pair production channels are shown in \Figref{fig:LQprod}. There are more pair-production channels; however, they rely on Yukawa couplings to the first generation of quarks, which are highly constrained by flavor physics, as we discussed in \Secref{sec:flavor2}. 
Besides pair production, the leptoquark $\phi_\alpha$ can also be singly produced via its Yukawa interactions $\lambda^{LQ,eu}$, but they are similarly suppressed compared to the pair-production channels relying on Yukawa couplings.

The leptoquark will decay into either a down-type, mostly $b$, quark and a neutrino or a top quark and a charged lepton. The decay into $b$ quark plus missing transverse energy (MET) resembles the standard direct sbottom pair production search, but with a reduced branching fraction for $m_{\phi_\alpha}>m_t$. Hence, the possible collider signatures are two $b$-jets plus MET, leptons and/or additional jets in the final state. 
The leptoquarks in this model differ from the standard assumption in leptoquark searches that the leptoquark couples to one generation only. Due to the flavor structure of the neutrino mass matrix, the leptoquark has to couple to the third generation of left-handed quarks as well as all three generations of left-handed leptons. The couplings to the right-handed particles as well as the first two generations of left-handed quarks are not required in order to generate neutrino masses and therefore can be arbitrarily small. In fact, the couplings to the first two generations of quarks are strongly constrained by collider searches and especially flavor physics constraints, as we discussed in \Secref{sec:flavor2}.

The CMS Collaboration has performed a search for two $b$-jets plus MET, leptons and/or additional jets in the final state, based on 4.7 fb$^{-1}$ of $\sqrt{s}=$7 TeV data~\cite{Chatrchyan:2012st}.  The analysis sets a lower limit on $m_{\phi}$ of 450 GeV assuming a 100\% branching fraction to a $b$-quark and a $\tau$-neutrino. However the limit from this search is much lower for realistic branching fractions, since the branching fraction decreases from $100\%$ for leptoquarks lighter than the top quark down to $50\%$ for a leptoquark mass much heavier than the top quark. In our model, we expect a limit slightly above the top mass.  Nonetheless this situation is certain to change with increased data and the LHC should be able to provide important constraints on this model. A detailed study is left for future work.


\subsection{Fermionic Colored Octet}

\begin{figure}\centering
\begin{subfigure}{0.24\linewidth}
\FMDG{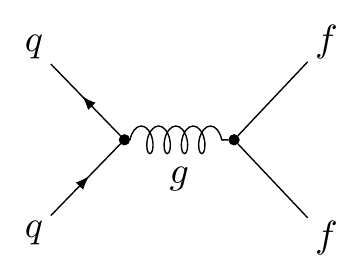}
\caption{}
\end{subfigure}
\begin{subfigure}{0.24\linewidth}
\FMDG{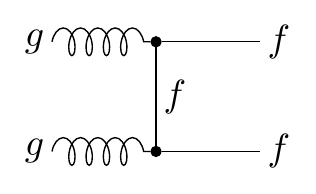}
\caption{}
\end{subfigure}
\begin{subfigure}{0.24\linewidth}
\FMDG{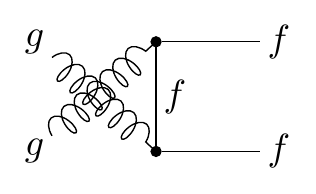}
\caption{}
\end{subfigure}
\begin{subfigure}{0.24\linewidth}
\FMDG{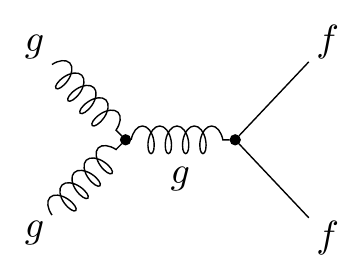}
\caption{}
\end{subfigure}
\caption{Pair production channels of the fermionic colored octet\label{fig:colOctProd}}
\end{figure}

The fermionic colored octet $f$ is pair-produced similarly to a gluino, dominantly via gluon fusion. The main production channels are shown in \Figref{fig:colOctProd}.
It decays via the coupling $\lambda^{df}$ into a down-type, mostly $b$-, quark and a leptoquark, which decays via the interactions $\lambda^{LQ,eu}$ to either a $b$-quark and a neutrino or a top quark and a charged lepton,
\begin{equation}
f\to b\phi^\dagger \to bb \nu\ , \quad\mathrm{and}\quad
f\to b\phi^\dagger \to bt \ell\ .
\end{equation}
The top quark decays in the usual way into a $W$ boson and a $b$-quark and therefore leads to another $b$-jet in the final state as well as additional jets, or a lepton and MET. Depending on the masses, the decays are via off-shell particles. Hence the final state contains four $b$-jets and depending on the decay channel MET, leptons and/or additional jets. 
As the fermionic colored octet is a Majorana fermion, it can lead to a fermion number violating process and lead to like-sign top as well as lepton pairs in the final state, which are ideal channels to search for new physics due to the small SM background.
Due to the similarity of production as well as some of the decay channels with a gluino, we expect that the LHC experiments should be already able to produce reasonable constraints on the colored fermionic octet. However, none of the current searches is directly applicable without a simulation.

In order to set a limit on $m_f$, a recent gluino search from the CMS Collaboration could be exploited~\cite{Chatrchyan:2012sa,*Chatrchyan:2012paa}. This result is useful as they present the raw numbers obtained at a number of intermediate steps in the analysis, which would allow the differences between our present model and the SUSY gluino scenario to be accounted for.  Despite this, we have not done this in the present paper for pragmatic reasons.  As seen in \Secref{sec:constraints}, the neutrino mass generated in our model is only weakly dependent on $m_f$, and derived limits are unlikely to meaningfully constrain the model.  Future searches may alter this conclusion, but the more interesting constraint colliders can place on this model is through $m_{\phi_\alpha}$.


\section{Naturalness Constraints}
\label{sec:naturalness}

As there are three scalars in the model, we have to discuss the effects of quadratic corrections to the scalar masses from other particles. We will firstly consider the contributions of the scalar leptoquarks as well as the fermionic colored octet to the SM Higgs and then calculate the quadratic correction to the leptoquark masses.

\subsection{Contributions to the SM Higgs}
\begin{figure}\centering
\begin{subfigure}{6cm}
\FMDG{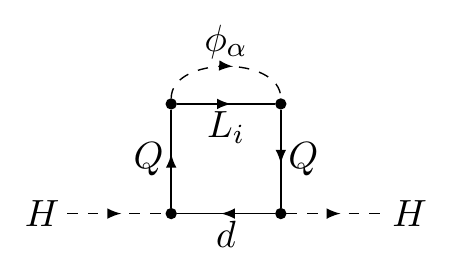}
\caption{Relevant contribution proportional to $(\lambda^{LQ})^2$.}
\end{subfigure}
\hspace{1cm}
\begin{subfigure}{6cm}
\FMDG{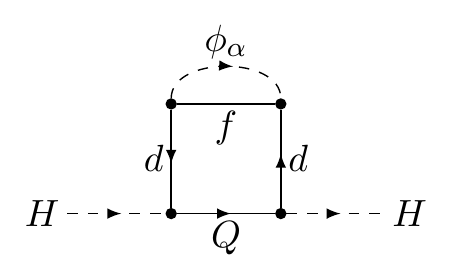}
\caption{Relevant contribution proportional to $(\lambda^{df})^2$.}
\end{subfigure}
\caption{Different contributions to the effective Higgs potential\label{fig:naturalnessDiagrams}}
\end{figure}

The newly introduced particles, the colored octet fermion as well as the two leptoquarks, contribute to the effective Higgs potential. For completeness, we give the tree-level Higgs potential
\begin{equation}
V=-\mu^2 H^\dagger H + \lambda (H^\dagger H)^2
\end{equation}
and therefore the VEV $\ev{H}$ and the Higgs mass $m_h^2$ are given by
\begin{equation}
\ev{H}^2 = \frac{\mu^2}{2\lambda}\ , \quad\mathrm{and}\quad
m_h^2 = 4 \mu^2=8\lambda \ev{H}^2\ .
\end{equation}
We are especially interested in corrections to the mass terms, since they are quadratically sensitive to the new mass scales. 
Even setting all new couplings which do not enter the neutrino mass contribution to zero at some scale, there is a contribution dependent on the parameters $\lambda^{LQ}_{i3\alpha}$ as well as $\lambda^{df}_{3\alpha}$ at two-loop order,
\begin{equation}\label{eq:natQuad}
-\frac{|y_b|^2}{(4\pi)^4}  \sum_{\alpha} \left[3\,\sum_i\left|\lambda^{LQ}_{i3\alpha} \right|^2  f_1(m_{\phi_\alpha}^2) 
+4\, \left|\lambda^{df}_{3\alpha}\right|^2 f_2(m_{\phi_\alpha}^2,m_f^2)\right] H^\dagger H\ ,
\end{equation}
where the functions $f_i$ encode the structure of the two-loop diagram.
The relevant diagrams are shown in \Figref{fig:naturalnessDiagrams}.
We will use this contribution to estimate the natural region of parameter space, where the contribution to the Higgs mass is less than $126\GeV$, the measured value of the resonance at the LHC:
\begin{subequations}
\begin{align}
\frac{12|y_b|^2}{(4\pi)^4} \sum_i\left|\lambda^{LQ}_{i3\alpha}\right|^2 m_{\phi_\alpha}^2&\lesssim 4 \mu^2 = m_h^2
&\Rightarrow &
m_{\phi_\alpha}\lesssim \frac{1900}{\sqrt{\sum_i\left|\lambda^{LQ}_{i3\alpha}\right|^2}} m_h \\
\frac{16|y_b|^2}{(4\pi)^4} \left|\lambda^{df}_{3\alpha}\right|^2 \max (m_{\phi_\alpha}^2,m_f^2)&\lesssim 4 \mu^2 = m_h^2 
&\Rightarrow &
m_{\phi_\alpha},m_f\lesssim \frac{1600}{\left|\lambda^{df}_{3\alpha}\right|}m_h\ ,
\end{align}
\end{subequations}
where we assumed that the loop functions $f_i(m^2)\sim m^2$.
Similarly, at three-loop order, there are contributions to the quartic Higgs coupling. However, they are too small to lead to any competitive constraints. Summarizing, naturalness prefers smaller masses of the new particles.

\subsection{Naturalness of the Leptoquark Masses}
\begin{figure}\centering
\begin{subfigure}{6cm}
\FMDG{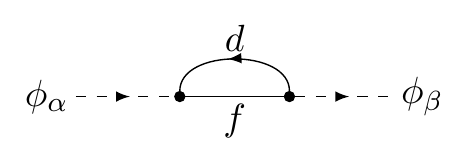}
\caption{Relevant contribution proportional to $(\lambda^{df})^2$.}
\end{subfigure}
\begin{subfigure}{6cm}
\FMDG{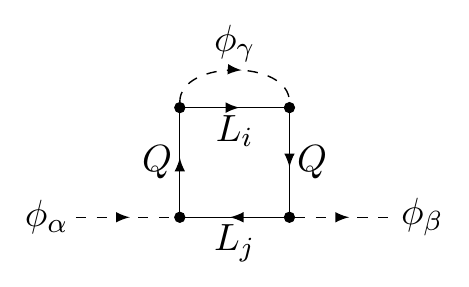}
\caption{Relevant contribution proportional to $(\lambda^{LQ})^4$.}
\end{subfigure}
\caption{Different contributions to the effective potential of the leptoquark\label{fig:naturalnessDiagramsLQ}}
\end{figure}

Similarly to the Higgs mass, the mass term of the leptoquark receives corrections to its mass. The most important diagrams are shown in \Figref{fig:naturalnessDiagramsLQ}. We can readily estimate the contribution due to the fermionic colored octet $f$
\begin{equation}
(\delta m_{\phi}^2)_{\alpha\beta}= -\frac43\frac{\lambda^{df}_{3\alpha}\lambda^{df*}_{3\beta}}{16\pi^2} A_0[m_f^2] \Rightarrow m_f\lesssim \frac{2\sqrt{3}\pi}{\left|\lambda^{df}_{3\alpha}\right|} m_{\phi_\alpha}
\end{equation}
and therefore the hierarchy between the leptoquark masses and the colored octet mass is limited by naturalness depending on the coupling $\lambda^{df}$. Similarly, the hierarchy between the leptoquark masses itself is limited by naturalness
\begin{equation}
(\delta m_\phi^2)_{\alpha\beta} = -\sum_{\gamma,i,j}\frac{\lambda^{LQ}_{j3\alpha}\lambda^{LQ}_{j3\beta}}{(4\pi)^4}   \left|\lambda^{LQ}_{i3\gamma}\right|^2 h_1(m_{\phi_\gamma}^2)\Rightarrow
m_{\phi_\gamma}\lesssim \frac{16\pi^2}{\sqrt{\sum_{i,j}|\lambda^{LQ}_{j3\alpha}\lambda^{LQ}_{i3\gamma}|^2}} m_{\phi_\alpha}\ ,
\end{equation}
where $h_1$ encodes the loop integral structure. 
We do not take into account any corrections to the quartic interactions, since they do not lead to any competitive constraints. Concluding, naturalness disfavors a large hierarchy in the masses of the new particles.
%

\section{Mathematica Code}
\label{sec:mathematica}

Together with this paper, we publish the Mathematica package \texttt{ANT}, which provides all Passarino-Veltman (PV) functions up to boxes in the limit of vanishing external momenta using the definitions of \texttt{FormCalc}/\texttt{LoopTools}~\cite{Hahn:1998yk}.
In addition, it includes all first derivatives with respect to the external momenta of the following PV functions
\begin{equation*}
B_0,\quad B_1,\quad C_0,\quad C_1,\quad C_2,\quad \mathrm{and}\quad C_{00}\;,
\end{equation*}
which were necessary for the calculations in this work. See \Appref{sec:appPV} for the list of all PV functions contained in the Mathematica package. After loading the Mathematica package, 
the function \texttt{ANT} is available, which can be used to evaluate an arbitrary expression containing PV functions given in the limit of vanishing external momenta. Besides the function \texttt{ANT}, it defines the functions \texttt{A0ant}, \texttt{B0ant}, \texttt{C0ant}, and \texttt{D0ant}, which directly evaluate to the result for the corresponding functions \texttt{A0}, \texttt{B0i}, \texttt{C0i}, and \texttt{D0i}, respectively, in the given limit.
The package \texttt{ANT} can be downloaded from \href{http://ant.hepforge.org}{http://ant.hepforge.org}.

In order to illustrate the use of \texttt{ANT}, we give one short example and refer the interested reader to the documentation for further information.
The following example code
\definecolor{shadecolor}{rgb}{0.9,0.9,0.9}
\begin{snugshade}
\begin{verbatim}
<< ANT.m;
expr=a + b C0i[cc0,0,0,0,m1s,m2s,0];
ANT[expr]
\end{verbatim}
\end{snugshade}
\noindent leads to the output
$$
a+b\, \frac{2\, \text{m1s}^2 \log (\text{m2s})+3\, \text{m1s}^2-2 \,\text{m1s}^2\, \log (\text{m1s})-4\, \text{m1s} \,\text{m2s}+\text{m2s}^2}{6 \,(\text{m1s}-\text{m2s})^3}\;.
$$


\section{Summary and Conclusions}
\label{sec:conc}

The origin of neutrino masses and mixings is one of the most important topics in physics beyond the standard model.  The three well-known see-saw models are tree-level scenarios for Majorana neutrino mass generation that lie at one end of a whole sequence of possible models.  Viewing these theories from the vantage point of gauge-invariant and baryon-symmetric effective operators that violate lepton number conservation by two units, the see-saw models are seen to be the simplest UV completions of the dimension-5 Weinberg operator $LLHH$.  The other models in the sequence are based on more complicated effective operators, and they necessarily feature radiative neutrino mass generation.  We have constructed and thoroughly analyzed the constraints on a new model of this kind, based on a certain UV completion of an effective operator with flavor structure $LLQd^cQd^c$.  This is the first time, to our knowledge, that a radiative neutrino mass model involving exotic fermions in addition to exotic scalars has been so exhaustively studied.  Neutrino masses and mixings are generated at 2-loop level using a Majorana color-octet fermion and two copies of a color-triplet, isosinglet, charge $-1/3$ leptoquark scalar.

Fitting the parameters to reproduce the observed neutrino masses and mixings, we then derived the constraints on the remaining parameter space from various, mainly flavor-changing processes.  We focused on the region of parameter space where the only relevant Yukawa couplings were to the third generation.  We found that the lepton-flavor-violating processes of $\mu$ to $e$ conversion in nuclei, $\mu \to e \gamma$ and $\mu \to eee$ provided the most stringent constraints, as summarized in Figs.~\ref{fig:LFVdecays}-\ref{fig:Mu2E}.  We also derived bounds from other processes -- exotic contributions to $Z$ and $t$ decay, neutral-meson mixing, $b \to s \gamma$ -- that turned out to less constraining on the parameters of this particular model.  Finally, we estimated bounds on Yukawa coupling constants that play no role in the generation of neutrino mass but are required by the theory to exist.  These included the couplings of the leptoquarks to right-handed up-type quarks and charged leptons, and other Yukawa couplings to the first two generations of quarks and leptons.

Collider constraints were also examined.  While the leptoquarks have the same quantum numbers as right-handed sbottoms, and similar states have in general have been searched for at the LHC, currently no useful collider bounds exist.  The reason is that the characteristic leptoquark decay branching ratios in this model, as driven by their role in neutrino mass generation, place them out of view of current searches.  This situation is expected to change in the future, however.  The exotic fermion resembles a gluino, but since the neutrino masses and mixings do not depend critically on its mass, current collider constraints are not significant.

The dashed black lines in Figs.~\ref{fig:LFVdecays}-\ref{fig:Mu2E} show expected sensitivities of future experiments.  Clearly, the best prospects lie in the quite significant improvements expected in $\mu$ to $e$ conversion measurements~\cite{Hungerford:2009zz,*Cui:2009zz}.  In addition, it would be very useful to extend the scope of LHC searches for leptoquark scalars to encompass a wider range of decay mode branching ratios.


\section*{Acknowledgements}
We thank K.~S.~Babu, Y.~Bai, I.~Baldes, J.~Barnard, T.~Gherghetta, A.~de Gouv\^{e}a, K.~McDonald, B.~McKellar and M.~White for many very useful discussions. We thank L. Silvestrini for providing a constraint on $C_{\Delta m_K}$ and explaining why it has been removed from the list. We thank D.~Straub for pointing out a mistake in the previous version. This work was supported in part by the Australian Research Council.


\appendix

\section{More Details of Flavor Physics}
\label{app:flavor}
In this section, we summarize general formula for the contributions to LFV rare decays involving left- as well as right-handed couplings.
\mathversion{bold}
\subsection{$l_i\to l_j\gamma$}
\mathversion{normal}
Contributions to $l_i\to l_j\gamma$ from left- as well as right-handed couplings to the third generation of quarks are given by
\begin{subequations}
\label{eq:sigmaRH}
\begin{align}
\nonumber
\sigma_{Lij} =& \frac{1}{16\pi^2} 
\sum_{a=1}^3\sum_{m=1}^2
 \frac{\lambda_{iam}^{LQ}\lambda_{jam}^{LQ\dagger} m_{l_i} +\lambda_{iam}^{eu}\lambda_{jam}^{eu\dagger} m_{l_j} }{m_{\phi_m}^2}
\frac{1+4t_{am}-5t_{am}^2+2t_{am}(2+t_{am})\ln t_{am}}{4 ( t_{am}-1)^4}
\\
+&\frac{1}{16\pi^2}\sum_{a=1}^3\sum_{m=1}^2 \frac{\lambda_{iam}^{LQ}\lambda_{jam}^{eu\dagger} m_{u_a}}{m_{\phi_m}^2}
\frac{7-8t_{am}+t_{am}^2+2(2+t_{am})\ln t_{am}}{2 (t_{am}-1)^3}\ ,
\label{eq:sigmaL}
\\
\nonumber
\sigma_{Rij}=& \frac{1}{16\pi^2} 
\sum_{a=1}^3\sum_{m=1}^2
 \frac{\lambda_{iam}^{LQ}\lambda_{jam}^{LQ\dagger} m_{l_j} +\lambda_{iam}^{eu}\lambda_{jam}^{eu\dagger} m_{l_i} }{m_{\phi_m}^2}
\frac{1+4t_{am}-5t_{am}^2+2t_{am}(2+t_{am})\ln t_{am}}{4 ( t_{am}-1)^4}
\\
+&\frac{1}{16\pi^2}\sum_{a=1}^3\sum_{m=1}^2 \frac{\lambda_{iam}^{eu}\lambda_{jam}^{LQ\dagger} m_{u_a}}{m_{\phi_m}^2}
\frac{7-8t_{am}+t_{am}^2+2(2+t_{am})\ln t_{am}}{2 (t_{am}-1)^3}\ ,
\label{eq:sigmaR}
\end{align}
\end{subequations}
where $t_{am} =m_{u_a}^2 / m_{\phi_m}^2$ and $m_{u_a}$ is the mass of the appropriate up-type quark mass.

\mathversion{bold}
\subsection{$\mu^- \to e^-e^+e^-$}
\mathversion{normal}

In this subsection, we list the different contributions to $\mu^-\to e^- e^+ e^+$ from left- as well as right-handed couplings to the third generation of quarks
\begin{subequations}
\begin{eqnarray}
\nonumber
A_1^{L,R} &=& \sum_{a=1}^3\sum_{m=1}^2 \frac{\lambda^{LQ,eu}_{2am}\lambda^{LQ,eu\dagger}_{1am} }{ 384 \pi^2m_{\phi_m}^2} \\
 &\times& \frac{ \left(t_{am}-1\right) \left(10+\left(-17+t_{am}\right)t_{am}\right)-2\left(4-6t_{am}-t_{am}^3\right)\ln t_{am}}{(t_{am}-1)^4}\ ,\\ 
A_2^{L,R}&=&\frac{\sigma_{L21,R21}}{m_\mu}\ ,\\
F_{L,R}&=& \sum_{a=1}^3\sum_{m=1}^2
-\frac{ 3 e \lambda^{LQ,eu}_{2am}\lambda^{LQ,eu\dagger}_{1am}}{32 \pi^3 \sin\theta_W\cos\theta_W}\frac{t_{am}(1-t_{am}+\ln t_{am})}{(t_{am}-1)^2}\ ,\\
B_1^{L,R}&=& \sum_{i,j,m,n}-\frac{3}{16\pi^2 e^2} \lambda^{LQ,eu}_{2im}\lambda^{LQ,eu\dagger}_{1in} \lambda^{LQ,eu\dagger}_{1jm} \lambda^{LQ,eu}_{1jn}
D_{00}\left[ m_{\phi_m}^2, m_{\phi_n}^2, m_{u_i}^2, m_{u_j}^2 \right]\ ,\\
B_2^{L,R}&=& \sum_{i,j,m,n}-\frac{3}{16\pi^2 e^2} \lambda^{LQ,eu}_{2im} \lambda^{LQ,eu\dagger}_{1in}\lambda^{eu,LQ\dagger}_{1jm}\lambda^{eu,LQ}_{1jn}
D_{00}\left[m_{\phi_m}^2, m_{\phi_n}^2, m_{u_i}^2, m_{u_j}^2 \right]\ ,\\
B_3^{L,R}&=& \sum_{i,j,m,n}-\frac{3m_{u_i} m_{u_j}}{16\pi^2 e^2} \lambda^{eu,LQ}_{2im} \lambda^{LQ,eu\dagger}_{1in}\lambda^{LQ,eu\dagger}_{1jm}\lambda^{eu,LQ}_{1jn}
D_0\left[m_{\phi_m}^2, m_{\phi_n}^2, m_{ u_i}^2, m_{u_j}^2 \right]\ , \\
B_4^{L,R}&=& 0\ ,
\end{eqnarray}
\end{subequations}
where again $t_{am} =m_{u_a}^2 / m_{\phi_m}^2$ and $\sigma_{L,R}$ are as defined in Eq.~\ref{eq:sigmaL} and \ref{eq:sigmaR}. 

\mathversion{bold}
\subsection{$\mu\leftrightarrow e$ Conversion in Nuclei}
\mathversion{normal}
In this subsection, we list the right-handed contributions to $\mu\leftrightarrow e$ conversion in nuclei
\begin{subequations}
\begin{align}
g_{RV(d)}^{box}=&
\frac{|V_{td}|^2}{64\pi^2} \Bigg\{
2\sum_{m,n,i}
\lambda ^{eu}_{23m}\lambda ^{eu*}_{13n}\lambda^{LQ}_{i3n} \lambda ^{LQ*}_{i3m} D_{00}\left(0,m_{\phi
   _m}^2,m_{\phi _n}^2,m_t^2\right)\\\nonumber
&+\sum_m\lambda _{23m}^{eu} \lambda _{13m}^{eu*}
\Big[2 y_t^2 D_{00}\left(m_W^2,m_{\phi _m}^2,m_t^2,m_t^2\right)
-g^2 m_t^2 D_0\left(m_W^2,m_{\phi _m}^2,m_t^2,m_t^2\right)\Big]\Bigg\}\ ,
\\
g_{RV(d)}^{\gamma}=&-\frac{\alpha }{144 \pi }\sum_m\frac{\lambda ^{eu}_{23m} \lambda ^{eu*}_{13m}}{m_{\phi _m}^2 }\frac{t_{3m}^3-18 t_{3m}^2+27 t_{3m}+2 \left(t_{3m}^3+6 t_{3m}-4\right) \ln \left(t_{3m}\right)-10}{\left(t_{3m}-1\right)^4}\ ,
\\
g_{RV(d)}^{Z}=& -\frac{g^2\left(4 s_w^2-3\right)}{128 \pi ^2 m_W^2}\sum_m \lambda ^{eu}_{23m} \lambda^{eu*}_{13m}
\frac{ t_{3m} \left(t_{3m}-\ln \left(t_{3m}\right)-1\right)}{\left(t_{3m}-1\right){}^2}\ , 
\end{align}
\end{subequations}
and the corresponding up-type contributions are 
\begin{align}
g_{RV(u)}^{box}=&0\ ,
&
g_{RV(u)}^{\gamma}=-2\,g_{RV(d)}^{\gamma}\ ,&
&
g_{RV(u)}^{Z}=-\frac{8s_W^2-3}{4s_W^2-3}\,g_{RV(d)}^{Z}\ .
\end{align}

\mathversion{bold}
\section{Calculation of $I_{ij\alpha\beta}$}
\mathversion{normal}
\label{sec:appint}

In this section we evaluate the integral $I_{ij\alpha\beta}$ from Eq.~\ref{eq:loopint-exact2} in general, and in the limit of vanishing quark masses.  To do this we will partly use the results of~\cite{vanderBij:1983bw,*Ghinculov:1994sd,*McDonald:2003zj}.  Firstly we can use partial fractions to re-express the integral in terms of four simpler terms
\begin{equation}
I_{ij\alpha\beta} = \frac{1}{m_f^2} \frac{1}{t_{\alpha}-r_i} \frac{1}{t_{\beta}-r_j} \left[ \hat{I} \left( t_{\alpha}, t_{\beta} \right) - \hat{I} \left( r_i, t_{\beta} \right) - \hat{I} \left( t_{\alpha}, r_j \right) + \hat{I} \left( r_i, r_j \right) \right]\ ,
\label{eq:partialfrac}
\end{equation}
with
\begin{equation}
\hat{I} \left( s, t \right) \equiv \mu^{\epsilon} \int d^dp \int d^dq \frac{1}{p^2-s} \frac{1}{q^2-t} \frac{1}{(p+q)^2-1}\ ,
\label{eq:Idef}
\end{equation}
where we have set $q \to -q$ and introduced the dimensionful parameter $\mu$ to facilitate dimensional regularization.  In contrast to~\cite{vanderBij:1983bw,*Ghinculov:1994sd,*McDonald:2003zj}, we will not use the partial p operation of~\cite{tHooft:1972fi} to transform these integrals into less divergent expressions involving four propagators.  It turns out here that neglecting this step leads to a simpler final result.  The momentum space integration in Eq.~\ref{eq:Idef} can be performed through the introduction of Feynman parameters, yielding
\begin{equation}
\hat{I} \left( s, t \right) = \pi^4 \left( \frac{\mu}{\pi^2} \right)^{\epsilon} \Gamma \left[ -1+2\epsilon \right] \int_0^1 dx \int_0^1 dy \frac{y^{\epsilon-1}}{\left(1-x\right)^{\epsilon} x^{\epsilon}} \left( (1-y) s + \frac{y}{x} t + \frac{y}{1-x} \right)^{1-2\epsilon}\ .
\label{eq:Idefsimp}
\end{equation}
Expanding in $\epsilon$ and substituting this into Eq.~\ref{eq:partialfrac} we obtain
\begin{equation}
I_{ij\alpha\beta} = \frac{\pi^4}{m_f^2} \frac{1}{t_{\alpha}-r_i} \frac{1}{t_{\beta}-r_j} \int_0^1 dx \int_0^1 dy \left[ h \left( t_{\alpha}, t_{\beta} \right) - h \left( r_i, t_{\beta} \right) - h \left( t_{\alpha}, r_j \right) + h \left( r_i, r_j \right) \right]\ ,
\label{eq:Iwithh}
\end{equation}
where we have defined
\begin{equation}
h \left(s,t \right) \equiv \frac{sx (1-x)(1-y)+(1-x)yt+xy}{x(1-x)y} \ln \left[ sx(1-x)(1-y)+(1-x)yt+xy \right]\ .
\label{eq:hdefn}
\end{equation}
The integration with respect to $y$ leads to an integral over a sum of dilogarithms
\begin{equation}
I_{ij\alpha\beta} = \frac{\pi^4}{m_f^2} \frac{1}{t_{\alpha}-r_i} \frac{1}{t_{\beta}-r_j} \left[ -g \left( t_{\alpha}, t_{\beta} \right) + g \left( r_i, t_{\beta} \right) + g \left( t_{\alpha}, r_j \right) - g \left( r_i, r_j \right) \right]\ ,
\label{eq:Iwithg}
\end{equation}
where
\begin{equation}
g \left(s,t \right) \equiv s \int_0^1 dx \operatorname{Li}_2 \left(1-\mu^2 \right)~~\textrm{with}~~s \mu^2 \equiv \frac{1}{(1-x)}+\frac{t}{x}\ .
\label{eq:gdefn}
\end{equation}
Using the results in~\cite{vanderBij:1983bw,*Ghinculov:1994sd,*McDonald:2003zj} the integral over $x$ can be performed to obtain:
\begin{multline}
g \left(s,t\right) = \frac{s}{2} \ln s \ln \frac{t}{s} + \sum \limits_{\pm} \pm \frac{s(1-s)+3st+2(1-t)x_{\pm}}{2w} \\
\times \left[ \operatorname{Li}_2 \left( \frac{x_{\pm}}{x_{\pm}-s} \right) - \operatorname{Li}_2 \left( \frac{x_{\pm}-s}{x_{\pm}} \right) + \operatorname{Li}_2 \left( \frac{t-1}{x_{\pm}} \right) - \operatorname{Li}_2 \left( \frac{t-1}{x_{\pm}-s} \right) \right]\ ,
\label{eq:geval}
\end{multline}
with
\begin{align}
x_{\pm}&=\frac12(-1+s+t\pm w)\ ,&
w&=\sqrt{1+s^2+t^2-2(s+t+s t)}\ .
\label{eq:xwdefn}
\end{align}
Looking back at Eq.~\ref{eq:Iwithg}, it can be seen that terms depending only on one ratio $t_{i,j,\alpha,\beta}$ will vanish.  Therefore, we can replace the function $g$ in Eq.~\ref{eq:Iwithg} with the slightly simplified function
\begin{multline}
\hat{g} \left(s,t\right) \equiv g(s,t) + \frac{s}{2} \ln^2s = \frac{s}{2} \ln s \ln t + \sum \limits_{\pm} \pm \frac{s(1-s)+3st+2(1-t)x_{\pm}}{2w} \\
\times \left[ \operatorname{Li}_2 \left( \frac{x_{\pm}}{x_{\pm}-s} \right) - \operatorname{Li}_2 \left( \frac{x_{\pm}-s}{x_{\pm}} \right) + \operatorname{Li}_2 \left( \frac{t-1}{x_{\pm}} \right) - \operatorname{Li}_2 \left( \frac{t-1}{x_{\pm}-s} \right) \right]\ .
\label{eq:ghateval}
\end{multline}
The combination of Eq.~\ref{eq:Iwithg} with $g\to \hat{g}$ and~\ref{eq:ghateval} provides the full analytic solution to $I_{ij\alpha\beta}$.  Nevertheless it is also useful to evaluate the integral in the case of vanishing $r_i$ and $r_j$, as the down-type quark masses are much lighter than the leptoquark and color octet masses.  Thus we also present the integral in this limit
\begin{equation}
I_{\alpha \beta} \equiv I_{00\alpha\beta} = \frac{\pi^4}{m_f^2} \frac{\hat{g} \left( t_{\alpha}, 0 \right) - \hat{g} \left( t_{\alpha}, t_{\beta} \right)}{t_{\alpha} t_{\beta}}\ ,
\label{eq:Inoquarkm}
\end{equation}
where the limit $\hat g(s,t)\stackrel{t\to0}{\longrightarrow}\hat g(s,0)$ is most easily obtained by taking the limit $t \to 0$ before performing the $x$-integration
\begin{equation}
\hat{g} \left(s,0 \right) = -s \frac{\pi^2}{6} - (1-s) \ln s \ln (1-s) - (1-s) \operatorname{Li}_2 (s)\ .
\label{eq:gnoquarkm}
\end{equation}
In order to verify our result, we compared each expression with the numerical integration of the integral Eq.~\ref{eq:loopint-exact2}.

\section{Analytic Expressions for certain Passarino-Veltman Integrals}
\label{sec:appPV}

Here we give analytic expressions for the N-point functions used in the calculation of rare processes in the main text.  All of the integrals are calculated in limit of zero external momentum.  For the definition of the integrals, see~\cite{Denner:1991kt,Hahn:1998yk}. The two-point functions and their derivatives are as follows
\begin{subequations}
\begin{align}
B_0[a_2]=&\Delta_\epsilon + 1-\frac{t \ln t}{t-1}\ ,\\
B_1[a_2]=&-\frac{1}{2}\Delta_\epsilon+\frac{1}{2}\ln\frac{m_2^2}{\mu^2} +\frac{-3 t^2+2 t^2 \ln t+4 t-1}{4 (t-1)^2}\ ,\\
m_2^{-2} B_{00}[a_2]=&\frac{1}{4}(t+1)\Delta_\epsilon+\frac{3-3 t^2+2 t^2 \ln t}{8(1- t)}\ ,\\
B_{11}[a_2]=&\frac{1}{3}\Delta_\epsilon+ \frac{11 t^3-6 t^3 \ln t-18 t^2+9 t-2}{18 (t-1)^3}\ ,
\end{align}
\end{subequations}
where $a_2= \{m_1^2,m_2^2\}$, $t=\frac{m_1^2}{m_2^2}$ and as we use dimensional regularization,
\begin{equation}
\Delta_{\epsilon} \equiv \frac{2}{\epsilon} - \gamma_E + \ln 4 \pi \ .
\end{equation}
Note that $B_{00}$ is defined in the units $m_2^2$, whereas the remaining two-point functions are dimensionless. The derivatives of the $B$s are
\begin{subequations}
\begin{align}
m_2^2\frac{\partial B_0[a_2]}{\partial p^2}=&\frac{t \left(t^2-2 t \ln t-1\right)}{2 (t-1)^3}\ ,\\
m_2^2\frac{\partial B_1[a_2]}{\partial p^2}=&-\frac{t \left(t^3-6 t^2+3 t+6 t \ln t+2\right)}{6 (t-1)^4}\ ,
\end{align}
\end{subequations}
and both carry units of $m_2^{-2}$. 

The three-point functions are given by
\begin{subequations}
\begin{align}
m_3^2 C_{0}[a_3] =&-\frac{t_1 \ln t_1}{\left(t_1-1\right) \left(t_1-t_2\right)}
+\frac{t_2 \ln t_2}{\left(t_1-t_2\right) \left(t_2-1\right)}\ ,\\
m_3^2 C_{1}[a_3]=&-\frac{t_2}{2 \left(t_1-t_2\right) \left(t_2-1\right)}
+\frac{t_1^2 \ln t_1}{2 \left(t_1-1\right) \left(t_1-t_2\right)^2}-\frac{t_2 \left(t_1 \left(t_2-2\right)+t_2\right) \ln t_2}{2
   \left(t_1-t_2\right)^2 \left(t_2-1\right)^2}\ , \\
m_3^2 C_{2}[a_3]=&-\frac{t_2-t_1}{2 \left(t_1-1\right) \left(t_1-t_2\right) \left(t_2-1\right)}
+\frac{t_1^2 \ln t_1}{2 \left(t_1-1\right)^2 \left(t_1-t_2\right)}   
-\frac{t_2^2 \ln t_2}{2 \left(t_1-t_2\right) \left(t_2-1\right)^2}\ ,\\
C_{00}[a_3]=& \frac{1}{4}\left(\Delta_\epsilon -\ln\frac{m_3^2}{\mu^2}\right)
+\frac{3}{8}+-\frac{t_1^2 \ln t_1}{4 \left(t_1-1\right) \left(t_1-t_2\right)}+
\frac{t_2^2 \ln t_2}{4 \left(t_1-t_2\right) \left(t_2-1\right)}\ ,\\
\nonumber
m_3^2 C_{11}[a_3]=&\frac{t_2 \bigg(\left(3 t_2-5\right) t_1^2-4 \left(t_2-2\right) t_2
   t_1+\left(t_2-3\right) t_2^2\bigg)}{6 \left(t_1-t_2\right){}^3
   \left(t_2-1\right)^2}\\
   &-\frac{t_1^3 \ln t_1}{3 \left(t_1-1\right) \left(t_1-t_2\right)^3}
   +\frac{t_2 \bigg(\left(t_2^2-3 t_2+3\right) t_1^2+\left(t_2-3\right) t_2
   t_1+t_2^2\bigg) \ln t_2}{3 \left(t_1-t_2\right)^3
   \left(t_2-1\right)^3}\ ,\\
\nonumber
m_3^2 C_{12}[a_3]=&\frac{\left(t_2^2+1\right) t_1^2-t_2 \left(t_2^2+t_2+2\right) t_1+t_2^2
   \left(t_2+1\right)}{6 \left(t_1-1\right) \left(t_1-t_2\right){}^2
   \left(t_2-1\right){}^2}\\
   &-\frac{t_1^3 \ln \left(t_1\right)}{6 \left(t_1-1\right){}^2 \left(t_1-t_2\right){}^2}
   +\frac{t_2^2 \left(t_1 \left(t_2-3\right)+2 t_2\right) \ln \left(t_2\right)}{6
   \left(t_1-t_2\right){}^2 \left(t_2-1\right){}^3}\ ,\\
 \nonumber
m_3^2  C_{22}[a_3]=&\frac{\left(3-5 t_2\right) t_1^2+\left(5 t_2^2-1\right) t_1-3 t_2^2+t_2}{6
   \left(t_1-1\right)^2 \left(t_1-t_2\right) \left(t_2-1\right)^2}
   -\frac{t_1^3 \ln t_1}{3 \left(t_1-1\right){}^3 \left(t_1-t_2\right)}  \\
&+\frac{t_2^3 \ln t_2}{3 \left(t_1-t_2\right) \left(t_2-1\right)^3}\ ,
\end{align}
\end{subequations}
where $a_3= \{m_1^2, m_2^2, m_3^2 \}$ and $t_{1,2}=\frac{m_{1,2}^2}{m_3^2}$.  $C_{00}$ has mass dimension 0, whilst the other three-point functions are defined with units of $m_3^{-2}$.

The calculation of the $\gamma$-penguin used in the text also requires the derivatives of several of the three-point functions.  Here we list the analytic expressions for those used:
\begin{subequations}
\begin{align}
\nonumber
m_3^4\frac{\partial C_0[a_3]}{\partial p_1^2}=&-\frac{-2 t_2 t_1^2+t_1^2+\left(2 t_1-1\right) t_2^2}{2 \left(t_1-1\right)
   \left(t_1-t_2\right)^3 \left(t_2-1\right)}\\
   =&-\frac{t_1 \left(t_1^2+t_2 t_1-2 t_2\right) \ln t_1}{2
   \left(t_1-1\right)^2 \left(t_1-t_2\right){}^3}
   +\frac{t_2 \left(t_2^2+t_1 \left(t_2-2\right)\right) \ln t_2}{2
   \left(t_1-t_2\right)^3 \left(t_2-1\right)^2}\ ,\\
\nonumber
m_3^4\frac{\partial C_0[a_3]}{\partial p_2^2}=&-\frac{\left(t_2+1\right) t_1^2-t_2 \left(t_2+3\right) t_1+2 t_2^2}{2
   \left(t_1-1\right) \left(t_1-t_2\right)^2 \left(t_2-1\right)^2}\\
   &+\frac{t_1^2 \ln t_1}{2 \left(t_1-1\right)^2 \left(t_1-t_2\right)^2}
   -\frac{t_2 \left(t_2^2+t_2-2 t_1\right) \ln t_2}{2
   \left(t_1-t_2\right)^2 \left(t_2-1\right)^3}\ ,\\
\nonumber   
m_3^4\frac{\partial C_0[a_3]}{\partial p_{12}^2}=&
\frac{\left(1-t_2\right) \left(-\left(t_2-2\right) t_1^2+\left(t_2-3\right) t_2
   t_1+t_2^2\right)}{2 \left(t_1-1\right){}^2 \left(t_1-t_2\right){}^2
   \left(t_2-1\right){}^2}\\
&  -\frac{t_1 \left(t_1^2+t_1-2 t_2\right) \ln t_1}{2 \left(t_1-1\right)^3
   \left(t_1-t_2\right)^2}
 +\frac{t_2^2 \ln t_2}{2 \left(t_1-t_2\right)^2 \left(t_2-1\right)^2}\ ,  \\
 \nonumber
m_3^4 \frac{\partial C_1[a_3]}{\partial p_1^2} =&
   \frac{\left(-5 t_2^2+9 t_2-2\right) t_1^2-t_2 \left(t_2^2-2 t_2+5\right) t_1+t_2^2
   \left(t_2+1\right)}{6 \left(t_1-1\right) \left(t_1-t_2\right){}^3
   \left(t_2-1\right){}^2}\\\
  \nonumber 
   &+\frac{t_1^2 \left(t_1^2+2 t_2 t_1-3 t_2\right) \ln t_1}{3
   \left(t_1-1\right)^2 \left(t_1-t_2\right)^4}\\
  & -\frac{t_2 \left(t_2^3+2 t_1 \left(t_2-2\right) t_2^2+t_1^2 \left(t_2^2-3
   t_2+3\right)\right) \ln t_2}{3 \left(t_1-t_2\right)^4
   \left(t_2-1\right)^3}\ ,\\
\nonumber   
m_3^4 \frac{\partial C_1[a_3]}{\partial p_2^2} =&
   \frac{\left(t_2^2-5 t_2-2\right) t_1^2+t_2 \left(t_2^2+2 t_2+9\right) t_1-t_2^2
   \left(t_2+5\right)}{6 \left(t_1-1\right) \left(t_1-t_2\right){}^2
   \left(t_2-1\right){}^3}\\
 \nonumber  
   & -\frac{t_1^3 \ln t_1}{3 \left(t_1-1\right)^2 \left(t_1-t_2\right)^3}\\
   &+\frac{t_2 \left(3 t_1^2+t_2 \left(t_2^2-4 t_2-3\right) t_1+t_2^2 \left(2
   t_2+1\right)\right) \ln t_2}{3 \left(t_1-t_2\right)^3
   \left(t_2-1\right)^4}\ ,\\
 \nonumber  
m_3^4 \frac{\partial C_1[a_3]}{\partial p_{12}^2} =&
 -\frac{\left(t_2^2-t_2+1\right) t_1^3-t_2 \left(t_2^2+2\right) t_1^2+t_2^2
   \left(t_2+2\right) t_1-t_2^3}{3 \left(t_1-1\right)^2 \left(t_1-t_2\right)^3
   \left(t_2-1\right)^2}\\
   &+\frac{t_1^2 \left(t_1^2+\left(t_2+1\right) t_1-3 t_2\right) \ln t_1}{6
   \left(t_1-1\right)^3 \left(t_1-t_2\right)^3}
   -\frac{t_2^2 \left(t_1 \left(t_2-3\right)+t_2 \left(t_2+1\right)\right) \ln
      t_2}{6 \left(t_1-t_2\right)^3 \left(t_2-1\right)^3}\ ,\\
\nonumber      
m_3^4 \frac{\partial C_2[a_3]}{\partial p_1^2} =&
   \frac{\left(-t_2^2+t_2-1\right) t_1^2+t_2 \left(t_2+1\right) t_1-t_2^2}{3
   \left(t_1-1\right){}^2 \left(t_1-t_2\right){}^2 \left(t_2-1\right){}^2}\\
   &+\frac{t_1^2 \left(t_1^2+\left(t_2+1\right) t_1-3 t_2\right) \ln t_1}{6
   \left(t_1-1\right)^3 \left(t_1-t_2\right)^3}
   -\frac{t_2^2 \left(t_1 \left(t_2-3\right)+t_2 \left(t_2+1\right)\right) \ln t_2}{6 \left(t_1-t_2\right)^3 \left(t_2-1\right)^3}\ ,\\
  \nonumber 
m_3^4 \frac{\partial C_2[a_3]}{\partial p_2^2} =&\frac{\left(2 t_2^2+5 t_2-1\right) t_1^2-\left(9 t_2^2+2 t_2+1\right) t_1+t_2 \left(5
   t_2+1\right)}{6 \left(t_1-1\right)^2 \left(t_1-t_2\right) \left(t_2-1\right)^3}\\
   &-\frac{t_1^3 \ln t_1}{3 \left(t_1-1\right){}^3 \left(t_1-t_2\right)^2}
   +\frac{t_2^2 \left(t_2 \left(t_2+2\right)-3 t_1\right) \ln t_2}{3
   \left(t_1-t_2\right)^2 \left(t_2-1\right)^4}\ ,\\
   \nonumber
m_3^4 \frac{\partial C_2[a_3]}{\partial p_{12}^2} =&
\frac{\left(-2 t_2^2+9 t_2-5\right) t_1^2+\left(-5 t_2^2+2 t_2-1\right) t_1+t_2
   \left(t_2+1\right)}{6 \left(t_1-1\right)^3 \left(t_1-t_2\right)
   \left(t_2-1\right)^2}\\
 & +\frac{t_1^2 \left(t_1^2+2 t_1-3 t_2\right) \ln t_1}{3
   \left(t_1-1\right)^4 \left(t_1-t_2\right)^2}
   -\frac{t_2^3 \ln t_2}{3 \left(t_1-t_2\right)^2 \left(t_2-1\right)^3}\ ,\\
\nonumber
 m_3^2\frac{\partial C_{00}[a_3]}{\partial p_1^2} =&
 \frac{\left(t_2-1\right) t_1^2+t_2^2 t_1-t_2^2}{12 \left(t_1-1\right)
   \left(t_1-t_2\right)^2 \left(t_2-1\right)}
   -\frac{t_1^2 \left(t_1 \left(2 t_2+1\right)-3 t_2\right) \ln t_1}{12
   \left(t_1-1\right)^2 \left(t_1-t_2\right)^3}\\
   &  +\frac{t_2^2 \left(t_2+t_1 \left(2 t_2-3\right)\right) \ln t_2}{12
   \left(t_1-t_2\right)^3 \left(t_2-1\right)^2}\ ,\\
  \nonumber
m_3^2 \frac{\partial C_{00}[a_3]}{\partial p_2^2} =&
 \frac{t_2 \left(t_2+1\right)-t_1 \left(t_2^2+1\right)}{12 \left(t_1-1\right)
   \left(t_1-t_2\right) \left(t_2-1\right)^2}
   +\frac{t_1^3 \ln t_1}{12 \left(t_1-1\right)^2 \left(t_1-t_2\right)^2}\\
&   -\frac{t_2^2 \left(t_1 \left(t_2-3\right)+2 t_2\right) \ln t_2}{12
   \left(t_1-t_2\right)^2 \left(t_2-1\right)^3}\ ,\\
  \nonumber
m_3^2 \frac{\partial C_{00}[a_3]}{\partial p_{12}^2} =& 
 \frac{\left(t_2-1\right) t_1^2-t_1+t_2}{12 \left(t_1-1\right)^2 \left(t_1-t_2\right)
   \left(t_2-1\right)}-\frac{t_1^2 \left(t_1 \left(t_2+2\right)-3 t_2\right) \ln t_1}{12
   \left(t_1-1\right)^3 \left(t_1-t_2\right)^2}\\ 
   &+\frac{t_2^3 \ln t_2}{12 \left(t_1-t_2\right)^2 \left(t_2-1\right)^2}\ ,
\end{align}
\end{subequations}
where the derivatives of $C_{00}$ carry units of $m_3^{-2}$ and the others carry $m_3^{-4}$.

Finally the analytical expressions for the four-point functions are the following:
\begin{subequations}
\begin{align}
\nonumber
m_4^{4}D_{0}[a_4]   =&   
-\frac{t_1 \ln t_1}{\left(t_1-1\right) \left(t_1-t_2\right)
   \left(t_1-t_3\right)}
+\frac{t_2 \ln t_2}{\left(t_1-t_2\right) \left(t_2-1\right)
   \left(t_2-t_3\right)}   \\
   &
 -\frac{t_3 \ln t_3}{\left(t_1-t_3\right) \left(t_2-t_3\right)
   \left(t_3-1\right)}\ ,  \\
\nonumber
m_4^{4}D_1[a_4]   =&  
-\frac{t_2}{2 \left(t_1-t_2\right) \left(t_2-1\right) \left(t_2-t_3\right)}
+\frac{t_1^2 \ln t_1}{2 \left(t_1-1\right) \left(t_1-t_2\right)^2 \left(t_1-t_3\right)}\\
\nonumber
 & +\frac{t_2 \Big(-t_2^3+t_3 t_2+t_1 \big(t_2 \left(t_3+1\right)-2 t_3\big)\Big)\ln t_2}
    {2 \left(t_1-t_2\right)^2 \left(t_2-1\right)^2\left(t_2-t_3\right)^2}   \\
& -\frac{t_3^2 \ln t_3}{2 \left(t_1-t_3\right) \left(t_2-t_3\right)^2 \left(t_3-1\right)}\ ,\\
\nonumber
m_4^{4}D_2 [a_4] =&
\frac{t_3}{2 \left(t_1-t_3\right) \left(t_2-t_3\right) \left(t_3-1\right)}
+\frac{t_1^2 \ln t_1}{2 \left(t_1-1\right) \left(t_1-t_2\right)
   \left(t_1-t_3\right)^2}\\
   \nonumber
& -\frac{t_2^2 \ln t_2}{2 \left(t_1-t_2\right) \left(t_2-1\right)
   \left(t_2-t_3\right)^2}   \\
&   
+\frac{t_3 \Big(t_1 \big(t_2 \left(t_3-2\right)+t_3\big)+t_3
   \left(t_2-t_3^2\right)\Big) \ln t_3}{2 \left(t_1-t_3\right)^2
   \left(t_2-t_3\right){}^2 \left(t_3-1\right)^2}\ ,   \\
\nonumber
m_4^{4}D_3[a_4]   =&
\frac{t_3-t_2}{2 \left(t_1-1\right) \left(t_2-1\right) \left(t_2-t_3\right)
   \left(t_3-1\right)}
+\frac{t_1^2 \ln t_1}{2 \left(t_1-1\right)^2 \left(t_1-t_2\right)
   \left(t_1-t_3\right)}\\
&-\frac{t_2^2 \ln t_2}{2 \left(t_1-t_2\right) \left(t_2-1\right)^2
   \left(t_2-t_3\right)}
-\frac{t_3^2 \ln t_3}{2 \left(t_1-t_3\right) \left(t_3-1\right)^2
   \left(t_3-t_2\right)}\ ,\\      
\nonumber
m_4^{2}D_{00}[a_4]   =&
-\frac{t_1^2 \ln t_1}{4 \left(t_1-1\right) \left(t_1-t_2\right)
   \left(t_1-t_3\right)}
+\frac{t_2^2 \ln t_2}{4 \left(t_1-t_2\right) \left(t_2-1\right)
   \left(t_2-t_3\right)}\\   
&+\frac{t_3^2 \ln t_3}{4 \left(t_1-t_3\right) \left(t_3-1\right)
   \left(t_3-t_2\right)}\ , \\
\nonumber   
m_4^{4}D_{11}[a_4]   =& 
     \Bigg(\left(t_2^2-3 \left(t_3+1\right) t_2+5 t_3\right)
   t_1^2+\left(t_2^3+t_2^2+3 t_3^2 t_2-5 t_3^2\right) t_1\\
\nonumber   
   &\;\;\;\;-t_2 t_3 \left(t_2^2+t_3 t_2+t_2-3 t_3\right)\Bigg)
    \times \frac{t_2} {6 \left(t_1-t_2\right)^2 \left(t_2-1\right)^2
   \left(t_1-t_3\right) \left(t_2-t_3\right)^2}\\
\nonumber  
 &-\frac{t_1^3 \ln t_1}{3 \left(t_1-1\right) \left(t_1-t_2\right)^3
   \left(t_1-t_3\right)}\\
\nonumber   
& +  \Bigg(\bigg(\left(t_3^2+t_3+1\right) t_2^2-3 t_3 \left(t_3+1\right) t_2+3
   t_3^2\bigg) t_1^2 \\
\nonumber   
&\;\;\;\; + t_2 \left(t_2^4-3 \left(t_3+1\right) t_2^3+6 t_3 t_2^2+t_3
   \left(t_3+1\right) t_2-3 t_3^2\right) t_1\\
\nonumber   
&\;\;\;\;+t_2^2 \left(\left(t_3+1\right) t_2^3-3 t_3
   t_2^2+t_3^2\right)\Bigg) \times  \frac{t_2 \ln t_2}{3 \left(t_1-t_2\right)^3
   \left(t_2-1\right)^3 \left(t_2-t_3\right)^3} \\
& +\frac{t_3^3 \ln t_3}{3 \left(t_1-t_3\right) \left(t_3-1\right)
   \left(t_3-t_2\right)^3}\ ,  \\
\nonumber  
m_4^{4}D_{12} [a_4]   =&   
\frac{t_2 t_3 \left(-2 t_3 t_2+t_2+t_3\right)+t_1 \left(\left(t_3-1\right) t_2^2+t_3^2
   t_2-t_3^2\right)}{6 \left(t_1-t_2\right) \left(t_2-1\right) \left(t_1-t_3\right)
   \left(t_2-t_3\right)^2 \left(t_3-1\right)}\\
\nonumber  
&-\frac{t_1^3 \ln t_1}{6 \left(t_1-1\right) \left(t_1-t_2\right)^2
   \left(t_1-t_3\right)^2}\\
\nonumber      
& +\frac{t_2^2 \Big(t_2 \left(t_2^2+t_3 t_2-2 t_3\right)-t_1 \left(t_2 \left(2
   t_3+1\right)-3 t_3\right)\Big) \ln t_2}{6 \left(t_1-t_2\right)^2
   \left(t_2-1\right)^2 \left(t_2-t_3\right)^3}  \\
&+\frac{t_3^2 \Bigg(t_3 \left(t_3^2+t_2 \left(t_3-2\right)\right)-t_1 \bigg(t_3+t_2
   \left(2 t_3-3\right)\bigg)\Bigg) \ln t_3 }{6 \left(t_1-t_3\right)^2
   \left(t_3-1\right)^2 \left(t_3-t_2\right)^3}\ ,  \\
\nonumber
m_4^{4}D_{13}[a_4] =&    
\frac{t_1 \left(\left(t_3-1\right) t_2^2-t_2+t_3\right)-t_2 \left(t_2
   \left(t_3-2\right)+t_3\right)}{6 \left(t_1-1\right) \left(t_1-t_2\right)
   \left(t_2-1\right)^2 \left(t_2-t_3\right) \left(t_3-1\right)}\\
\nonumber
&-\frac{t_1^3 \ln t_1}{6 \left(t_1-1\right)^2 \left(t_1-t_2\right)^2
   \left(t_1-t_3\right)}   \\
\nonumber
&+\frac{t_2^2 \left(t_2 \left(t_2^2+t_2-2 t_3\right)-t_1 \left(t_2 \left(t_3+2\right)-3
   t_3\right)\right) \ln t_2}{6 \left(t_1-t_2\right)^2
   \left(t_2-1\right)^3 \left(t_2-t_3\right)^2}   \\
&+\frac{t_3^3 \ln t_3}{6 \left(t_1-t_3\right) \left(t_2-t_3\right)^2
   \left(t_3-1\right)^2}\ ,   \\
\nonumber
m_4^{4}D_{22}[a_4] =&
-\frac{t_3 \left(t_3-t_2\right) \Bigg(t_1 \left(t_2 \left(5-3
   t_3\right)+\left(t_3-3\right) t_3\right)+t_3 \Big(t_2 \left(t_3-3\right)+t_3
   \left(t_3+1\right)\Big)\Bigg) }{6 \left(t_1-t_3\right)^2
   \left(t_2-t_3\right)^3 \left(t_3-1\right)^2}\\
\nonumber
&-\frac{t_1^3 \ln t_1}{3 \left(t_1-1\right) \left(t_1-t_2\right)
   \left(t_1-t_3\right)^3}   
   +\frac{t_2^3 \ln t_2}{3 \left(t_1-t_2\right) \left(t_2-1\right)
   \left(t_2-t_3\right)^3}\\
 \nonumber  
&   \Bigg(\Big(\left(t_3^2-3 t_3+3\right) t_2^2+\left(t_3-3\right) t_3
   t_2+t_3^2\Big) t_1^2\\
 \nonumber  
& \;\;\;\;+ t_3 \left(\left(t_3-3\right) t_3^3+t_2^2
   \left(t_3-3\right)+t_2 \left(-3 t_3^3+6 t_3^2+t_3\right)\right) t_1\\
&\;\;\;\;+t_3^2
   \left(t_3^3+t_2 \left(t_3-3\right) t_3^2+t_2^2\right)\Bigg) \times
   \frac{t_3 \ln t_3}{3 \left(t_1-t_3\right)^3 \left(t_3-1\right)^3
   \left(t_3-t_2\right)^3}\ , \\
\nonumber   
m_4^{4}D_{23}[a_4]=&
-\frac{t_1 \left(t_2 \left(t_3^2+1\right)-t_3 \left(t_3+1\right)\right)-t_3 \left(t_3
   t_2+t_2-2 t_3\right)}{6 \left(t_1-1\right) \left(t_2-1\right) \left(t_1-t_3\right)
   \left(t_2-t_3\right) \left(t_3-1\right)^2}\\
 \nonumber
&-\frac{t_1^3 \ln t_1}{6 \left(t_1-1\right){}^2 \left(t_1-t_2\right)
   \left(t_1-t_3\right)^2}  
  + \frac{t_2^3 \ln t_2}{6 \left(t_1-t_2\right) \left(t_2-1\right)^2
   \left(t_2-t_3\right)^2}\\
  &+ \frac{t_3^2 \Big(t_3 \left(t_3^2+t_3-2 t_2\right)-t_1 \left(t_2 \left(t_3-3\right)+2
   t_3\right)\Big) \ln t_3}{6 \left(t_1-t_3\right)^2
   \left(t_2-t_3\right)^2 \left(t_3-1\right)^3}\ ,\\
\nonumber
m_4^{4}D_{33}[a_4]=&
-\frac{\left(t_3-t_2\right) \Big(-3 t_3 t_2+t_2+t_3+t_1 \left(-3 t_3+t_2 \left(5
   t_3-3\right)+1\right)+1\Big)}{6 \left(t_1-1\right){}^2 \left(t_2-1\right)^2
   \left(t_2-t_3\right) \left(t_3-1\right)^2} \\
\nonumber   
&-\frac{t_1^3 \ln t_1}{3 \left(t_1-1\right)^3 \left(t_1-t_2\right)
   \left(t_1-t_3\right)}
   +\frac{t_2^3 \ln t_2}{3 \left(t_1-t_2\right) \left(t_2-1\right)^3
   \left(t_2-t_3\right)}\\
&+\frac{t_3^3 \ln t_3}{3 \left(t_1-t_3\right) \left(t_3-1\right)^3
   \left(t_3-t_2\right)}\ ,
\end{align}
\end{subequations}
where $a_4 = \{m_1^2, m_2^2, m_3^4, m_4^2\}$ and $t_i = \frac{m_i^2}{m_4^2}$ for $i=1,2,3$.  Here $D_{00}$ is defined with units of $m_4^{-2}$ and all other four-point functions carry units of $m_4^{-4}$.

\section{Effective Operator Approach to Neutrino Masses}
\label{app:decoupling}
In terms of an effective field theory language using dimensional regularization and $\overline{\text{MS}}$, the dominant contribution to neutrino masses does not originate from $\mathcal{O}_{11b}$, but from $\mathcal{O}_1$. We will argue this in the following. Assuming that $m_f\sim m_\phi\sim M\gg v$, we can match the full theory to the effective theory at the scale $M$. In particular, we generate a dimension-5 operator $\mathcal{O}_1$ by matching at two loop and the operator $\mathcal{O}_{11b}$ via tree-level matching. The Wilson coefficients $C_1$ ($C_{11b}$) of the two operators $\mathcal{O}_{1}$ ($\mathcal{O}_{11b}$) are given by
\begin{subequations}
\begin{align}
C_1&=4 \frac{m_f y_b^2}{(2\pi)^8} \sum_{\alpha,\beta=1}^{N_\phi}
\left(\lambda_{i3\alpha}^{LQ} \lambda_{3\alpha}^{df} \right)
\left( I_{\alpha\beta} \right)
\left(\lambda_{j3\beta}^{LQ} \lambda_{3\beta}^{df}  \right)\ ,\\
C_{11b}&= \sum_{\alpha,\beta=1}^{N_\phi} \frac{y_b^2}{m_{\phi_\alpha}^2 m_{\phi_\beta}^2 m_f}
\left(\lambda_{i3\alpha}^{LQ} \lambda_{3\alpha}^{df} \right)
\left(\lambda_{j3\beta}^{LQ} \lambda_{3\beta}^{df}  \right)\ .
\end{align}
\end{subequations}
After EW symmetry breaking, $\mathcal{O}_{11b}$ also contributes to neutrino masses, but its contribution is suppressed compared to the contribution of $\mathcal{O}_1$ by the bottom quark Yukawa coupling squared
\begin{equation}
m_\nu=\left(C_1 + C_{11b} \hat A_0(m_b)^2 \right) v^2\ ,
\end{equation}
where $\hat A_0$ denotes the finite part of $A_0$. Hence the ratio of the two contributions is of the order of 
\begin{equation}
\frac{C_{11b}\hat A_0^2(m_b)}{C_1}\sim \frac{m_b^4}{m_\phi^4}\ ,
\end{equation}
and the contribution of $\mathcal{O}_{11b}$ to neutrino masses can be safely neglected for $M>v$, since $y_b\ll 1$. Relaxing the assumption that $m_f\sim m_\phi$, we still arrive at the result that $\mathcal{O}_{11b}$ can be neglected compared to $\mathcal{O}_1$. However the resulting Wilson coefficients receive additional corrections due to running between the different scales $m_\phi$ and $m_f$.

Using a cutoff regularization scheme, the estimate of the different contributions to neutrino masses is different, since $A_0\sim M^2$ and the effective operator $\mathcal{O}_{11b}$ can be the dominant contribution to neutrino mass. 

A similar discussion applies to other $\Delta L=2$ operators. In fact, it is straightforward to generalize the argument to all $\Delta L=2$ operators.
\bibliography{draft}

\end{document}